\documentclass[a4paper,11pt]{article}
\pdfoutput=1
\usepackage{heppub}
\usepackage{amsthm}
\usepackage{amsmath}
\usepackage{bbm}
\usepackage{graphicx}
\usepackage{slashed}
\usepackage{amssymb}
\usepackage{footmisc}
\usepackage{enumitem}
\usepackage{xcolor}

\newcommand*\diff{d}

\newcommand{\nn}{\nonumber}

\newcommand{\be}{\begin{eqnarray}}
\newcommand{\ee}{\end{eqnarray}}
\newcommand{\ma}{\mathrm}
\newcommand{\ml}{\mathcal}
\newcommand{\T}{\mathcal{T}}
\newcommand{\bs}{\boldsymbol}

\newcommand{\W}{\mathcal{W}}

\DeclareMathOperator{\tr}{Tr}

\preprint{MIT-CTP/5545, IQuS@UW-21-049}

\title{Chromoelectric field correlator for quarkonium transport in the strongly coupled $\mathcal{N}=4$ Yang-Mills plasma from AdS/CFT}
\author[a]{Govert Nijs}
\author[a]{Bruno Scheihing-Hitschfeld}
\author[a,b]{and Xiaojun Yao}

\affiliation[a]{Center for Theoretical Physics, Massachusetts Institute of Technology, Cambridge, MA  02139, USA}

\affiliation[b]{InQubator for Quantum Simulation, Department of Physics, University of Washington, Seattle, WA 98195, USA}

\abstract{Previous studies have shown that a gauge-invariant correlation function of two chromoelectric fields connected by a straight timelike adjoint Wilson line encodes crucial information about quark-gluon plasma (QGP) that determines the dynamics of small-sized quarkonium in the medium. Motivated by the successes of holographic calculations to describe strongly coupled QGP, we calculate the analog gauge-invariant correlation function in strongly coupled $\mathcal{N}=4$ supersymmetric Yang-Mills theory at finite temperature by using the AdS/CFT correspondence. Our results indicate that the transition processes between bound and unbound quarkonium states are suppressed in strongly coupled plasmas, and moreover, the leading contributions to these transition processes vanish in both the quantum Brownian motion and quantum optical limits of open quantum system approaches to quarkonia. 
}

\emailAdd{govert@mit.edu}
\emailAdd{bscheihi@mit.edu}
\emailAdd{xjyao@uw.edu}

\begin{document}

\maketitle

\section{Introduction}
Heavy quarkonium is a bound state containing a heavy quark-antiquark pair ($Q\bar{Q}$)\@. Due to the large heavy quark mass, the spectra of the ground and the first few excited quarkonium states can be studied by solving a Schr\"odinger equation with a potential model such as the Cornell potential~\cite{PhysRevLett.34.369}\@. When a quarkonium state is placed inside a hot and/or dense nuclear environment, such as the quark-gluon plasma (QGP) produced in heavy ion collisions, the attractive potential can be significantly screened. As a result, at high temperatures a heavy quark-antiquark pair can no longer form a bound state, i.e., quarkonium melts~\cite{Matsui:1986dk,Karsch:1987pv}.

This static screening picture has motivated using quarkonium as a signature of the QGP formation in heavy ion collisions and, more generally, as a probe of the QGP\@. However, dynamical processes such as dissociation~\cite{Laine:2006ns,Beraudo:2007ky,Brambilla:2008cx,Brambilla:2010vq,Brambilla:2011sg,Brambilla:2013dpa} and recombination~\cite{Thews:2000rj,Andronic:2003zv,Andronic:2007bi} complicate the description of quarkonium inside the QGP\@. At the lowest order in a weak coupling picture, dissociation occurs when a quarkonium state absorbs a gluon from the medium and the bound state is excited to the continuum~\cite{Brambilla:2011sg}, while recombination happens when an unbound $Q\bar{Q}$ pair radiates out a gluon that takes away enough energy so that the unbound pair forms a bound state~\cite{Yao:2018sgn}\@. Many studies have been devoted to calculate these dynamical processes in a weak coupling picture, which can be dated back to the early work by Bhanot and Peskin~\cite{Peskin:1979va,Bhanot:1979vb}\@. However, it is known from current heavy ion collision experiments that the QGP created in such collisions is actually a strongly coupled fluid, which means that calculations derived from a weakly coupled description are not always reliable for practical phenomenology. More specifically, the issue of a weak coupling picture is that the QGP medium relevant for quarkonium dissociation and recombination is far away from being a gas of weakly interacting light quarks and gluons. 

The question then becomes how quarkonium dissociation and recombination can be formulated for a strongly coupled medium, where perturbative techniques may not apply. Thanks to recent developments using the open quantum system framework and nonrelativistic effective field theories of QCD, both quantum and classical transport equations to describe quarkonium in-medium dynamics have been derived~\cite{Akamatsu:2011se,Akamatsu:2012vt,Rothkopf:2013kya,Akamatsu:2014qsa,Akamatsu:2015kaa,Katz:2015qja,Blaizot:2015hya,Brambilla:2016wgg,Blaizot:2017ypk,Kajimoto:2017rel,Akamatsu:2018xim,Brambilla:2017zei,Yao:2018nmy,Blaizot:2018oev,Sharma:2019xum,Miura:2019ssi,Akamatsu:2021vsh,Blaizot:2021xqa,Brambilla:2020qwo,Brambilla:2021wkt,Miura:2022arv,Brambilla:2022ynh,Xie:2022tzs,Brambilla:2023hkw} (see recent reviews~\cite{Rothkopf:2019ipj,Akamatsu:2020ypb,Yao:2021lus,Sharma:2021vvu})\@. These developments allow one to factorize out the nonperturbative soft physics of the QGP from the rest of the heavy quark physics~\cite{Yao:2020eqy}\@. More specifically, the nonperturbative soft physics that originates from the strongly coupled QGP is encoded in terms of a gauge-invariant correlator of chromoelectric fields dressed with Wilson lines~\cite{Yao:2020eqy} (see also Ref.~\cite{Brambilla:2016wgg,Brambilla:2017zei} for a different perspective)\@. So far, this chromoelectric field correlator is only known up to next-to-leading (NLO) order in perturbation theory at finite temperature~\cite{Binder:2021otw}, which already shows a difference from a similar but different chromoelectric field correlator that characterizes single heavy quark diffusion~\cite{Eller:2019spw,Scheihing-Hitschfeld:2022xqx}\@. In spite of the fact that this correlation function has been previously considered in vacuum and Euclidean QCD (see, for instance, Refs.~\cite{DiGiacomo:1998nu,DiGiacomo:2000irz,Eidemuller:1997bb,Shifman:1978bx,Voloshin:1978hc,Leutwyler:1980tn,Dosch:1994wj}), not much is known about the chromoelectric field correlator for quarkonium in-medium dynamics beyond NLO. With the proper setup, one may be able to calculate the zero frequency limit of the correlator by using Euclidean lattice QCD methods and analytically continuing the result to real time. Studying the correlator at finite frequency directly in real time is intractable numerically due to the notorious sign problem.

In this paper, we calculate this chromoelectric field correlator in strongly coupled $\mathcal{N}=4$ Yang-Mills theory at both zero and finite frequency using the AdS/CFT correspondence~\cite{Maldacena:1997re,casalderrey2014gauge}\@. The general strategy of our studies parallels the approaches used in studies of jet quenching and heavy quark diffusion, where effective field theory was first applied to describe the strongly coupled physics inside the QGP in terms of gauge-invariant objects that require a nonperturbative determination, which were then subsequently calculated via the AdS/CFT holographic duality. In the case of jet quenching, Soft-Collinear Effective Theory~\cite{Bauer:2000ew,Bauer:2000yr,Bauer:2001ct,Bauer:2001yt,Bauer:2002nz} had been used to formulate the jet quenching parameter as a Wilson loop consisting of two light-like Wilson lines, and was later calculated using the AdS/CFT correspondence in Refs.~\cite{Liu:2006ug,Liu:2006he,DEramo:2010wup}\@. In the case of heavy quark diffusion, heavy quark effective theory was applied to define the heavy quark diffusion coefficient as a chromoelectric field correlator in Ref.~\cite{Casalderrey-Solana:2006fio}, which was then calculated in the same work via the holographic principle (see Ref.~\cite{Herzog:2006gh} for a different way of studying heavy quark diffusion in the AdS/CFT approach)\@. In this work, we will use the AdS/CFT technique to calculate the chromoelectric field correlator relevant for quarkonium transport, which has been defined by using potential nonrelativistic QCD (pNRQCD)~\cite{Brambilla:1999xf,Brambilla:2004jw}\@. This correlator is given by~\cite{Brambilla:2016wgg,Brambilla:2017zei,Yao:2020eqy,Binder:2021otw,Scheihing-Hitschfeld:2022xqx}
\begin{equation}
   \big\langle \hat{\ml{T}} E^a_i(t) \ml{W}^{ab}_{[t,0]} E^b_i(0) \big\rangle \, , \label{eq:EE-intro}
\end{equation}
where $E_i^a$ is the chromoelectric field, $\ml{W}^{ab}_{[t,0]}$ is an adjoint representation Wilson line and $\hat{\T}$ denotes the time ordering operator.
Our approach will be to first rewrite the chromoelectric field correlator as a path variation of a Wilson loop, which defines a contour on the boundary of an AdS black hole spacetime. Then, using the holographic correspondence, we will calculate the expectation value of the Wilson loop by finding the extremal surface in the bulk of the AdS spacetime that hangs from the contour defined by the Wilson loop. Finally, we will obtain the expectation value of the chromoelectric field correlator by taking the path variation, which amounts to solving linear equations for fluctuations that propagate on the extremal surface.

The result presented here is important for quarkonium phenomenology, because it provides the first nonperturbative picture of the in-medium quarks and gluons that are relevant for quarkonium dynamics in the QGP and goes beyond the assumption of a weakly interacting gas. Crucially, the nonperturbative distributions of in-medium quarks and gluons can be process-dependent, as in the case of deep-inelastic scattering (inclusive versus semi-inclusive, polarized versus unpolarized, and so on)\@. In general, it is not expected that the in-medium quarks and gluons relevant for jet quenching would have the same distribution as those affecting heavy quarks.  

The paper is organized as follows: We will first review the transport equations for quarkonium in-medium dynamics and the relevant chromoelectric field correlator in Section~\ref{sec:Quarkonia}\@. The setup of the AdS/CFT calculation will be given in Section~\ref{sect:setup_begin}, followed by details of the calculation in sections~\ref{sec:HQ-setup} and~\ref{sec:QQ-setup} for two well-motivated setups. We will discuss the AdS/CFT result with a focus on its implications for quarkonium transport in the QGP in Section~\ref{sec:results-applications}\@. Finally, we will draw our conclusions in Section~\ref{sec:conclusions}\@.

\section{Quarkonium transport at high and low temperatures} \label{sec:Quarkonia}
Three energy scales are used to describe heavy quarkonium in vacuum: the heavy quark mass $M$, the inverse of quarkonium size $\frac{1}{r}$ and the binding energy $|E_b|$\@. Nonrelativistically, these three scales form a hierarchy $M\gg \frac{1}{r}\gg |E_b|$\@. Depending on where the plasma temperature $T$ fits into the hierarchy, we may have different descriptions of quarkonium in-medium dynamics. We will always consider the case of $M\gg T$, since the highest temperature achieved in current heavy ion collision experiments is on the order of $500$\,MeV, which is smaller than the charm quark mass $M_c\approx1.3$\,GeV and much smaller than the bottom quark mass $M_b\approx4.2$\,GeV\@. Furthermore, throughout this paper, we will focus on heavy quarks and quarkonia at low transverse momenta.

When the plasma temperature is very high $M\gg T \gg \frac{1}{r} \gg |E_b|$, which is the case in the early stage of heavy ion collisions, the interaction between a heavy quark-antiquark ($Q\bar{Q}$) pair is significantly screened. As a result, the in-medium dynamics of a $Q\bar{Q}$ pair can be described in terms of two independent heavy quarks that diffuse and dissipate in the plasma. This dynamics can be approximately described by a Langevin equation with drag and diffusion and the heavy quark diffusion coefficient has been calculated nonperturbatively via lattice methods~\cite{Banerjee:2011ra,Francis:2015daa,Brambilla:2020siz,Altenkort:2020fgs}\@. One can improve the Langevin equation by adding an attractive potential effect that is screened when the $Q\bar{Q}$ pair is far away in space. This potential effect can only last for a time scale given by the imaginary potential obtained from lattice studies~\cite{Rothkopf:2011db,Larsen:2019zqv,Larsen:2019bwy,Bala:2021fkm,Petreczky:2021zmz}\@.

After the formation of the QGP in heavy ion collisions, the plasma expands quickly and cools down. When the temperature drops to the region where $T\sim\frac{1}{r}$, the interaction between a $Q\bar{Q}$ pair can no longer be neglected and must be included in the Langevin description. When the temperature further drops, $M\gg  \frac{1}{r} \gg T \gg |E_b|$, a different description comes into play, in which the color correlation between the $Q\bar{Q}$ pair becomes important. The in-medium dynamics of a $Q\bar{Q}$ pair can be described by a Lindblad equation in the quantum Brownian motion limit~\cite{Brambilla:2016wgg,Brambilla:2017zei} (the subscript ``$S$'' stands for ``system'' in the open quantum system framework, as contrary to the environment, i.e., the QGP)
\begin{align}
\frac{d\rho_S(t)}{d t} &= -i\big[ H_S + \Delta H_S,\, \rho_S(t) \big] + \kappa_{\rm adj} \Big( L_{\alpha i} \rho_S(t) L^\dagger_{\alpha i} - \frac{1}{2}\big\{  L^\dagger_{\alpha i}L_{\alpha i},\, \rho_S(t)\big\} \Big)\,,
\end{align}
where the Hamiltonian is given by
\begin{align}
H_S = \frac{{\bs p}_{\text{rel}}^2}{M} + \begin{pmatrix} 
- \frac{C_F\alpha_s}{r} & 0 \\
 0 & \frac{\alpha_s}{2N_cr}
\end{pmatrix} \,, \qquad\quad \Delta H_S = \frac{\gamma_{\rm adj}}{2} r^2 \begin{pmatrix}
1 & 0\\
0 & \frac{N_c^2-2}{2(N_c^2-1)}
\end{pmatrix} \,,
\end{align}
and the density matrix is assumed to be block diagonal in the color singlet and octet basis
\begin{align}
\rho_S(t) = \begin{pmatrix}
\rho_S^{(s)}(t) & 0 \\
0 & \rho_S^{(o)}(t)
\end{pmatrix} \,.
\end{align}
The Lindblad operators are given by
\begin{align}
L_{1i} &= \Big(r_i + \frac{1}{2MT}\nabla_i - \frac{N_c}{8T}\frac{\alpha_s r_i}{r} \Big)\begin{pmatrix}
0 & 0\\
1 & 0
\end{pmatrix}  \\
L_{2i} &= \sqrt{\frac{1}{N_c^2-1}}\Big(r_i + \frac{1}{2MT}\nabla_i + \frac{N_c}{8T}\frac{\alpha_s r_i}{r} \Big)\begin{pmatrix}
0 & 1\\
0 & 0
\end{pmatrix}  \\
L_{3i} &= \sqrt{\frac{N_c^2-4}{2(N_c^2-1)}}
\Big(r_i + \frac{1}{2MT}\nabla_i \Big)\begin{pmatrix}
0 & 0\\
0 & 1
\end{pmatrix} \,,
\end{align}
where $i=x,y,z$.
The transport coefficients $\kappa_{\rm adj}$ and $\gamma_{\rm adj}$ are defined in terms of chromoelectric field correlators
\begin{align}
\label{eqn:kappa_gamma_adj}
\kappa_{\rm adj} &= \frac{g^2 T_F }{3 N_c} {\rm Re} \int d t\, \big\langle \hat{\ml{T}} E^a_i(t) \ml{W}^{ab}(t,0) E^b_i(0) \big\rangle_T \\
\gamma_{\rm adj} &= \frac{g^2 T_F }{3 N_c} {\rm Im} \int d t\, \big\langle \hat{\ml{T}} E^a_i(t) \ml{W}^{ab}(t,0) E^b_i(0) \big\rangle_T \,,
\end{align}
where $T_F$ is defined by ${\rm Tr}(T^a_FT^b_F)=T_F\delta^{ab}$ with $T_F^a$ being the generator in the fundamental representation and $\ml{W}^{ab}(t,0)$ denotes a time-like Wilson line in the adjoint representation from time $0$ to $t$:
\be
\ml{W}(x,y) = {P} \exp \left( ig \int_y^x \!\! d z^\mu A_\mu^a(z) T^a_A \right) \,,
\ee
in which $x$ and $y$ are Minkowski position 4-vectors connected by a straight path. The expectation value of an operator $O$ is defined as $\langle O \rangle_T \equiv \tr(O e^{-\beta H_E})/\tr(e^{-\beta H_E})$ where $\beta = 1/T$ is the inverse of the plasma temperature and $H_E$ denotes the Hamiltonian of light quarks and gluons in the QGP\@. As can be seen, in this high temperature limit where the quantum Brownian motion limit of the open quantum system framework is valid, it is the zero frequency limit of the correlator shown in Eq.~\eqref{eq:EE-intro} that is relevant for quarkonium in-medium dynamics.

As the QGP temperature continues dropping and finally becomes of the same order as the binding energy, $M\gg  \frac{1}{r} \gg T \sim |E_b|$, we need another description that is based on a classical Boltzmann equation which can be derived by using the open quantum system framework in the quantum optical limit, pNRQCD, the Wigner transform and semiclassical gradient expansion~\cite{Yao:2020eqy} (a subtlety of using the quantum optical limit can be resolved by working in the semiclassical limit, as explained in Ref.~\cite{Yao:2021lus})\@. If we further integrate over the momentum distribution of the phase space distribution, we will arrive at a rate equation for the density of a quarkonium state $n_b$ with the quantum number $b$,
\begin{align}
\label{eqn:rate}
\frac{d n_b(t,{\bs x})}{d t} = -\Gamma\, n_b(t,{\bs x}) + F(t,{\bs x}) \,,
\end{align}
where $\Gamma$ is the dissociation rate and $F$ denotes the contribution of quarkonium formation (in-medium recombination). They are given by
\begin{align}
\label{eqn:disso}
\Gamma &=  \frac{g^2T_F}{3N_c}\int \frac{d^3p_{\ma{rel}}}{(2\pi)^3} 
| \langle \psi_b | {\bs r} | \Psi_{{\bs p}_\ma{rel}} \rangle |^2 [g^{++}_E]^{>}\Big(-|E_b| - \frac{p^2_\ma{rel}}{M}\Big) \\
\label{eqn:reco}
F(t,{\bs x}) &=  \frac{g^2T_F}{3N_c} \int \frac{d^3p_{\ma{cm}}}{(2\pi)^3}  \frac{d^3p_{\ma{rel}}}{(2\pi)^3} 
| \langle \psi_b | {\bs r} | \Psi_{{\bs p}_\ma{rel}} \rangle |^2 \nn\\
& \quad \times
[g^{--}_E]^{>}\Big(\frac{p^2_\ma{rel}}{M}+|E_{b}|\Big) f_{Q\bar{Q}}(t, {\bs x}, {\bs p}_{\ma{cm}}, {\bs x}_{\rm rel}=0, {\bs p}_{\ma{rel}}) \,, 
\end{align}
where $\langle \psi_b | {\bs r} | \Psi_{{\bs p}_\ma{rel}} \rangle$ is the dipole transition amplitude between a bound quarkonium state wavefunction $\psi_b$ and an unbound $Q\bar{Q}$ wavefunction $\Psi_{{\bs p}_\ma{rel}}$ that is a scattering wave with momentum ${\bs p}_{\rm rel}$\@. The two-particle phase space distribution $f_{Q\bar{Q}}(t, {\bs x}, {\bs p}_{\ma{cm}}, {\bs x}_{\rm rel}=0, {\bs p}_{\ma{rel}})$ is for an unbound $Q\bar{Q}$ pair with the center-of-mass (cm) position ${\bs x}_{\rm cm}={\bs x}$, cm momentum ${\bs p}_{\ma{cm}}$, relative position ${\bs x}_{\rm rel}=0$ and relative momentum ${\bs p}_{\ma{rel}}$\@. The relative position is fixed to be $0$ which is a result of a gradient expansion used in taking the semiclassical limit. The $Q\bar{Q}$ phase space distribution does not factorize into the product of two single particle distributions
\be
f_{Q\bar{Q}}(t, {\bs x}, {\bs p}_{\ma{cm}}, {\bs x}_{\rm rel}=0, {\bs p}_{\ma{rel}} ) \neq f_Q(t, {\bs x}, {\bs p}_Q) f_{\bar{Q}}(t, {\bs x}, {\bs p}_{\bar{Q}})\,,
\ee
which means that the formation term $F$ in the rate equation can account for both correlated and uncorrelated recombination~\cite{Yao:2020xzw}\@. The Boltzmann and rate equations have been extensively used in phenomenological studies of quarkonium and exotics production in heavy ion collisions~\cite{Du:2017qkv,Yao:2017fuc,Yao:2018zze,Zhao:2021voa,Du:2022uvj,Wu:2023djn}\@.

In the rate equation, we follow the notation of Ref.~\cite{Binder:2021otw} and write the physical (Wightman) correlation functions that govern quarkonium transport as
\begin{align}
[g_{E}^{++}]^>(t) &= \frac{1}{Z} {\rm Tr}_{\mathcal{H}} \left[ E_i^a(t) \W^{ac}(t,+\infty) \W^{cb}(+\infty,0) E_i^b(0) e^{-\beta H} \right] \label{eq:gE++>} \\
[g_{E}^{++}]^<(t) &= \frac{1}{Z} {\rm Tr}_{\mathcal{H}} \left[ \W^{cb}(+\infty,0) E_i^b(0) E_i^a(t) \W^{ad}(t,+\infty)  e^{-\beta H} \W^{dc}(+\infty -i\beta,+\infty) \right] \label{eq:gE++<} \\
[g_{E}^{--}]^>(t) &= \frac{1}{Z} {\rm Tr}_{\mathcal{H}} \left[ \W^{cb}(-\infty,t) E_i^b(t) E_i^a(0) \W^{ad}(0,-\infty)  e^{-\beta H} \W^{dc}(-\infty -i\beta,-\infty) \right] \label{eq:gE-->} \\
[g_{E}^{--}]^<(t) &= \frac{1}{Z} {\rm Tr}_{\mathcal{H}} \left[ E_i^a(0) \W^{ac}(0,-\infty) \W^{cb}(-\infty,t) E_i^b(t) e^{-\beta H} \right] \, , \label{eq:gE--<}
\end{align}
where $H$ is the environment (QGP) Hamiltonian, ${\rm Tr}_{\mathcal{H}}$ denotes a trace over states in the Hilbert space of the theory, and $Z = {\rm Tr}_{\mathcal{H}} \left[ e^{-\beta H} \right]$ is the partition function of the QGP\@. As opposed to Ref.~\cite{Binder:2021otw}, here we have written the thermal averages explicitly, with the (Euclidean) adjoint Wilson line at ${\rm Re}\{t\} = -\infty$ in the definition of $[g_{E}^{--}]^>(t)$ accounting for the fact that in the corresponding physical situation, the QGP and the point adjoint color charge have thermalized together.

In frequency space, we define the correlators as
\begin{align}
    [g_E^{\pm \pm'}]^{\lessgtr}(\omega) = \int_{-\infty}^\infty \!\! dt \, e^{i \omega t} [g_E^{\pm \pm'}]^{\lessgtr}(t) \, ,
\end{align}
in terms of which the KMS relations are given by
\begin{align}
[g_{E}^{++}]^>(\omega) &= e^{\omega/T} [g_{E}^{++}]^<(\omega) \\
[g_{E}^{--}]^>(\omega) &= e^{\omega/T} [g_{E}^{--}]^<(\omega) \, ,
\end{align}
and are related by the combined action of parity and time-reversal
\be
\label{eqn:kms}
[g_{E}^{++}]^>(\omega) = [g_E^{--}]^<(-\omega) \,,
\ee
where we have assumed the QGP is symmetric under these discrete symmetries. A short proof of the KMS relations is given in Appendix~\ref{sec:App-KMS}, where we give a derivation that is more clearly indicative of the physics at work than in Ref.~\cite{Binder:2021otw}\@. From this proof it is also clear that the normalization factor $Z$ of the KMS conjugates $[g_E^{\pm \pm'}]^{\lessgtr}(\omega)$ must be the same. Perturbative calculations~\cite{Binder:2021otw} verify that the QGP partition function adequately serves this purpose.

Following~\cite{Binder:2021otw}, one may then define spectral functions that respect the KMS relations:
\begin{align}
\rho_{E}^{++}(\omega) & = [g_{E}^{++}]^>(\omega) - [g_{E}^{++}]^<(\omega) \\
\rho_{E}^{--}(\omega) & = [g_{E}^{--}]^>(\omega) - [g_{E}^{--}]^<(\omega) \, .
\end{align}
Contrary to typical spectral functions, the ones we have just introduced are not guaranteed to have a definite parity under $\omega \to -\omega$. Rather, because of parity and time-reversal, they are related to each other via:
\begin{align}
    \rho_{E}^{++}(\omega) = - \rho_{E}^{--}(-\omega) \, ,
\end{align}
so it is still true that complete knowledge of one spectral function fully determines all correlation functions introduced above. Conversely, complete knowledge of one of the physical (Wightman) correlation functions also fully determines all the rest of the correlation functions.

In this paper, we will use the AdS/CFT correspondence to evaluate the time-ordered version of the correlation function:
\begin{align}
    [g_E^{{\mathcal{T}}}](t) = \langle \hat{\mathcal{T}} E_i^a(t) \mathcal{W}^{ab}_{[t,0]} E_i^b(0) \rangle_T \, ,
\end{align}
which can be written as
\begin{equation}
    [g_E^{{\mathcal{T}}}](t) = \theta(t) [g_{E}^{++}]^>(t) + \theta(-t) [g_{E}^{++}]^>(-t) \, . 
\end{equation}
Furthermore, with $[g_E^{\mathcal{T}}](t)$ in hand we also have access to the anti-time-ordered correlator through complex conjugation (denoted by a star ${}^*$)
\begin{align}
    [g_E^{{\overline{\mathcal{T}}}}](t) = \langle \hat{\overline{\mathcal{T}}} E_i^a(t) \mathcal{W}^{ab}_{[t,0]} E_i^b(0) \rangle_T = \left\{ [g_E^{{\mathcal{T}}}](t) \right\}^* \, .
\end{align}
Then, we can also express the Wightman function $[g_{E}^{++}]^>(t)$ in terms of its (anti)time-ordered counterparts:
\begin{align}
\label{eq:>fromT}
    [g_{E}^{++}]^>(t) = \theta(t) [g_E^{{\mathcal{T}}}](t) + \theta(-t) [g_E^{{\overline{\mathcal{T}}}}](t) \, .
\end{align}
This means that once we obtain the time-ordered correlator from the AdS/CFT correspondence, we can evaluate all the other correlation functions, in particular the physical (Wightman) correlators that enter the Boltzmann and rate equations.

We can use all of the above to write Eq.~\eqref{eq:>fromT} in frequency space:
\begin{equation}
    [g_{E}^{++}]^>(\omega) = {\rm Re}\!\left\{ [g_E^{{\mathcal{T}}}](\omega) \right\} + \frac{1}{\pi} \int_{-\infty}^\infty d p_0 \, \mathcal{P} \left( \frac{1}{p_0} \right) {\rm Im}\! \left\{ [g_E^{{\mathcal{T}}}](\omega + p_0) \right\} \, ,
\end{equation}
where $\ml{P}$ denotes the Cauchy principal value.
The inverse map is given by\footnote{To prove that they are the inverse of each other, one needs to use a particular representation of the Dirac delta:
\begin{equation}
    \int_{-\infty}^\infty dx \, \mathcal{P} \left( \frac{1}{(x-1) (x^2 - a^2)} \right) = \frac{\pi^2}{2} \delta(|a| - 1) \, ,
\end{equation} which may be verified by direct action of this distribution on functions whose arguments are the parameter $a$.}
\begin{equation}
    [g_E^{{\mathcal{T}}}](\omega) = \frac12 \left( [g_{E}^{++}]^>(\omega) + [g_{E}^{++}]^>(-\omega) \right) + \frac{1}{2\pi i} \int_{-\infty}^\infty dp_0 \, \mathcal{P} \left( \frac{2 p_0}{p_0^2 - \omega^2} \right) [g_{E}^{++}]^>(p_0) \, .
\end{equation}

Once we obtain $[g_{E}^{++}]^>(\omega)$ from $[g_E^{{\mathcal{T}}}](\omega)$, the spectral functions follow immediately. As such, all we need is to calculate the time-ordered correlator. In the next section we describe the theoretical foundations we will employ to evaluate it.

\section{Field strength correlators from Wilson loops}
\label{sect:setup_begin}

In this section we will describe the setup of the calculation of the non-Abelian electric field correlator we wish to obtain. We will start in Subsection~\ref{sec:setup} by describing how such a correlation function can be obtained by taking variations of a Wilson loop in a purely field-theoretic setup. Then, in Subsection~\ref{sec:W-loop-AdS} we will proceed to describe how we can evaluate Wilson loops at strong coupling in $\mathcal{N}=4$ supersymmetric Yang–Mills (SYM) theory using the AdS/CFT correspondence. Here we will discuss the role of the additional parameter $\hat{n} \in S_5$ when constructing the supersymmetric Wilson loop that preserves the features of an adjoint Wilson line. The subsequent Subsection~\ref{sec:EE-setup-AdS} discusses how to take variations of a Wilson loop (i.e., how to introduce field strength insertions) along the contour that defines it from the point of view of the dual gravitational theory. We will also establish the prescriptions necessary to fully define the correlation function for quarkonium in-medium dynamics, and discuss the differences with the correlation function that determines the heavy quark diffusion coefficient. 

We will motivate two different holographic configurations that determine \textit{different} correlation functions, and describe the setup for the calculation of each object in separate sections~\ref{sec:HQ-setup} and~\ref{sec:QQ-setup}\@. As we will shortly discuss, the configuration we present in Section~\ref{sec:QQ-setup} possesses features that make it a better proxy for the correlation function we seek to calculate~\eqref{eq:EE-corr-from-variations} than the one in Section~\ref{sec:HQ-setup}. Nonetheless, to be thorough, we will conclude which correlator is a better $\mathcal{N}=4$ SYM proxy for the description of two non-Abelian electric fields connected by an adjoint Wilson line only after obtaining explicit results for both of them.

\subsection{Wilson loops in gauge theory and their variations}
\label{sec:setup}

As we pointed out in the previous section, the transport properties of quarkonia are governed by the correlation functions of chromoelectric fields dressed by Wilson lines, calculated inside a medium. Properties of the medium are encoded in terms of expectation values, which can be determined by a thermal density matrix $Z^{-1} \exp (-\beta H)$\@. However, for the purposes of this subsection we will not need to make reference to the nature of the medium. Rather, we will only discuss how to construct the expectation value we are interested in, by starting from another class of observables that has a well-defined prescription for evaluation at strong coupling using the gauge-gravity duality, which we will discuss explicitly in Subsection~\ref{sec:W-loop-AdS}\@.

Concretely, it is possible to study the correlation functions of gauge theory field strengths $F_{\mu \nu}$ dressed by Wilson lines starting from another class of gauge-invariant operators, namely, Wilson loops $W[\mathcal{C}]$. They are defined as
\begin{align} \label{eq:W-loop}
    W[\mathcal{C}] = \frac{1}{N_c} {\rm Tr}_{\rm color} \! \left[ U_{\ml{C}} \right] = \frac{1}{N_c} {\rm Tr}_{\rm color} \! \left[ {P} \exp \left( ig \oint_{\ml{C}} T^a A^a_\mu dx^\mu  \right) \right] \,,
\end{align}
where $A_\mu^a$ is the SU($N_c$) non-Abelian gauge potential, $T^a$ denotes the generator matrix of the group, $g$ is the coupling constant, and ${P}$ denotes path ordering in the product of group elements $A_\mu = A_\mu^a T^a$\@.
The study of these operators is a cornerstone of much of our understanding of heavy or highly-energetic quarks, and in particular for their dynamics inside a thermal medium. The heavy quark-antiquark interaction potential~\cite{Wilson:PhysRevD.10.2445,Maldacena:1998im,Rey:1998ik}, the jet quenching parameter~\cite{Baier:1996sk,Liu:2006ug}, and the heavy quark diffusion inside a thermal medium~\cite{Casalderrey-Solana:2006fio,Caron-Huot:2009ncn,Burnier:2010rp} have all been formulated and studied through Wilson loops. Crucially, all of them admit a holographic description in $\mathcal{N}=4$ supersymmetric Yang-Mills theory.

As it turns out, one can connect the expectation values of Wilson loops and that of field strengths dressed by Wilson lines by considering functional variations of the path on which the Wilson loop is defined. Concretely, we consider a Wilson loop defined by a path $\mathcal{C}$, and let $\gamma^\mu(s)$ be a parametrization of the path, with $s \in [0, 1]$, and $\gamma^\mu(0) = \gamma^\mu(1)$ for a closed path. Then we consider a deformation of the path, which we denote by $\mathcal{C}_f$ and parametrize by $\gamma_f^\mu(s) = \gamma^\mu(s) + f^\mu(s)$\@. It is then a textbook exercise~\cite{Polyakov:1987ez} to show that
\begin{align} \label{eq:F-insertion-1}
\left.  \frac{\delta}{\delta f^{\mu}(s) } W[\mathcal{C}_f] \right|_{f = 0}  = \frac{(ig)}{N_c} {\rm Tr}_{\rm color} \! \left\{ U_{[1,s]} F_{\mu \rho}[\gamma(s)]\dot{\gamma}^\rho(s) U_{[s,0]}  \right\} \,,
\end{align}
where $U_{[s',s]}$ denotes a Wilson line from $\gamma(s)$ to $\gamma(s')$ in the same representation as $W[\ml{C}_f]$ and $\dot{\gamma}^\rho(s) = d\gamma^\rho(s)/ds$\@. We will abuse the notation a bit to use $U_{[\gamma(s'),\gamma(s)]}$ and $U_{[s',s]}$ interchangeably.
The order of the operators in this expression, and of those in the present discussion before Section~\ref{sec:W-loop-AdS}, only refers to the SU($N_c$) matrix product. Operator ordering in the sense of the order in which they act on a state in the Hilbert space of the theory has yet to be specified (this will be done below, and further developed in Appendix~\ref{sec:App-W-ordering})\@.

This property provides us with a tool to generate as many field strength insertions as we want along the path of the Wilson loop. By acting on $W[\mathcal{C}]$ with one more derivative and assuming $s_2 > s_1$, 
\begin{align} \label{eq:F-insertion-2}
    & \left.  \frac{\delta}{\delta f^{\mu}(s_2) } \frac{\delta}{\delta f^{\nu}(s_1) } W[\mathcal{C}_f] \right|_{f = 0}  \nonumber \\ 
    & \quad \quad \quad \quad = \frac{(ig)^2}{N_c} {\rm Tr}_{\rm color} \! \left\{ U_{[1,s_2]} F_{\mu \rho}[\gamma(s_2)] \dot{\gamma}^\rho(s_2) U_{[s_2,s_1]} F_{\nu \sigma}[\gamma(s_1)] \dot{\gamma}^\sigma(s_1) U_{[s_1,0]} \right\} \, .
\end{align}
We see explicitly that we can obtain correlation functions of field strength operators $F_{\mu \nu}$ dressed with Wilson lines by taking derivatives with respect to the path on which the Wilson loop~\eqref{eq:W-loop} is defined.\footnote{\label{fn:delta-disc} The only subtlety in this expression is that if $s_1=s_2$, then there is another term on the right hand side, due to the action of $\delta/\delta f$ on the $\dot{\gamma}$ present in Eq.~\eqref{eq:F-insertion-1}\@. This term is naturally proportional to derivatives of the delta function $\delta(s_2-s_1)$, and can be easily isolated from the rest of the correlator by looking at positions $s_2 > s_1$ and continuously extending the result to $s_2 = s_1$ (whenever possible)\@. If one looks at the Fourier transform of the left hand side of Eq.~\eqref{eq:F-insertion-2} with respect to $s_2-s_1$, as we will find most natural to do later on, then there will be a contribution from points with $s_2 = s_1$ in the form of a polynomial of positive powers of their Fourier conjugate variable, which we will have to subtract to obtain the correlation function of interest.} It is of course possible to continue beyond two field strength insertions, but for our present purposes it is sufficient to evaluate the two-point deformations.

Specifically, our correlator of interest can be obtained by taking $\mathcal{C}$ to be a closed loop parametrized by $\gamma^{\mu}(s)$, where $s \in [-\mathcal{T}/2, 3\mathcal{T}/2]$, as
\begin{align} \label{eq:C-contour-0}
\gamma^\mu(s) = \begin{cases} s t^\mu & -\frac{\T}2 < s < \frac{\T}2 \\ (\T - s) t^\mu &  \frac{\T}2 < s < \frac{3\T}2  \end{cases} \,,
\end{align}
with $t^\mu = (1,0,0,0)$ being a unit vector in the positive time direction. This describes a timelike loop that backtracks upon itself after reaching a maximal value for the time coordinate $t = \T/2$\@. We note that in our setup we must have $W[\mathcal{C}_{f=0}] = 1$, even for time-ordered operators (see Appendix~\ref{sec:App-W-ordering} for details)\@.

Then, taking variations of the path with respect to perturbations in a spatial direction $f^i(s)$ leads to
\begin{align} \label{eq:E-insertion-2}
    &\left.  \frac{\delta}{\delta f^{i}(s_2) } \frac{\delta}{\delta f^{j}(s_1) } W[\mathcal{C}_f] \right|_{f = 0}  =   \nonumber \\
    &\frac{(ig)^2}{N_c} \begin{cases}
    {\rm Tr}_{\rm color} \! \left\{ U_{[3\T/2,\T/2]} U_{[\T/2,s_2]} E_{i}[\gamma(s_2)] U_{[s_2,s_1]} E_j[\gamma(s_1)] U_{[s_1,-\T/2]} \right\} & {\T}/2 > s_2 > s_1 \\ 
    - {\rm Tr}_{\rm color} \! \left\{ U_{[3\T/2, s_2]} E_{i}[\gamma(s_2)] U_{[s_2,\T/2]} U_{[\T/2,s_1]} E_j[\gamma(s_1)] U_{[s_1,-\T/2]} \right\} & s_2 > {\T}/2 > s_1  \\ 
    {\rm Tr}_{\rm color} \! \left\{ U_{[3\T/2, s_2]}  E_{i}[\gamma(s_2)] U_{[s_2,s_1]} E_j[\gamma(s_1)] U_{[s_1,\T/2]} U_{[\T/2,-\T/2]} \right\}  &  s_2 > s_1 >\T/2 \end{cases} \,,
\end{align}
where we have assumed that $s_2 > s_1$\@.
For our purposes, we will take all quantum mechanical operators to be time-ordered, which is exactly what one obtains from the path integral formulation of QFT\@. Some comments on the operator ordering can be found in Appendix~\ref{sec:App-W-ordering}, where we discuss similarities and distinctions with other types of ordering, as well as explain why the time-ordered correlator describes quarkonium dynamics. 
 
We can summarize all of these possibilities as
\begin{align}
\left. \frac{\delta}{\delta f^{i}(s_2) } \frac{\delta}{\delta f^{j}(s_1) } W[\mathcal{C}_f] \right|_{f = 0} = {\rm sgn}[({\T}/2 -s_1)({\T}/2 - s_2)]  \frac{(ig)^2}{N_c} T_F E_i^a(t_2) \mathcal{W}^{ab}_{[t_2,t_1]} E_j^b(t_1) \,,
\end{align}
provided that we choose $s_2 \neq s_1$ such that $\gamma^\mu(s_2) = (t_2,0,0,0)$ and $\gamma^\mu(s_1) = (t_1,0,0,0)$, and $-\frac{\T}2 < t_1<t_2 < \frac{\T}2$\@. In deriving this last expression, we have also used the gauge theory identity
\begin{equation}
\label{eq:adj_fund}
    \W^{ab}_{[t_2,t_1]} = \frac{1}{T_F} {\rm Tr}_{\rm color}  \left[ \hat{\T} T^a U_{[t_2,t_1]} T^b U_{[t_2,t_1]}^\dagger \right] \,,
\end{equation}
where $\hat{\mathcal{T}}$ denotes a time-ordering symbol.

We can further simplify this expression by restricting ourselves to contour deformations of the form
\begin{align} \label{eq:f-antisymm}
f^\mu(s) = \begin{cases} h^\mu(s) & -\frac{\T}2 < s < \frac{\T}2 \\ - h^\mu (\T - s) &  \frac{\T}2 < s < \frac{3\T}2  
\end{cases} \,,
\end{align}
where $h^\mu(t)$ is the independent function with respect to which we will take a variation. Going through the same arguments given as above, one finds
\begin{align} \label{eq:EE-from-variations}
\left. \frac{\delta}{\delta h^{i}(t_2) } \frac{\delta}{\delta h^{j}(t_1) } W[\mathcal{C}_f] \right|_{h = 0} = 4\frac{(ig)^2}{N_c} T_F E_i^a(t_2) \mathcal{W}^{ab}_{[t_2,t_1]} E_j^b(t_1) \, .
\end{align}
The factor of $4$ is simply a consequence of doubling the size of the deformation to the contour. Intuitively, the reason why the correlator is determined by the antisymmetric deformation is that the symmetric contour deformation, i.e., $f^\mu$ of the form
\begin{align}
f^\mu(s) = \begin{cases} g^\mu(s) & -\frac{\T}2 < s < \frac{\T}2 \\  g^\mu (\T - s) &  \frac{\T}2 < s < \frac{3\T}2  \end{cases} \, ,
\end{align}
with $g^\mu(t)$ as an independent function, gives a vanishing variation
\begin{equation}
    \left. \frac{\delta}{\delta g^{i}(t_2) } \frac{\delta}{\delta g^{j}(t_1) } W[\mathcal{C}_{f}] \right|_{ g = 0} = 0 \, ,
\end{equation}
which is a consequence of the more fundamental statement that $W[\mathcal{C}_{f}] = 1$ for an arbitrary path change given by $g^\mu(t)$ that can be even non-infinitesimal. The reason behind this is that two anti-parallel fundamental Wilson lines $U, V$ that make up the Wilson loop as $W = \frac{1}{N_c} {\rm Tr}_{\rm color} [U V] $ will still satisfy $V = U^{-1}$, even when the path over which they are laid is deformed non-infinitesimally by $f$\@. On the other hand, the antisymmetric deformation described by $h^\mu(t)$ gives a nontrivial result, precisely because in the language we just introduced, we have $V \neq U^{-1}$.

With all of this, we can state that the correlation function we want to calculate is given in terms of path variations of a Wilson loop through the following expression:
\begin{align} \label{eq:EE-corr-from-variations}
\left. \frac{\delta}{\delta h^{i}(t_2) } \frac{\delta}{\delta h^{j}(t_1) } \left\langle W[\mathcal{C}_f] \right\rangle \right|_{h = 0} = 4\frac{(ig)^2}{N_c} T_F \left\langle E_i^a(t_2) \mathcal{W}^{ab}_{[t_2,t_1]} E_j^b(t_1) \right\rangle \, ,
\end{align}
where $\langle\cdots\rangle$ denotes the expectation value.
Having formulated a way to obtain the desired correlation function in terms of operations upon Wilson loops, we now turn to the computational tool that we will use to evaluate this correlation function in $\mathcal{N} = 4$ SYM theory.

\subsection{Wilson loops in AdS/CFT} \label{sec:W-loop-AdS}

The AdS/CFT correspondence has proven to be an invaluable tool to gain insight into the strongly coupled regime of non-Abelian gauge theories, by casting a potentially intractable non-perturbative quantum mechanical problem for a conformal field theory (CFT) in terms of a purely classical problem in a concrete gravitational setup of higher dimensionality. In what follows, we will use the real-time formulation of the duality~\cite{casalderrey2014gauge} between $\mathcal{N}=4$ supersymmetric Yang-Mills theory in the Minkowski four dimensional spacetime (Mink${}_4$) in the large $N_c$ limit and type IIB string theory in an (asymptotically) AdS${}_5 \times S_5$ spacetime. The asymptotic boundary of AdS${}_5$ is identified as the Mink${}_4$ on which the supersymmetric Yang-Mills theory lives. At strong coupling in the SYM theory, the dual description on the string theory side reduces to classical string dynamics in a curved spacetime.

The task now is to describe the expectation value of Wilson loops using the duality. This was first done by Maldacena in Ref.~\cite{Maldacena:1998im}\@. However, due to the supersymmetric nature of $\mathcal{N}=4$ SYM, the CFT object that has a simple gravitational dual description in AdS${}_5$ is the locally $1/2$ BPS\footnote{The state that develops this (straight) Wilson line as a phase factor in time evolution is called a $1/2$ Bogomol’nyi-Prasad-Sommerfield (BPS) state, and so the Wilson line in Eq.~\eqref{eq:W-loop-S} is referred to as locally $1/2$ BPS\@. The factor of $1/2$ comes from the fact that the Wilson line commutes with half (eight of the sixteen) supercharges.} Wilson loop
\begin{align} \label{eq:W-loop-S}
W_{\rm BPS}[\mathcal{C};\hat{n}] = \frac{1}{N_c} {\rm Tr}_{\rm color} \! \left[ \mathcal{P} \exp \left( ig \oint_{\ml{C}} ds \, T^a \left[  A^a_\mu \, \dot{x}^\mu + \hat{n}(s) \cdot {\vec{\phi}}^a \sqrt{\dot{x}^2} \right] \right) \right] \, ,
\end{align}
where $\dot{x}^\mu(s) = dx^\mu(s)/ds$ and $\vec{\phi} = (\phi^1, \ldots, \phi^6)$ are the six Lorentz scalar fields in the adjoint representation of SU($N_c$) intrinsic to $\mathcal{N}=4$ SYM. These scalars enter the Wilson loop coupled to a direction $\hat{n}(s) \in S_5$ that specifies the direction along which the string (to be introduced in the next paragraph) ``pulls'' the heavy quark (the string lives in a 10 dimensional space AdS${}_5 \times S_5$ and has a string tension). It can be thought of as an additional property that the heavy quark carries as it propagates through Mink${}_4$, and in general, it depends on the coordinate $s$ along the path $\mathcal{C}$\@. 

Specifically, the AdS/CFT duality gives an explicit prescription to evaluate the expectation value of these generalized Wilson loops. It is given by
\begin{align} \label{eq:duality}
\left\langle W_{\rm BPS}[\mathcal{C};\hat{n}] \right \rangle = \exp \left\{ i  \mathcal{S}_{\rm NG}[\Sigma(\mathcal{C};\hat{n})] - i\mathcal{S}_{0}[\mathcal{C};\hat{n}]  \right\} \, ,
\end{align}
where
\begin{align} \label{eq:NG-action}
\mathcal{S}_{\rm NG}[\Sigma] = - \frac{1}{2\pi \alpha'} \int d\sigma \, d\tau \sqrt{- \det \left( g_{\mu \nu} \partial_\alpha X^\mu \partial_\beta X^\nu \right) } \,,
\end{align}
is the Nambu-Goto action of a string configuration $\Sigma$ described by $X^\mu(\tau,\sigma) \in {\rm AdS}_5 \times S_5$, with $\mu \in \{0,1,\ldots,9\}$\@. $\Sigma(\mathcal{C}; \hat{n})$ is the surface (also referred to as a ``worldsheet'') that extremizes the Nambu-Goto action $\mathcal{S}_{\rm NG}$ with Dirichlet boundary conditions given by $\mathcal{C}$ and $\hat{n}$ at the asymptotic boundary of AdS${}_5$\@. It is in this sense that $\hat{n}$ defines the direction along which the string ``pulls'' the heavy quark.\footnote{To picture this, it is helpful to note that, asymptotically as $z \to 0$, the part of the metric in Eq.~\eqref{eq:Schwarzschild-AdS} involving $z$ and $\hat{n}$ is proportional to $dz^2 + z^2 d\Omega_5^2$, and as such, $(z, \hat{n})$ may be thought of as a 6-component vector along the direction $\hat{n}$ with length $z$.} The subtraction by $\mathcal{S}_{0}[\mathcal{C};\hat{n}]$ is necessary to regularize the result by subtracting the energy associated to the mass of the heavy quark propagating along $\mathcal{C}$, and is also useful because by subtracting the rest mass of the heavy quark it isolates the ``energy'' associated to the interactions described by the Wilson loop at hand. (In QCD, the subtraction also contains the mass renormalization caused by the heavy quark self interaction that is part of the physics contained in the Wilson loop. However, in $\ml{N}=4$ SYM the self interaction diagrams are ultraviolet (UV) finite, and thus the subtraction only contains the bare heavy quark mass~\cite{Drukker:1999zq}.) Only after this subtraction has taken place, may one have an action with a finite value that allows for a comparison of the ``energy'' of different string configurations determined by $\mathcal{S}_{\rm NG}$\@. In the case where one may equate the expectation value of a Wilson loop to that of a time evolution operator over a time period $\T$ for a fixed set of boundary conditions $\mathcal{C}$ and $\hat{n}$, it is the lowest energy $E[\Sigma(\mathcal{C};\hat{n})] =  \left( \mathcal{S}_{0}[\mathcal{C};\hat{n}] - \mathcal{S}_{\rm NG}[\Sigma(\mathcal{C};\hat{n})] \right) /\T$ configuration that determines the expectation value mentioned above.

We also need to specify the background metric $g_{\mu \nu}$ that describes the dual ${\rm AdS}_5 \times S_5$ spacetime. Because of our interest in thermal physics, we use the metric for the dual description of a field theory at finite temperature $T$\@. In terms of Poincar\'{e} coordinates, it is given by~\cite{casalderrey2014gauge}
\begin{align} \label{eq:Schwarzschild-AdS}
ds^2 = \frac{R^2}{z^2} \left[ - f dt^2 + d{\bf x}^2 + \frac{dz^2}{f} + z^2 d\Omega_5^2 \right] \, ,
\end{align}
where $f = 1 - (\pi T z)^4$, $R$ is the curvature radius of the AdS metric, $z$ is the radial AdS coordinate, with the asymptotic boundary at $z=0$ and the black hole horizon at $z = (\pi T)^{-1}$\@. The coordinates $t$ and ${\bf x}$ describe Minkowski spacetime at $z=0$, and the $S_5$ coordinates describe a sphere of radius $R$\@. This is the Schwarzschild-AdS metric that is dual to $\mathcal{N}=4$ SYM at finite temperature. The duality also prescribes $R^2/\alpha' = \sqrt{\lambda} = \sqrt{g_{\rm YM}^2 N_c}$, and this single-handedly determines how the coupling constant appears in the results for the strongly coupled limit. In the language of Refs.~\cite{Skenderis:2008dg,Skenderis:2008dh}, this corresponds to a single copy of a Lorentzian manifold in the gravity side of the duality. Therefore, all calculations done in this setup will give time-ordered quantities in terms of the action of operators on the Hilbert space of the quantum theory. If we needed to consider more complicated operator orderings we would be forced to introduce a larger manifold on the gravity side, containing more than one copy of AdS$_{5} \times S_5$\@. This is manifest in the holographic descriptions of heavy quark diffusion and jet quenching~\cite{Casalderrey-Solana:2006fio,DEramo:2010wup}, and is discussed at length in Ref.~\cite{Skenderis:2008dg}\@.

\subsubsection{The role of \texorpdfstring{$\hat{n}$} {} } \label{sec:nhat}

The classic results for Wilson loops in the strong coupling limit that are obtained from the gauge/gravity duality feature a constant value of $\hat{n}$ throughout the quark trajectory~\cite{Maldacena:1998im,Liu:2006ug,Casalderrey-Solana:2006fio}\@. This is a natural choice because, following the discussion of Ref.~\cite{Maldacena:1998im}, the value of $\hat{n}$ is determined by the vacuum expectation value of a Higgs field that has undergone spontaneous symmetry breaking. Given that this is tantamount to a choice of vacuum, the low-energy physics will not modify the value of $\hat{n}$ throughout the trajectory of a heavy W-boson (which is later referred to as ``quark,'' because the Wilson loop also describes the evolution of a heavy quark.)\@. However, inspecting the expression for the $1/2$ BPS Wilson loop~\eqref{eq:W-loop-S}, one realizes that keeping a constant value of $\hat{n} = \hat{n}_0$ throughout both sides of the contour shown in Eq.~\eqref{eq:C-contour-0} violates our expectation for conventional Wilson lines that a closed Wilson loop consisting of two overlapping anti-parallel Wilson lines has $ W[\mathcal{C}] = 1$\@. 
Indeed, the famous result for the heavy quark interaction potential~\cite{Maldacena:1998im}, determined by evaluating a rectangular Wilson loop $\ml{C}_L$ of temporal extent $\T$ and spatial size $L$ (after subtracting the heavy quark mass contribution $\mathcal{S}_{0}[\mathcal{C}_L;\hat{n}=\hat{n}_0]$, and assuming $\T \gg L $) gives
\begin{equation}
    \left\langle W_{\rm BPS}[\mathcal{C}_L;\hat{n}=\hat{n}_0] \right \rangle = \exp \left( i \T \frac{4\pi^2}{\Gamma^4\big(\frac14\big)} \frac{\sqrt{\lambda}}{L} \right) \, ,
\end{equation}
which does \textit{not} satisfy $\lim_{L \to 0} W[\mathcal{C}_L] = 1$\@. The reason behind this is that the contributions from the scalar fields to the locally $1/2$ BPS Wilson loop do not cancel if $\hat{n}$ is held constant~\cite{Maldacena:1998im,Drukker:1999zq,Zarembo:2002an}\@. In other words, this way of approaching a Wilson loop made of coincident anti-parallel fundamental Wilson lines does not have the ``zig-zag'' symmetry~\cite{Polyakov:1997tj,Polyakov:2000ti} that the standard Wilson loop~\eqref{eq:W-loop} respects.\footnote{One may wonder whether there are any mass renormalization effects induced by the self interactions of the $1/2$ BPS Wilson lines at each side of the loop, which would also have to be factored out from the Wilson loop if one wants to isolate the interaction energy between the heavy quarks. However, the supersymmetric nature of $\mathcal{N}=4$ SYM sets these corrections to zero, which may be verified perturbatively~\cite{Drukker:1999zq}, and hence the bare mass subtraction is equivalent to subtracting twice the physical mass of the heavy quark from the energy of the quark-string configuration.}

This is an unavoidable obstacle if one attempts to interpret the variations of the Wilson loop that leads to the heavy quark potential as the dual object of a correlation function of two chromoelectric fields connected by an adjoint Wilson line. Concretely, the standard gauge theory identity for an adjoint representation Wilson line in terms of fundamental representation Wilson lines, i.e., Eq.~\eqref{eq:adj_fund},
is not satisfied if one tries to build such an object by taking variations of a loop at constant $\hat{n}$, because the two anti-parallel fundamental Wilson lines that enter the loop at constant $\hat{n}$ ($U_{[t_2,t_1]}$ and $U_{[t_1,t_2]}$) are not each other's inverse. This had already been noted and discussed previously in Refs.~\cite{Maldacena:1998im,Drukker:1999zq,Zarembo:2002an}\@.

At this point, the appropriate Wilson loop to use in $\mathcal{N} = 4$ SYM is apparent: We have to construct a locally $1/2$ BPS Wilson loop that has two timelike links $U, V$ of temporal extent $\T$ that satisfy $V = U^{-1}$, such that $W_{\rm BPS} = \frac{1}{N_c} {\rm Tr} \left[ U V \right] = 1$\@. This is realized when the $S_5$ coordinates of the Wilson lines are at antipodal points on opposite sides of the contour. In our setup introduced in Section~\ref{sec:setup}, we may choose $\hat{n} = \hat{n}_0$ for $s \in (-\T/2, \T/2)$, and $\hat{n} = -\hat{n}_0$ for $s \in (\T/2, 3\T/2)$. This is a perfectly sensible configuration on the $\mathcal{N}=4$ SYM side, and it actually preserves a maximal number of supersymmetry charges~\cite{Maldacena:1998im,Drukker:1999zq,Zarembo:2002an}\@. Furthermore, it immediately satisfies $W[\mathcal{C}] = 1$, with $\mathcal{C}$ the contour introduced in Eq.~\eqref{eq:C-contour-0}, and it respects the ``zig-zag'' symmetry, in the sense that if we extend the two timelike segments by an arbitrary extent, the contributions from each side of the contour cancel each other. 

Perhaps the most intuitive argument for the time evolution of a heavy quark-antiquark pair to be represented by two Wilson lines with antipodal positions on S${}_5$ comes from inspecting their purported equations of motion.\footnote{This is simply illustrative, as there are no fermions in the fundamental representation in the $\mathcal{N}=4$ SYM Lagrangian. However, if one were to construct a theory with heavy quarks coupled to the $\mathcal{N}=4$ SYM fields, they should follow Eqs.~\eqref{eq:Q-eom-prop} and~\eqref{eq:Qbar-eom-prop}\@.} Namely, we can define the notion of a heavy antiquark as the object that transforms in the representation conjugate to that of a heavy quark, and therefore follows a evolution equation conjugate to that of the heavy quark. This means that if we take the evolution equation for a heavy quark $Q$ (with its mass already subtracted) to be
\begin{equation}
    (\overrightarrow{\partial}_0 - i A_0 - i \hat{n} \cdot \phi ) Q = 0 \, , \label{eq:Q-eom-prop}
\end{equation}
then the evolution of a heavy antiquark follows
\begin{equation}
    \bar{Q} (\overleftarrow{\partial}_0 + i A_0 + i \hat{n} \cdot \phi ) = 0 \, , \label{eq:Qbar-eom-prop}
\end{equation}
where the arrows on top of $\partial_0$ indicate the directions for it to act.
If one then constructs the supersymmetric Wilson loop~\eqref{eq:W-loop-S} that describes the joint evolution of this heavy quark-antiquark pair, the sign flip in front of $A_0$ is accounted for by flipping the sign of $\dot{x}^\mu$ along the same path, and the sign flip in front of $\phi$ is accounted by inverting the direction $\hat{n}$.

Therefore, based on these considerations, we have a good candidate for the analogous object to the QCD chromoelectric correlator~\eqref{eq:EE-corr-from-variations} to calculate in the $\mathcal{N} = 4$ SYM theory, namely, the correlation function obtained from a $1/2$ BPS Wilson loop that has two timelike lines at antipodal positions on $S_5$, which has desirable properties distinct from the loop with constant $\hat{n}$\@. That being said, we will nonetheless evaluate the resulting correlation functions from both of them in the following sections, for two reasons:
\begin{enumerate}
    \item Even if the limit $L \to 0$ of the expectation value of the Wilson loop with constant $\hat{n}$ does not satisfy our expectations, it might still prove instructive to study non-Abelian electric field insertions in a Wilson loop that describes the interaction energy between a pair of heavy quarks;
    \item Ultimately, the motivation behind this calculation is the phenomenology we want to extract for heavy particle pairs in a thermal medium. Since we are using this holographic setup as a model of QCD, we should evaluate all objects that have a reasonable chance to resemble our correlation function of interest, and judge them by their phenomenological implications. 
\end{enumerate}

Before proceeding, we also want to comment on a more recent prescription~\cite{Alday:2007he,Polchinski:2011im} to evaluate the standard Wilson loop~\eqref{eq:W-loop}, which arguably should take precedence in our analysis over the locally $1/2$ BPS Wilson loop because it is constructed from exactly the same fields as in the standard Yang-Mills Wilson loop. This prescription states that the strong coupling limit of Eq.~\eqref{eq:W-loop} is given by extremizing the Nambu-Goto action with Neumann boundary conditions on the $S_5$\@. In practice, this means that most results for the standard Wilson loop at strong coupling are the same as those for the Wilson loop~\eqref{eq:W-loop-S} with constant $\hat{n}$, and they only start to differ at the next order in $1/\sqrt{\lambda}$~\cite{Drukker:2000ep}. This is true because, incidentally, a constant $\hat{n}$ configuration is consistent with Neumann boundary conditions on $S_5$\@.

However, the situation for the limit of our interest, namely, the construction of a standard Wilson loop with two overlapping anti-parallel timelike Wilson lines such that $W[\ml{C}]=1$, using the Neumann boundary condition prescription, might be more subtle. To see this, let us re-examine the arguments that gave rise to this prescription, and to the conclusion that the strong-coupling results of Eqs.~\eqref{eq:W-loop} and~\eqref{eq:W-loop-S} are the same for a general Wilson loop. To facilitate the references to previous works, we will work in Euclidean signature for the remainder of this subsection, unless otherwise noted or if it is explicit from the discussion (e.g., if we refer to lightlike or timelike)\@.

The first reference to Neumann boundary conditions was given by Drukker, Gross and Ooguri in Ref.~\cite{Drukker:1999zq}, where they proposed a boundary condition for an even more general Wilson loop:
\begin{equation}
    W_{\rm DGO}[\mathcal{C};\hat{n}] = \frac{1}{N_c} {\rm Tr}_{\rm color} \! \left[ \mathcal{P} \exp \left( g \oint_{\ml{C}} ds \, T^a \left[  i A^a_\mu \, \dot{x}^\mu + \dot{y}^i \phi_i^a  \right] \right) \right] \, , \label{eq:W-loop-DGO}
\end{equation}
where $y^i = y^i(s) \in \mathbb{R}^6$ is now a general vector. Concretely, the boundary condition they prescribed was
\begin{align}
    X^\mu(\sigma_1,\sigma_2=0) &= x^\mu(\sigma_1) & & {\rm for} \, \mu \in \{0,1,2,3\} \, , \label{eq:bdy-cond-x} \\
    \frac{1}{\sqrt{h}} h_{1b} \epsilon^{bc} \partial_c Y^i(\sigma_1,\sigma_2=0) &= \dot{y}^i(\sigma_1) &  & {\rm for} \, i \in \{1,\ldots,6\} \, , \label{eq:bdy-cond-y}
\end{align}
where $X^\mu$ denotes the usual Mink${}_4$ coordinates  and $Y^i$ is a 6-dimensional vector with the magnitude given by the AdS${}_5$ radial coordinate $z$, and direction specified by $\hat{n} \in S_5$, $h_{ab} = \partial_a X^\mu \partial_b X^\nu g_{\mu \nu}$ is the induced metric on the worldsheet, $\sigma_1$ is the coordinate parametrizing the boundary contour, and $\sigma_2$ is a coordinate that runs into the worldsheet.

An immediate consequence of these boundary conditions, which is discussed explicitly in Ref.~\cite{Drukker:1999zq}, is that the area-minimizing extremal surface reaches the boundary $z = 0$ of AdS${}_5$ if and only if the loop variables obey the constraint $\dot{x}^2 = \dot{y}^2$\@. That is to say, only in this case $\sigma_2 = 0$ corresponds to $z = 0$\@. Once this constraint is incorporated, the inhomogeneous Neumann conditions in Eq.~\eqref{eq:bdy-cond-y} become Dirichlet conditions on the $S_5$ that select $\hat{n}^i(\sigma_1,z=0) = \dot{y}^i/|\dot{y}|$\@.

The boundary condition first proposed in Ref.~\cite{Alday:2007he} for the pure gauge Wilson loop~\eqref{eq:W-loop}, i.e., homogeneous Neumann boundary conditions for the $S_5$ coordinates, is exactly of the same form as in Eq.~\eqref{eq:bdy-cond-y} for the $S_5$ variables with the right hand side vanishing, but prescribes $z=0$ as a Dirichlet condition. This is more explicitly written in Ref.~\cite{Polchinski:2011im}, where the boundary conditions are written as $z = 0$ and $n_a h^{ab} \partial_b U^i = 0$, where, in their notation, $U^i \in S_5$ is the unit vector that we call $\hat{n}$, and $n_a$ is a unit vector normal to the worldsheet boundary.\footnote{Choosing coordinates such that $n_a \leftrightarrow \partial/\partial \sigma_2$, one can show that $n_a h^{ab} \propto h_{1a} \epsilon^{ab}$, and therefore both ways of writing down the Neumann boundary condition are equivalent.}

The motivation behind this prescription of homogeneous boundary conditions for the $S_5$ variables in Ref.~\cite{Alday:2007he} was not disconnected from the preceding discussion of the inhomogeneous Neumann boundary condition shown in Eq.~\eqref{eq:bdy-cond-y} introduced in Ref.~\cite{Drukker:1999zq}\@. Indeed, part of the reasoning that led to this prescription in Ref.~\cite{Alday:2007he} was that when we go to the real time description of Eq.~\eqref{eq:W-loop-DGO} (i.e., to Minkowski signature), provided the condition $\dot{y}^2 = \dot{x}^2$ is met, the coupling to the scalars disappears if we choose $x^\mu$ to be a lightlike path $\dot{x}^2 = 0$, because in this way we force $\dot{y}^i = 0$, and as such, the original Wilson loop~\eqref{eq:W-loop} is recovered.\footnote{See Section 4 of Ref.~\cite{Alday:2007he}\@.} The boundary conditions induced on the $S_5$ variables by this boundary contour are exactly homogeneous Neumann conditions, thus substantiating the proposal in Ref.~\cite{Alday:2007he}\@. It then also follows that, if the path of a supersymmetric Wilson loop at constant $\hat{n}$ can be approximated by a lightlike path in such a way that the extremal worldsheets of both configurations are approximately the same, converging onto each other when the lightlike approximation of the original path becomes better and better (e.g.~by using many small lightlike segments with neighboring segments perpendicular to approximate a straight line), then the expectation values of both Wilson loops in Eqs.~\eqref{eq:W-loop} and~\eqref{eq:W-loop-S} calculated with their respective prescriptions will agree, since a constant $\hat{n}$ fulfills the homogeneous Neumann condition.

However, for our setup where the two anti-parallel Wilson lines are coincident in space, if $\hat{n}$ is held constant, it is not clear whether a lightlike deformation of the (timelike) boundary contour produces a controllable approximation to the worldsheet obtained from the original undeformed contour. Indeed, given that the radial AdS${}_5$ extent of the worldsheet we will find in Section~\ref{sec:HQ-setup} is proportional to the distance $L$ between the two Wilson lines, which we want to take to zero $L \to 0$, any deformation (lightlike or otherwise) will qualitatively affect the resulting extremal surface, and thus there is no guarantee that the generalized area of the worldsheet with the deformed contour will have the same value as the original undeformed one. Furthermore, for the proper definition of our observable, the first limit we should take is $L \to 0$, because it is only in this limit where we are free to modify the temporal extent of the contour along the time direction without changing the result. This motivates us to look more closely at how the homogeneous Neumann condition can be obtained as a dual description of the standard Wilson loop~\eqref{eq:W-loop} and the role of $\hat{n}$\@.

Another way of arguing for the Neumann prescription for the pure gauge Wilson loop~\eqref{eq:W-loop} has been given in Ref.~\cite{Polchinski:2011im}\@. 
Namely, by considering the value of $\hat{n}$ on the boundary contour as another dynamical variable to be integrated over in the path integral, one can write
\begin{equation}
    \int D \! \left[\hat{n}(s)\right] \langle W_{\rm BPS}[\mathcal{C};\hat{n}(s)] \rangle = \int D \! \left[\hat{n}(s)\right] e^{i\mathcal{S}_{\rm NG} [\Sigma(\mathcal{C};\hat{n})]} \, , \label{eq:Polckinski-proposal}
\end{equation}
where both $\langle W_{\rm BPS}[\mathcal{C};\hat{n}(s)] \rangle$ and $e^{i\mathcal{S}_{\rm NG} [\Sigma(\mathcal{C};\hat{n})]}$ are obtained from a path integral in $\mathcal{N}=4$ SYM and 10-dimensional Supergravity, respectively. On the right hand side of this equality, treating $\hat{n}(s)$ as a dynamical variable gives the Neumann condition as an equation of motion. On the left hand side, one can argue as in Ref.~\cite{Polchinski:2011im} that this path integral gives the pure gauge Wilson loop~\eqref{eq:W-loop}\@. This is achieved by expanding the Wilson loop in a power series in $\hat{n}(s)$, noting that the first term is exactly the Wilson loop~\eqref{eq:W-loop}, and the rest of the terms either vanish by symmetry or are irrelevant operators. 

In fact, in both ways of arguing for the Neumann prescription, no reference to the constant $\hat{n}$ solution is made in its formulation. The connection to the constant $\hat{n}$ solution only appears by noting that, since a constant $\hat{n}$ satisfies the Neumann condition, it follows that the dual descriptions of Eqs.~\eqref{eq:W-loop} and~\eqref{eq:W-loop-S} are governed by the same saddle points, provided that this saddle point is the dominant one, which is usually the case. 

However, direct inspection of the left hand side of Eq.~\eqref{eq:Polckinski-proposal} reveals that there is another saddle point for our contour of interest $\mathcal{C}$, given in Eq.~\eqref{eq:C-contour-0}\@. Namely, the equations of motion that extremize the left hand side of Eq.~\eqref{eq:Polckinski-proposal} have a solution given by $\hat{n}(s) = \hat{n}_0$ for $s \in (-\T/2,\T/2)$ and $\hat{n}(s) = -\hat{n}_0$ for $s \in (\T/2,3\T/2)$, where $\hat{n}_0$ is a fixed direction on $S_5$\@. This is easy to see: any first-order variation of the Wilson loop will give zero because varying it with respect to any field in it will result in an operator insertion along the loop, proportional to an SU$(N_c)$ generator matrix $T^a$\@. For the antipodal $\hat{n}(s)$ configuration that we claim as a solution, in the equation of motion for the $\hat{n}(s)$ variable, the Wilson lines cancel and all that remains is proportional to ${\rm Tr}_{\rm color}[T^a] = 0$\@. The rest of the saddle point configuration is determined by the variations of the $\mathcal{N}=4$ SYM action, which provides its standard equations of motion (i.e., the same equations as in the absence of a Wilson loop)\@.

Furthermore, when we select $\hat{n}$ with antipodal positions on the $S_5$, because the Wilson lines are at separate, antipodal coordinates, it seems plausible that the contour $\mathcal{C}$ as given in Eq.~\eqref{eq:C-contour-0} does admit a lightlike approximation that modifies the extremal surface in such a way that it converges to the unmodified solution as the approximation is made finer. This is so because, as opposed to the case of Section~\ref{sec:HQ-setup}, the size of the deformation here can indeed be taken to be much smaller than the radial extent of the unperturbed worldsheet we will find in Section~\ref{sec:QQ-setup}, which is $(\pi T)^{-1}$\@. Proceeding in this way, we will find a saddle point that should\footnote{We say ``should'' because, as will be made apparent by our discussion in Section~\ref{sec:QQ-setup}, at present we have no explicit extremal surface solution with which to verify the homogeneous Neumann solution at the endpoints of the contour shown in Eq.~\eqref{eq:C-contour-0}, i.e., at $s = \pm \T/2$\@. This is an interesting direction that we leave open to future work.} respect the homogeneous Neumann boundary condition (it is guaranteed to respect it if the proposal of Ref.~\cite{Polchinski:2011im}, expressed through Eq.~\eqref{eq:Polckinski-proposal}, holds), and yields a result $\langle W[\mathcal{C}] \rangle = 1$\@. Therefore, because the pure gauge Wilson loop~\eqref{eq:W-loop} satisfies the unitarity bound $\langle W[\mathcal{C}] \rangle \leq 1$, we would have necessarily found an extremal solution of minimal energy, and consequently, this saddle point would be the one that provides a dual description of the Wilson loop on the contour shown in Eq.~\eqref{eq:C-contour-0}\@. We stop short of claiming a proof of this result because an explicit verification should also provide the extremal worldsheet that is dual to the Wilson loop on the path $\mathcal{C}$ given by Eq.~\eqref{eq:C-contour-0} and explicitly verify all of the boundary conditions discussed above. However, we do conjecture that the pure gauge adjoint Wilson line has the antipodal $\hat{n}$ configuration as its gravitational dual through the gauge/gravity duality.

To close this section, we note that when there is a nonvanishing spatial separation between the Wilson lines that comprise a Wilson loop, as in the case of the heavy quark interaction potential, one can indeed use the same solution as at constant $\hat{n}$ to describe the extremal worldsheet that gives the expectation value for the pure gauge Wilson loop~\eqref{eq:W-loop}, in accordance to the conjecture laid out in Ref.~\cite{Alday:2007he}\@. One may then worry about the unitarity bound $\langle W[\mathcal{C}_L] \rangle \leq 1$ being violated. However, the situation here is rather similar to that of QCD: only after regularizing and renormalizing does the statement $\log \langle W[\mathcal{C}_L] \rangle \propto 1/L $ make sense. For concreteness, we consider QCD on a 4-dimensional Euclidean lattice, characterized by a lattice spacing $a$\@. If one considers simply the expectation value $\langle W[\mathcal{C}_L] \rangle$ in the limit $L\to0$ at fixed $a$, the value will only converge to $1$ if lattice artifacts are taken into account, which happen when $L\lesssim a$\@. 
This means that to access $\langle W[\mathcal{C}_{L=0}] \rangle$ on the lattice, one has to take $a\to0$ first before taking $L\to0$\@. At finite $L$ the expectation value $\langle W[\mathcal{C}_L] \rangle$ contains both an $L$-dependent term and an $L$-independent diverging term, and they cancel at $L=0$\@.
However, the ratio $\langle W[\mathcal{C}_{L_1}] \rangle/\langle W[\mathcal{C}_{L_2}] \rangle$ will be finite as $a\to 0$ with $L_1, L_2 > 0$ and will exhibit features of a Coulomb potential at distances small compared to the non-perturbative scale $\Lambda_{\rm QCD}^{-1}$\@. While it is true that $\langle W[\mathcal{C}] \rangle = 1$ for the contour shown in Eq.~\eqref{eq:C-contour-0}, its ratio with $\langle W[\mathcal{C}_L] \rangle$ at $L>0$ in the continuum limit is formally infinite. To give meaning to the Wilson loop at a finite spatial separation $L$ and extract energy differences from $\langle W[\mathcal{C}_L] \rangle$, renormalization and regularization is required. Taking ratios achieves this immediately. The need for extra regularization is also clear from calculations in the gravitational description where the boundary contour is modified to be lightlike, see, e.g., Eq.~(3.7) in Ref.~\cite{Alday:2007he}\@.

\subsection{Variations of Wilson loops in AdS/CFT} \label{sec:EE-setup-AdS}

Having introduced the necessary concepts to formulate the calculation of the expectation value of a Wilson loop and its path variations (i.e., functional derivatives) that provide the non-Abelian electric field correlator we want to calculate in Yang-Mills theory, we now proceed to describe how the calculation of these path variations takes place in the gravitational description of a Wilson loop in $\mathcal{N}=4$ SYM\@. Concretely, we will first lay out the steps one needs to follow to extract correlation functions from the calculation of an extremal worldsheet in a gravitational description and its derivatives. We will defer their evaluation to the concrete setups we discuss in sections~\ref{sec:HQ-setup} and~\ref{sec:QQ-setup}\@.

Secondly, since we are interested in the real-time correlation functions of operators in a thermal ensemble, we will discuss the setup of our calculation on the Schwinger-Keldysh contour and how it is realized holographically in the dual gravitational description in the last part of this section. Specifically, we will discuss the $i\epsilon$ prescription appropriate for time-ordered quantities, which is exactly the nature of the correlation function we want to calculate. We will also discuss the qualitative differences between the observable we will calculate and the correlation function that defines the heavy quark diffusion coefficient.

\subsubsection{Generating non-Abelian electric field insertions} \label{sec:generating-EE}

Our goal is to insert field strength operators along the boundary contour $\mathcal{C}$\@. As we described in Section~\ref{sec:setup}, this is achieved by taking functional derivatives with respect to deformations of the contour $\mathcal{C}$, parametrized by $h^i(t)$\@. The corresponding operation in the gravitational description is to take functional derivatives with respect to the boundary conditions of the extremal surface. Operationally, we can achieve this by: 
\begin{enumerate}
    \item introducing fluctuation fields $y_i(\tau,\sigma)$ in all directions $i \in \{0,1,2,3,4\}$ on the worldsheet, removing the linear combinations that constitute a coordinate reparametrization, 
    \item expanding the Nambu-Goto action up to a given power $p$ in these perturbations $y_i$,
    \item solving the equations of motion of $y_i$ up to the same order $p$ as a function of arbitrary boundary conditions $h^i(t)$,
    \item and finally, evaluating the Nambu-Goto action expanded up to order $p$, i.e., $\mathcal{S}^{(p)}_{\rm NG}[\Sigma]$ on the worldsheet solution $\Sigma = \Sigma^{(p)}(\ml{C};h)$ obtained up to order $p$ in step 3.
\end{enumerate}
This can be done systematically, starting from the lowest order up to the desired number of powers in the perturbation. The result may be organized as
\begin{align}
    \mathcal{S}_{\rm NG}^{(p)}[\mathcal{C};h] &= \mathcal{S}_{\rm NG}[\mathcal{C};h=0] + \sum_{n=2}^p \frac{1}{n!} \int dt_1 \ldots dt_n \left. \frac{\delta^n \mathcal{S}[\mathcal{C};h]}{\delta h^{i_1}(t_1) \cdots \delta h^{i_n}(t_n)} \right|_{h=0} \, h^{i_1}(t_1) \cdots h^{i_n}(t_n)  \, ,
\end{align}
where the kernel $\left. \frac{\delta^n \mathcal{S}_{\rm NG}[\mathcal{C};h]}{\delta h^{i_1}(t_1) \cdots \delta h^{i_n}(t_n)} \right|_{h=0}$ can be obtained from the four steps listed above. With this definition, $\mathcal{S}_{\rm NG}^{(p)}[\mathcal{C};h]$ is the generating functional for (non-Abelian electric) field strength insertions along the contour $\mathcal{C}$ up to order $p$, which allows one to evaluate correlation functions with up to $p$ insertions of operators.
This object fully characterizes the $n$-point functions of non-Abelian electric field strength insertions along the contour $\mathcal{C}$ at leading order in the large 't Hooft coupling limit, as discussed in the paragraph before Eq.~\eqref{eq:Schwarzschild-AdS}\@.
To marginally ease the notation, let us introduce
\begin{align}
    \Delta_{ij}(t_1,t_2) = - \left. \frac{\delta^2  \mathcal{S}_{\rm NG}[\mathcal{C};h] }{\delta h^i(t_1) \delta h^j(t_2)} \right|_{h=0} \, ,
\end{align}
which we will determine by explicit calculation in the following sections.

From the definition of our operator of interest in terms of Wilson loop variations~\eqref{eq:EE-from-variations}, it should be clear that, \textit{if} the Nambu-Goto action gave a dual description of the pure gauge Wilson loop~\eqref{eq:W-loop}, then its linear response kernel $\Delta_{ij}^{EE}$ would be equal (up to an overall factor) to the chromoelectric field correlation function of our interest dressed by the respective Wilson loop:
\begin{align}
4(ig)^2 T_F \frac{1}{N_c} \frac{\langle \hat{\mathcal{T}} E_i^a(t_2) \mathcal{W}^{ab}_{[t_2,t_1]} E_j^b(t_1) \rangle_T}{\langle \hat{\mathcal{T}} W[\mathcal{C}] \rangle} &=
\left. \frac{1}{\langle \hat{\mathcal{T}} W[\mathcal{C}_h] \rangle}  \frac{\delta}{\delta h^{i}(t_2) } \frac{\delta}{\delta h^{j}(t_1) } \langle \hat{\mathcal{T}} W[\mathcal{C}_h] \rangle \right|_{ h = 0}  \nonumber \\
&= \left. \frac{1}{e^{i \mathcal{S}_{\rm NG}[\mathcal{C};h] }} \frac{\delta}{\delta h^{i}(t_2) } \frac{\delta}{\delta h^{j}(t_1) } e^{i \mathcal{S}_{\rm NG}[\mathcal{C};h] } \right|_{h = 0}  \nonumber \\
&= -i  \Delta^{EE}_{ij}(t_2 - t_1)  \, , \label{eq:EE-true-from-Delta}
\end{align}
where $\ml{C}_h$ denotes the contour $\ml{C}$ perturbed by $h$\@.

We note, however, that the Wilson loop in the duality~\eqref{eq:duality} is \textit{not} the pure gauge loop~\eqref{eq:W-loop}, because it also involves scalar fields~\eqref{eq:W-loop-S}, and so its path variations also give rise to a  contribution from the scalars that modifies the chromoelectric field operators. Namely, assuming that the fluctuations $h^i$ are only in the spatial directions, the non-Abelian field strength insertions are modified to
\begin{equation}
    \dot{x}^\nu F_{i\nu}^a(t) \to \dot{x}^\nu F_{i\nu}^a(t) + \hat{n} \! \cdot \! \big[ D_i \vec{\phi} \, \big]^a \, ,
\end{equation}
where $\hat{n}$ is the direction on the $S_5$ that appears in~\eqref{eq:W-loop-S}, and $\left[ D_i \phi \right]^a = \partial_i \phi^a + g f^{abc} A_i^b \phi^c$ is the gauge covariant derivative of the scalar field. Therefore, after defining 
\begin{equation}
    \tilde{E}_i^a(t) = E_i^a(t) - {\rm sgn}(\dot{x}^0) \, \hat{n} \! \cdot \! \big[ D_i \vec{\phi} \, \big]^a \, ,
\end{equation}
what we can calculate using the duality~\eqref{eq:duality} is
\begin{align}
4(ig)^2 T_F \frac{1}{N_c} \frac{\langle \hat{\mathcal{T}} \tilde{E}_i^a(t_2) \mathcal{W}^{ab}_{[t_2,t_1]} \tilde{E}_j^b(t_1) \rangle_T}{\langle \hat{\mathcal{T}} W_{\rm BPS}[\mathcal{C}] \rangle} &=
\left. \frac{1}{\langle \hat{\mathcal{T}} W_{\rm BPS}[\mathcal{C}_h] \rangle}  \frac{\delta}{\delta h^{i}(t_2) } \frac{\delta}{\delta h^{j}(t_1) } \langle \hat{\mathcal{T}} W_{\rm BPS}[\mathcal{C}_h] \rangle \right|_{ h = 0}  \nonumber \\
&= \left. \frac{1}{e^{i \mathcal{S}_{\rm NG}[\mathcal{C};h] }} \frac{\delta}{\delta h^{i}(t_2) } \frac{\delta}{\delta h^{j}(t_1) } e^{i \mathcal{S}_{\rm NG}[\mathcal{C};h] } \right|_{h = 0}  \nonumber \\
&= -i  \Delta_{ij}(t_2 - t_1)  \,.
\label{eq:EE-from-Delta}
\end{align}
In this expression, $\hat{n}$ takes the sign that it has on the part of the contour where the derivative is taken.\footnote{We note that since the sign of $\dot{x}^\mu$ flips for the two timelike segments of our contour $\mathcal{C}$, for the operator insertions to be equal it is also necessary to flip the sign of $\hat{n}$, as prescribed by our setup in Section~\ref{sec:QQ-setup}\@. In the setup of Section~\ref{sec:HQ-setup}, operators inserted on opposite sides of the contour will have a different relative sign between the $E_i^a$ fields and the scalars $\phi^a$\@.} If our conjecture at the end of Section~\ref{sec:nhat} holds true, then we also have $\Delta_{ij}(t) = \Delta^{EE}_{ij}(t)$, and the distinction becomes idle. In this case, according to the prescription presented in Ref.~\cite{Alday:2007he}, the contribution from the scalars disappears.

In these expressions, we have included an extra normalization factor given by the unperturbed Wilson loop. As explained before, we should have $\langle \hat{\mathcal{T}} W[\mathcal{C}] \rangle = 1$ for a Wilson loop in pure Yang-Mills theory (or QCD), and also $\langle \hat{\mathcal{T}} W_{\rm BPS}[\mathcal{C}] \rangle = 1$ for the configuration we discuss in Section~\ref{sec:QQ-setup}, but, as we will remind the reader later, this is not the case for the loop that describes the heavy quark interaction potential, which we discuss in Section~\ref{sec:HQ-setup}\@. In this situation it is appropriate to normalize the correlator by the expectation value of the ``background'' Wilson loop.

When we do have $\langle \hat{\mathcal{T}} W[\mathcal{C}] \rangle = 1$, we can summarize the above result as
\begin{align} \label{eq:EE-Delta}
\frac{g^2}{N_c} [g_E^{\T}]_{ij}(t_2 - t_1) = \frac{g^2}{N_c} \langle \hat{\mathcal{T}} E_i^a(t_2) \mathcal{W}^{ab}_{[t_2,t_1]} E_j^b(t_1) \rangle_T = \frac{i}{2}  \Delta_{ij}(t_2 - t_1) \, ,
\end{align}
where we have used the standard normalization $T_F = 1/2$ for the fundamental representation of SU($N_c$)\@. Therefore, the kernel $\Delta_{ij}$ is exactly the object we are interested in, for each worldsheet configuration of interest. Before proceeding to calculate them, we will take a small digression to discuss aspects of the worldsheet near the turning points at $t = \pm \T/2$\@. In particular, we will discuss how the time-ordering $i\epsilon$ prescription emerges in this setup, and we will also comment on the interplay between the chosen form for the boundary conditions $h^i$ and how the fluctuations behave near $t = \pm \T/2$\@. 

\subsubsection{The Schwinger-Keldysh contour in AdS/CFT} \label{sec:ends-matching}

The fact that we want to calculate a thermal expectation value requires us to introduce the Schwinger-Keldysh (SK) contour~\cite{keldysh1965diagram} in order to represent the observable of interest through path integrals. The holographic realization of the Schwinger-Keldysh contour in AdS/CFT dates back to the early work of Herzog, Son and Starinets~\cite{Son:2002sd,Herzog:2002pc}, which was more recently expanded and refined by Skenderis and van Rees~\cite{Skenderis:2008dg,Skenderis:2008dh,vanRees:2009rw}\@. In a nutshell, each segment of the Schwinger-Keldysh contour is the boundary of an asymptotically AdS${}_5$ bulk geometry, and these bulk spacetimes are glued together according to appropriate matching conditions, discussed at length in Ref.~\cite{Skenderis:2008dg}\@. This allows one to formulate the Schwinger-Keldysh contour (and, consequently, the bulk geometries) in the complex time plane holographically. As in any quantum-mechanical theory, the thermal nature of the average is dictated by the fact that modes with energy $\omega$ are coupled to themselves across a Euclidean time direction of extent $\beta$, which gives rise to the characteristic thermal statistics through factors of $e^{-\beta \omega}$, and this is realized in the holographic setup by matching the bulk manifolds accordingly. Importantly, the contour in the complex time plane also defines the $i\epsilon$ prescriptions necessary to define the correlation functions in the limit where we take $\T \to \infty$\@. This limit is convenient to do calculations because it restores local translational invariance in the time direction, which is also necessary if one wants to extract the Fourier components of a correlation function at an arbitrary frequency $\omega$, and crucially, to have a continuous limit of it as $\omega \to 0$\@. Because of this, and as we will state explicitly later on, we will indeed take the limit $\T \to \infty$ for both of the configurations that we have introduced and will study in Sections~\ref{sec:HQ-setup} and~\ref{sec:QQ-setup}\@. 

We will derive the appropriate $i\epsilon$ prescription for our setup. Before proceeding, it is also important to note that not all observables are equally sensitive to the thermal nature of the contour. In particular, for our correlation function of interest, which is defined through a Wilson loop that ``backtracks'' over the path that it traverses to cover the distance between the two electric field insertions, the extremal surface that defines the expectation value of this Wilson loop will lie only inside the bulk manifold that has the time-ordered part of the Schwinger-Keldysh contour as its boundary. This is so because, crucially, all operators are time-ordered in our setup. The main consequence of this is that the fluctuations that propagate on top of this extremal worldsheet will lose sensitivity to the Euclidean (thermal) part of the Schwinger-Keldysh contour. This is one of the most important qualitative differences between our setup and the heavy quark diffusion calculation in Ref.~\cite{Casalderrey-Solana:2006fio}, where such a distinction is necessary. We will elaborate further on this at the end of this section, after establishing the prescriptions to select the mode solutions in our setup, in accordance with the boundary conditions that the fluctuations must satisfy.

\paragraph{The $i\epsilon$ prescription for time-ordered correlation functions} \hspace{\fill}

The following discussion applies to both configurations to be discussed in Sections~\ref{sec:HQ-setup} and~\ref{sec:QQ-setup}\@. We will first make use of the deformability of the Schwinger-Keldysh contour to derive the $i\epsilon$ prescription that is appropriate for the time-ordered correlation function we want to calculate. Secondly, and as a consistency check, we will verify that this prescription is the same as that one would get by considering the contributions from the turnaround points ($t = \pm \T/2$) of the Wilson loop where the path $\mathcal{C}$ becomes spacelike.

To carry out this derivation, it is first necessary to make some remarks about the nature of the Schwinger-Keldysh contour. In principle, any contour starting at $t = t_i$ and ending at $t = t_i -i\beta$ in the complex time plane is adequate in order to evaluate thermal expectation values. The only requisites are that:
\begin{enumerate}
    \item The operators of interest are placed at some point along this contour, and
    \item The contour itself never goes upward in the complex time plane, i.e., its tangent vector always has a non-positive imaginary part. This is necessary for the path integral to be evaluated by a saddle point approximation.
\end{enumerate}
The standard Schwinger-Keldysh contour achieves this by first going from $t=t_i$ along the real axis to some final time $t_f$ (which defines its time-ordered segment), then turning around and going from $t_f$ to $t_i$ along the real time axis (infinitesimally displaced by $-i\epsilon$, which defines the anti-time-ordered segment), and finally going from $t_i$ to $t_i - i\beta$ to close the contour. In fact, the second point of our requisites suggests that we can tilt the time-ordered and the anti-time-ordered parts of the Schwinger-Keldysh contour slightly: Instead of drawing the time-ordered contour exactly along the real axis, parametrized by $t_c = t'$, where $t' \in [0, t_f - t_i]$ is the parameter running along the contour, the natural way to have a well-defined saddle point is to take $t_c = t_i + t'(1 - i\epsilon)$ for the time-ordered branch, with $\epsilon$ a small (infinitesimal) positive number. The anti-time-ordered branch, going back to $t_i$ must then be parametrized by $t_c = t_f - i\epsilon - t''(1 + i\epsilon)$, with $t'' \in (0, t_f - t_i)$\@. The contour is then closed by a straight line along the imaginary time axis at ${\rm Re}(t_c) = t_i$ downward until reaching $t = t_i - i\beta$. As we will see momentarily, these $i\epsilon$ deformations provide exactly the standard time-ordered and anti-time ordered prescriptions to evaluate correlation functions.

With this setup, we can consider the Nambu-Goto action that describes the worldsheet in the spacetime that has the time-ordered segment of the Schwinger-Keldysh contour as its boundary. In terms of the parameter $t'$, which we write as $t$ in what follows, the metric reads as
\begin{equation}
    ds^2 = \frac{1}{z^2} \left[ - f(z) \, dt^2 (1 - i\epsilon) + d{\bf x}^2 + \frac{1}{f(z)} dz^2 + z^2 d\Omega_5^2 \right] \, .
\end{equation}
The more general case we will consider for our purposes is that of a worldsheet parametrized by
\begin{equation}
    X^\mu = ( t(s,z), x(s,z) , y(s,z) , 0, z, \hat{n}(s,z)) ) \, ,
\end{equation}
where $s$ is a worldsheet coordinate that parametrizes the Wilson loop, which we may define at $z=0$ to be the arc length on the loop. The spacetime coordinates $t, x, z, \hat{n}$ describe the background solution, while $y$ describes the fluctuations we seek to solve for. In our setup, $\hat{n}$ is completely determined by a large circle angle $\phi(s,z)$ due to symmetry considerations. In what follows, we will denote the derivatives of a coordinate $a$ by $\frac{da}{dz} = a'$ and $\frac{da}{ds} = \dot{a}$\@.

Since we will be considering essentially infinitesimal perturbations $y$, whether the worldsheet is spacelike or timelike is wholly determined by the background solution. Expanding up to quadratic order on $y$, one obtains the following action:
\begin{align} \label{eq:quadratic-action-general}
    \mathcal{S}_{\rm NG}[\Sigma] &= - \frac{\sqrt{\lambda}}{2\pi} \left[ S_{\rm NG}^{(0)}[\Sigma_0] + S_{\rm NG}^{(2)}[\Sigma_0;y] + \ldots \right] \\
    S_{\rm NG}^{(0)}[\Sigma_0] &= \int \frac{\diff s \, \diff z}{z^2} \bigg\{ \left[ \dot{t}^2  + f \big( \dot{t} x' - t' \dot{x} \big)^2  + f z^2 \big( \dot{t} \phi' - \dot{\phi} t' \big)^2 \right] (1 - i\epsilon) \nonumber \\ & \quad \quad \quad \quad \quad \quad \quad \quad \quad \quad \quad \quad \quad  - \frac{\dot{x}^2}{f} - \frac{z^2 \dot{\phi}^2 }{f} - z^2 \big( \dot{x} \phi' - x' \dot{\phi} \big)^2 \bigg\}^{1/2} \label{eq:bkg-action-general} \\
    S_{\rm NG}^{(2)}[\Sigma_0;y] &= \int \frac{\diff s \, \diff z}{2 z^2} \left[ f \big(\dot{t} y' - t' \dot{y} \big)^2 (1 - i\epsilon) - \big( \dot{x} y' - x' \dot{y} \big)^2 - z^2 \big( \dot{y} \phi' - y' \dot{\phi} \big)^2 - \frac{\dot{y}^2}{f} \right] \nonumber \\ & \quad \quad \quad \quad \times \bigg\{ \left[ \dot{t}^2  + f \big( \dot{t} x' - t' \dot{x} \big)^2  + f z^2 \big( \dot{t} \phi' - \dot{\phi} t' \big)^2 \right] (1 - i\epsilon) \nonumber \\ & \quad \quad \quad \quad \quad \quad \quad \quad \quad \quad \quad \quad \quad  - \frac{\dot{x}^2}{f} - \frac{z^2 \dot{\phi}^2 }{f} - z^2 \big( \dot{x} \phi' - x' \dot{\phi} \big)^2 \bigg\}^{-1/2} \label{eq:fluct-action-general} \,.
\end{align}
Moreover, Eq.~\eqref{eq:fluct-action-general} provides a prescription that affects the equations of motion of $y$\@. To see this in a concrete setup (which will be relevant in Section~\ref{sec:QQ-setup}), we consider the case of a radially infalling background worldsheet at constant $x$ and $\hat{n}$, with $t = s$\@. The action for the fluctuations simplifies to
\begin{equation}
    S^{(2)}_{\rm NG}[\Sigma_0;y] = \int \frac{\diff s \,\diff z}{2 z^2} \left[ f y'{}^2 (1-i\epsilon)^{1/2} - \frac{1}{f} \dot{y}^2 (1-i\epsilon)^{-1/2} \right] \, ,
\end{equation}
which implies that at the level of the equations of motion, the frequency $\omega$ of the mode solutions will always appear as $\omega^2 (1 + i\epsilon)$. This in turn defines the pole prescription to evaluate the propagator. As we will see later in Section~\ref{sec:EE-calculation-QQ}, this also determines which mode solution should be used when calculating correlation functions.

One may also wonder whether the behavior of the worldsheet around the turnaround times $t = \pm \T/2$ affects this conclusion. Specifically, one can wonder whether one can get extra imaginary terms in the equations of motion by having a transition where the induced metric on the worldsheet goes from having Minkowski signature (i.e., timelike) to having Euclidean signature (i.e., spacelike)\@.

To have control over the behavior of the worldsheet at the turnaround times $t = \pm \T/2$, we need to regulate the backtracking of the loop in a way that its tangent vector is continuous throughout, such that if we look closely enough, the extremal surface will still be smooth. Our choice of regulator for the present purpose is to introduce a small spatial separation $L$ between the two lines, which is compatible with the discussion in the previous sections. Conversely, the only way to smoothly turn from a timelike tangent vector $\dot{x}^\mu$ going in the future direction to one going in the past direction is by having a segment where it is spacelike. As such, our choice for a regulator is actually generic.

This motivates studying the behavior of a worldsheet close to a spacelike boundary segment. To gain intuition, let us first discuss a few examples. A family of solutions that is easily obtained at $T=0$ is $z(t,x) = \sqrt{t^2 - x^2 - \rho_0^2}$, either for $t \geq \sqrt{x^2 + \rho_0^2}$ or for $t \leq - \sqrt{x^2 + \rho_0^2}$\@. These solutions are spacelike surfaces that satisfy the Euler-Lagrange equations obtained from the Nambu-Goto action that are bounded by the hyperbola $t^2 - x^2 = \rho_0^2$ at $z = 0$\@. Another family of solutions to the Euler-Lagrange equations is given (implicitly) by the integral $\int_0^{z(t)/z_c} \frac{u^2 \diff u}{\sqrt{1+u^4}} = \frac{t}{z_c}$, where $z_c$ is a parameter defining different solutions, all of which are bounded by the line $t=0$ at $z=0$, valid in a small neighborhood of a spatial Wilson line segment with varying $x$ and all else held constant (to fully determine a unique solution it is necessary to specify how the surface is closed, or equivalently, how the Wilson loop path closes itself, far away from the region we just studied)\@. All of these have the crucial property that they are spacelike surfaces, which motivates investigating whether our previous conclusion is affected when we deform the contour slightly by introducing a spatial separation.

Note that the $i\epsilon$ in Eq.~\eqref{eq:bkg-action-general} also provides a prescription to evaluate the action in the case of a spacelike worldsheet. Specifically, it determines that a spacelike worldsheet has a Nambu-Goto action determined by the substitution
\begin{equation} \label{eq:spacelike-prescription}
    \sqrt{- \det \left( \partial_\alpha X^\mu \partial_\beta X^\nu g_{\mu \nu} \right) } \to -i \sqrt{\left| \det \left( \partial_\alpha X^\mu \partial_\beta X^\nu g_{\mu \nu} \right) \right| } \, ,
\end{equation}
which, satisfactorily, is exactly what we would get by demanding that whenever the worldsheet is spacelike, the Nambu-Goto action should be the same as if we had started in Euclidean signature from the beginning.

Now we may ask what happens if we include perturbations on top of a background worldsheet that features a transition from spacelike to timelike and vice-versa. Given that these perturbations are introduced on top of a solution that extremizes the action, the action for the fluctuations in a spacelike region should be real and positive definite. We will verify this explicitly in what follows, as it will be crucial to our results that the $i\epsilon$ prescription would not be modified by contributions from a spacelike region.

When the background worldsheet is spacelike, the argument of the square root in Eq.~\eqref{eq:bkg-action-general} becomes negative, and we must therefore use Eq.~\eqref{eq:spacelike-prescription} to get
\begin{equation}
    i S_{\rm NG}^{(0)}[\Sigma_0] = \! \int \frac{\diff s \,  \diff z}{z^2} \sqrt{ \frac{\dot{x}^2}{f} + \frac{z^2 \dot{\phi}^2 }{f}  - \dot{t}^2 - f \big( \dot{t} x' - t' \dot{x} \big)^2 - f z^2 \big( \dot{t} \phi' - \dot{\phi} t' \big)^2 + z^2 \big( \dot{x} \phi' - x' \dot{\phi} \big)^2 } \,, \label{eq:bkg-action-imag}
\end{equation}
which for the fluctuations read
\begin{equation}
    i S_{\rm NG}^{(2)}[\Sigma_0;y] \!= \!\! \int \frac{\diff s \,  \diff z}{2 z^2} \frac{- f \big(\dot{t} y' - t' \dot{y} \big)^2 + \big( \dot{x} y' - x' \dot{y} \big)^2 + z^2 \big( \dot{y} \phi' - y' \dot{\phi} \big)^2 + \frac{\dot{y}^2}{f} }{\sqrt{ \frac{\dot{x}^2}{f} + \frac{z^2 \dot{\phi}^2 }{f}  - \dot{t}^2 - f \big( \dot{t} x' - t' \dot{x} \big)^2 - f z^2 \big( \dot{t} \phi' - \dot{\phi} t' \big)^2 + z^2 \big( \dot{x} \phi' - x' \dot{\phi} \big)^2 }} \,, \label{eq:fluct-action-imag}
\end{equation}
As written, this is a general expression. However, the expression is explicit enough for us to make generic statements about how the fluctuations $y(s,z)$ behave on a background specified by $X^\mu = (t(s,z), x(s,z),0,0,z,\hat{n}(s,z))$\@. The key observation is that the quadratic form in the numerator of the integrand in Eq.~\eqref{eq:fluct-action-imag} can be written as
\begin{equation}
    \begin{pmatrix}
        y' & \dot{y} 
    \end{pmatrix} 
    \begin{pmatrix}
        \dot{x}^2 + z^2 \dot{\phi}^2 - f \dot{t}^2 & f \dot{t} t' - \dot{x} x' - z^2 \dot{\phi} \phi' \\
        f \dot{t} t' - \dot{x} x' - z^2 \dot{\phi} \phi' &  \frac{1}{f} + x'{}^2 + z^2 \phi'{}^2 - f t'{}^2
    \end{pmatrix}
    \begin{pmatrix}
        y' \\ \dot{y} 
    \end{pmatrix} \, ,
\end{equation}
and noting that the $2 \times 2$ matrix in the middle of this expression is equal, component by component, to $ \partial_\alpha X^\mu \partial_\beta X^\nu g_{\mu \nu}$, where $X^\mu = (t(s,z), x(s,z),0,0,z,\hat{n}(s,z))$ describes the background solution, with the first component of the matrix (for the indices $\alpha, \beta$) corresponding to a derivative with respect to $s$, and the second with respect to $z$. Therefore, if the background worldsheet is spacelike, it follows that both eigenvalues of the induced metric $ \partial_\alpha X^\mu \partial_\beta X^\nu g_{\mu \nu}$ are positive, and hence, that it is a positive definite matrix. Consequently, the action for the fluctuations~\eqref{eq:fluct-action-imag} is positive definite whenever the background worldsheet is spacelike.

Then, extending the boundary contour as $\T \to \infty$, the contributions of these regions will be of the form (for definiteness, consider $\tau = \T/2 \to +\infty$)
\begin{equation}
    i S_{\rm NG}^{(2)}[y]_{\rm spacelike} = \int_0^{(\pi T)^{-1}} \!\!\!\!\!\! \diff z \, \vec{y}\, {}^T(\tau = +\infty,z)  \cdot \Sigma(z,z') \cdot \vec{y}(\tau = +\infty,z)
\end{equation}
for some positive definite quadratic form $\Sigma$ (note that the minus sign in the definition~\eqref{eq:quadratic-action-general} means that the Gaussian integral over the fluctuations $y$ is convergent)\@. The positive definiteness of $\Sigma$ is guaranteed by the fact that it is determined by the induced metric of a spacelike surface, as discussed below Eq.~\eqref{eq:fluct-action-imag}\@. The same is true for the region at $\tau = - \infty$\@. The net effect of this on the action, after decomposing $y$ in terms of the mode functions at intermediate times, is to add an $i\epsilon$ to the $\omega^2$ coming from the temporal derivatives in the action, with $\epsilon > 0$\@.\footnote{This is similar to the derivation of how the $i\epsilon$ appears in the free Feynman propagator in a quantum field theory (see, e.g.,~\cite{Weinberg:1995mt})\@.} This $i\epsilon$ modifies the mode equations by effectively shifting $\omega^2 \to \omega^2(1 + i\epsilon)$, in the same way as the modification induced by the Schwinger-Keldysh contour tilts on the complex plane. As such, the prescription is unambiguously determined.

As a side note, we remark that the above discussion did not address how the solutions on the different sides of the turnaround region are coupled to each other. Continuity of the fluctuations and of their (appropriately normalized) derivatives is a first requirement, but, as hinted from our previous discussions, closer inspection from the field theory perspective reveals that the solutions must actually be more constrained than that. To see this, if instead of considering perturbations as given by Eq.~\eqref{eq:f-antisymm}, we consider
\begin{align} \label{eq:f-symm}
f^\mu(s) = \begin{cases} g^\mu(s) & -\frac{\T}2 < s < \frac{\T}2 \\  g^\mu (\T - s) &  \frac{\T}2 < s < \frac{3\T}2  \end{cases} \, ,
\end{align}
we then obtain $W[\mathcal{C}_{f}] = 1$ for all $f$, since the loop consists of a single line that was traveled back and forth on top of each other. Hence, taking derivatives with respect to this kind of contour deformations gives zero, and as such, the corresponding response kernel for Wilson loop variations evaluated with AdS/CFT techniques must also be identically zero, whenever the loop satisfies $W[\mathcal{C}_f] = 1$ to begin with. We stress that this is the case for the loop with $\hat{n}$ at antipodal positions on the $S_5$, but it will not necessarily be the case when $\hat{n}$ is constant (even though, as we will see later, deformations as in Eq.~\eqref{eq:f-symm} do not contribute in this case as well)\@. That being said, since $W[\mathcal{C}_f] = 1$ is a property of the Wilson loop~\eqref{eq:W-loop}, the correlator we are after is unequivocally determined by the antisymmetric deformations, as presented in Eq.~\eqref{eq:f-antisymm}\@.

\paragraph{Differences with the heavy quark diffusion coefficient} \hspace{\fill}

Finally, let us comment on how this calculation differs from the one for the heavy quark diffusion coefficient~\cite{Casalderrey-Solana:2006fio}\@. To do this, it is most helpful to use the construction put forth by Skenderis and van Rees~\cite{Skenderis:2008dg,Skenderis:2008dh,vanRees:2009rw}, where the Schwinger-Keldysh contour has a concrete holographic realization. This is done by constructing a bulk manifold made up of several submanifolds satisfying appropriate matching conditions, where the boundary of each of these submanifolds is identified with the lower-dimensional spacetime that corresponds to a given segment of the Schwinger-Keldysh contour in the boundary CFT\@. This can be done in the same way for the correlator we are presently considering and the one that determines the heavy quark diffusion coefficient.

\begin{figure}
    \centering
    \includegraphics[width=\textwidth]{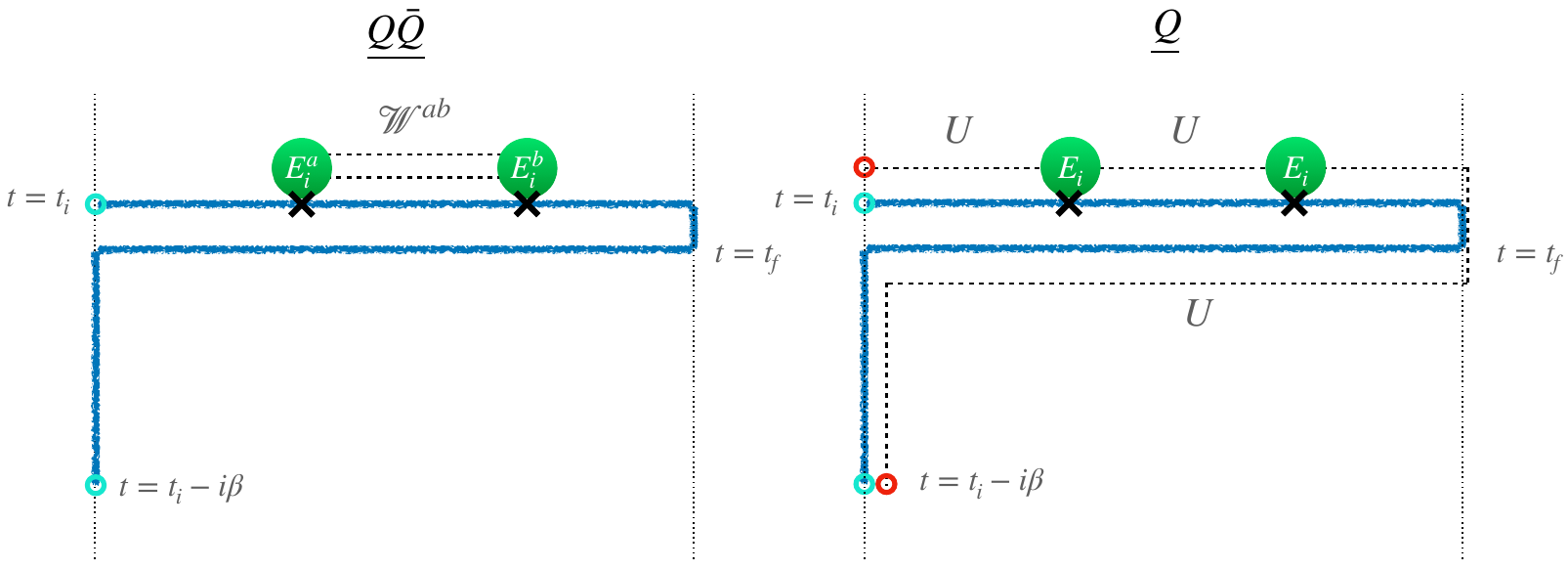}
    \caption{Schematic representation of the chromoelectric field correlators relevant for quarkonium transport (left) and for heavy quark diffusion (right)\@. The open cyan and red circles reflect that the corresponding ends of the contours should be identified. The adjoint Wilson line is denoted by $\W^{ab}$, and the fundamental lines by $U$. The difference in the Wilson line configuration reflects the different natures of the initial state of the QGP: in the quarkonia case, it is taken to be $\rho = \frac{1}{Z_{\rm QGP}} e^{-\beta H_{\rm QGP} } $, while in the single heavy quark case it is $\rho = \frac{1}{Z_{\rm HQ}} \sum_{Q} \langle Q | e^{-\beta H_{\rm tot}} | Q \rangle$, where $H_{\rm tot} = H_{\rm QGP} + H_Q$ and the sum over $Q$ runs over all states $| Q \rangle$ containing one heavy quark~\cite{McLerran:1981pb,Casalderrey-Solana:2006fio}\@.}
    \label{fig:EE-difference-SK}
\end{figure}

However, the manifold on which the fluctuations that are of interest to us propagate is not the full manifold associated to the Schwinger-Keldysh contour of the full CFT\@. Rather, the fluctuations propagate on a lower-dimensional manifold given by the background solution for the worldsheet configuration. Whether this worldsheet configuration spans every region of the Schwinger-Keldysh contour is solely determined by the shape of the boundary Wilson loop. In the case of the heavy quark diffusion coefficient setup, the corresponding Wilson loop consists of a single Wilson line that winds around the Schwinger-Keldysh contour once. This means that the background worldsheet configuration that defines the manifold on which fluctuations will propagate spans the whole bulk space at a single fixed position coordinate on the boundary ${\bs x}$, and is parametrized by a radial AdS coordinate and a temporal coordinate that goes over both Minkowski and Euclidean regions of the bulk geometry. The topology of the boundary manifold is that of a circle, and the end points of the real-time segments have to be matched with those of the imaginary-time segments, in consistency with the Schwinger-Keldysh contour (see Fig.~\ref{fig:EE-difference-SK})\@. Then, if one seeks for mode solutions for the fluctuations on top of this 2-dimensional geometry, the matching conditions discussed in Refs.~\cite{Skenderis:2008dg,Skenderis:2008dh,vanRees:2009rw} imply that Fourier modes $e^{-i\omega t} F_\omega(z)$ on the real-time segments (where $F_\omega(z)$ is the radial AdS profile of a solution with a frequency $\omega$ on the boundary) have to be matched with solutions of the form $e^{\pm \beta \omega} F_\omega(z)$ on the Euclidean segments. Therefore, factors of $e^{\beta \omega}$ naturally appear in the response functions. This is what gives rise to the KMS relations between the different types of correlation functions that can be calculated by introducing fluctuations and evaluating the response functions on different segments of the SK contour.\footnote{We note that for the heavy quark diffusion coefficient setup, the operator orderings of the chromoelectric field correlators that are related via KMS relations refer only to the chromoelectric field insertions, and do not affect the operator orderings of the Wilson lines (see also Appendix~\ref{sec:App-W-ordering})\@. This is a consequence of this correlator being derived from correlation functions of quark currents and subsequently integrating out the massive quark~\cite{Caron-Huot:2009ncn}\@. The fact that the heavy quark is present at all times means that, when one integrates it out, one should actually regard the Wilson line as a modification to the bath Hamiltonian enforcing a modified Gauss' law due to the presence of the static point color charge, which is felt by the bath at all times.}

On the other hand, the Wilson loop that defines the correlation function we are presently interested in does \textit{not} wind around the SK contour. That means that the ``thermal'' contributions $e^{\beta \omega}$ that come from matching the fluctuations around the contour will not be present in this case, and therefore all of the temperature dependence that will appear is going to be due to temperature effects on the bulk geometry of the Minkowski part of the manifold that holographically realizes the path integral associated to the Schwinger-Keldysh contour. Hence, the way in which both observables are defined has manifestly distinct effects in the values that the correlation functions take. After discussing the calculations in detail throughout the following sections, we will see how these differences manifest themselves in the final result.

\section{The Wilson loop with constant \texorpdfstring{$S_5$} {} coordinate} \label{sec:HQ-setup}
As we discussed in Section~\ref{sec:nhat}, the standard choice to do calculations of Wilson loops using the AdS/CFT correspondence in strongly coupled $\mathcal{N}=4$ SYM is to set a constant value for $\hat{n}$ throughout the Wilson loop. This is indeed the setup used in the celebrated paper by Maldacena~\cite{Maldacena:1998im} to calculate the heavy quark interaction potential at strong coupling. Since our interest is to describe the dynamics of a heavy quark-antiquark pair close together, we find this is a natural starting point that warrants exploration, regardless of our previous observation that this choice for $\hat{n}$ does not preserve all properties we expect from the Yang-Mills Wilson loop. We will consider the situation where the $\hat{n}$ coordinates are at antipodal points on the $S_5$ for the heavy quark and the heavy antiquark respectively in Section~\ref{sec:QQ-setup}\@.

The calculation consists of three steps. First, in Section~\ref{sec:HQ-bkg} we will discuss the ``background'' worldsheet solution that hangs from the unperturbed Wilson loop (i.e., without the deformations that give rise to the field strength insertions), thus establishing the geometry on which fluctuations can propagate. Secondly, in Section~\ref{sec:HQ-fluct} we will discuss the action for the perturbations on top of this background solution, and how to extract the correlation function of interest for the specific geometry we describe in the first step. Finally, in Section~\ref{sec:EE-calculation-HQ} we will calculate the correlation function as prescribed by the previous steps. We will provide more details for fluctuations that are transverse to the worldsheet, and discuss longitudinal fluctuations (to be defined in what follows) in a more succinct way. We will also check our results by a numerical calculation of the background extremal surface and the correlation function in Euclidean signature.

\subsection{Background} \label{sec:HQ-bkg}

The heavy quark interaction potential can be extracted from a rectangular Wilson loop of temporal extent $\T$ and spatial separation $L$, with $\T \gg L$\@. Its calculation using supersymmetric Wilson loops has been discussed many times in the literature.
The original papers discussed this at zero temperature~\cite{Maldacena:1998im,Rey:1998ik}\@. More general setups were later discussed including finite temperature effects and a relative velocity between the heavy-quark pair and the medium~\cite{Liu:2006ug,Peeters:2006iu,Chernicoff:2006hi,Avramis:2006em,Argyres:2006vs,Liu:2006nn}\@. The same loop has also been considered in AdS/QCD to calculate the characteristic correlation lengths of field strength correlators~\cite{Andreev:2010zg} in the limit $L \gg \T$. In what follows, we review the main features of the extremal surface that appears in the AdS/CFT calculation of the static heavy-quark potential in $\mathcal{N} = 4$ SYM. Our goal is to study it in the limit $L\to 0$, where the two parallel Wilson lines that construct the timelike segments of the loop get pulled close together.
Because the solution for this Wilson loop has been well-studied in the literature, we will only briefly review the results and highlight their most important features for our purposes.

\begin{figure}
    \centering
    \includegraphics[width=0.7\textwidth]{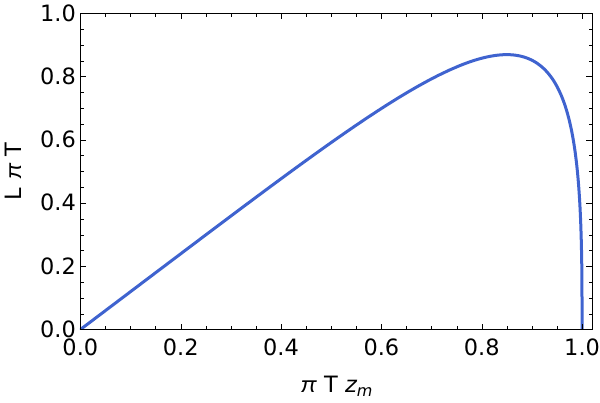}
    \caption{Solutions to Eq.~\eqref{eq:zm-solutions} in the $z_m$-$L$ plane. For each $L$ below a threshold value, there are two solutions $z_m(L)$\@.}
    \label{fig:zm-solutions}
\end{figure}

In the presence of a black hole described through the metric~\eqref{eq:Schwarzschild-AdS}, the minimal area worldsheet configuration that hangs from a rectangular contour of size $\T \times L$ on the boundary, with $\T \gg L$, can be parametrized by
\begin{align}
X^\mu(\tau,\sigma) = (\tau, \sigma, 0 , 0, z(\sigma), \hat{n}_0) \, ,
\end{align}
where $\sigma \in [0,L]$ and $\tau \in [-\T, \T]$\@. For such a parametrization, the Nambu-Goto action reads
\begin{equation}
    \mathcal{S}_{\rm NG} [\Sigma] = - \frac{\sqrt{\lambda}}{2\pi} \int d\tau \, d\sigma \frac{\sqrt{f + z'{}^2}}{z^2} \, .
\end{equation}
Because the action does not depend on $\sigma$ explicitly, there is a conserved quantity, which is the analog of the Hamiltonian $\mathcal{H}$ in standard classical mechanics, with $\mathcal{H} = p \dot{q} - \mathcal{L}$ and $p = \partial \ml{L}/\partial \dot{q}$\@. Using this conserved quantity, one finds that the background worldsheet satisfies
\begin{align} \label{eq:z-eom}
    \sqrt{z'{}^2 + f} = \frac{z_m^2 f }{z^2 \sqrt{f_m}} \iff z' = \pm \sqrt{\frac{f}{f_m}} \sqrt{\frac{z_m^4 - z^4}{z^4}} \, ,
\end{align}
where we have denoted $f_m = f(z_m)$\@. This equation can be integrated to find an implicit solution for $z(\sigma)$, which is given by
\begin{align} \label{eq:implicit-wx}
    1 - \frac{z^3(\sigma)}{z_{m}^3} \frac{\Gamma(5/4) F_1 \! \left( \frac{3}{4}, \frac12, \frac12, \frac74, \frac{z^4(\sigma)}{z_{m}^4}, \pi^4 T^4 z^4(\sigma) \right) }{\sqrt{\pi} \, \Gamma(7/4) {}_2 F_1 \! \left( \frac12, \frac34, \frac54, \pi^4 T^4 z_{m}^4  \right) } = 2\left| \frac{\sigma}{L} - \frac12 \right|,
\end{align}
where $z_m$ is the maximum value of the radial AdS coordinate $z(\sigma)$ and $F_1$ is the Appell hypergeometric function. It is in turn given by
\begin{align} \label{eq:zm-solutions}
    2 \pi T z_{m} \sqrt{1 - \pi^4 T^4 z_{m}^4} \frac{\sqrt{\pi} \, \Gamma(7/4) }{3 \, \Gamma(5/4)} {}_2 F_1 \! \left[ \frac12, \frac34, \frac54, \pi^4 T^4 z_{m}^4 \right] = \pi T L \, .
\end{align}
This equation has two solutions for any given value of $\pi T L < \pi T L_{\rm max} \approx 0.86912$, corresponding to a value of $z_m$ given by $z_m^{\rm crit} \approx 0.84978$\@. This is depicted in Fig.~\ref{fig:zm-solutions}\@. As discussed in Refs.~\cite{Friess:2006rk,Avramis:2006nv}, the solutions with $z_m > z_m^{\rm crit}$ are unstable, and beyond this value the preferred configuration is that of two disconnected, radially infalling surfaces from two parallel Wilson lines. Since we will be interested in the $L \to 0$ limit, namely, $L \pi T \ll 1$, the solution we have to consider is always in the branch with $z_m < z_m^{\rm crit}$\@.

The energy of this configuration\footnote{Because the energy of a configuration is dynamically reflected on the time evolution factor as $e^{-iE (2\T) }$, the energy associated to a Wilson loop configuration in the AdS language after subtracting the mass of the heavy quarks is given by $E = -(\mathcal{S}_{\rm NG}[\Sigma] - \mathcal{S}_0[\mathcal{C},\hat{n}] )/(2\T)$\@.} is given by
\begin{align}
E^c(L) &= - \frac{\sqrt{2\pi \lambda}}{z_m(L) \Gamma(1/4)^2 } (1 - (\pi T z_m(L))^4 ) {}_2 F_1 \! \left[ \frac12, \frac34, \frac14, (\pi T z_{m}(L))^4 \right] + \sqrt{\lambda} T \nonumber \\
&= - \frac{4\pi^2}{\Gamma(1/4)^4} \frac{\sqrt{\lambda}}{L} + \sqrt{\lambda} T + \ml{O}((\pi T L)^3) \, .
\end{align}
This is a Coulomb-like potential, which diverges in the short-distance limit $L \to 0$\@. The constant term proportional to $T$ comes from subtracting the area of the disconnected worldsheet that hangs only down to the horizon $z = (\pi T)^{-1}$, instead of all the way to $z \to \infty$\@. This term $\sqrt{\lambda}T$ corresponds to twice the thermal correction to heavy quark mass. The above equation means that, in the absence of another extra normalization factor in the RHS of Eq.~\eqref{eq:EE-corr-from-variations}, the result of taking the limit $L \to 0$ will be ill-defined. As such, to extract a finite correlation function from here we must also divide by the expectation value of the unperturbed Wilson loop, $\langle W[\mathcal{C}] \rangle_T = \exp ( - i \T E^c(L) )$, as anticipated in Eq.~\eqref{eq:EE-from-Delta}\@.

With the background solution in hand, we can now consider perturbations on top of it. By evaluating the second derivative of the action with respect to the perturbations, we can extract the non-Abelian electric field correlation function we seek.

\subsection{Fluctuations} \label{sec:HQ-fluct}

We now describe the dynamics induced on the worldsheet by small deformations on the contour that  bounds it. Following the discussion we presented in Section~\ref{sec:generating-EE}, one arrives at the conclusion that this is achieved by introducing small fluctuation fields on the string, which capture how the boundary perturbations propagate into the string. These fields obey second-order partial differential equations that are determined from the Nambu-Goto action. Once one solves the corresponding differential equations, one has to evaluate the Nambu-Goto action ``on-shell,'' i.e., on the solution to the equations of motion, as a function of a general boundary condition $h^i(t)$\@. Then, by taking functional derivatives with respect to $h^i(t)$ of the on-shell action, one can extract the correlation function we are interested in. Because of notational clarity, we will give the results in terms of the kernel $\Delta_{ij}$, which will be different for this configuration than that for the configuration in Section~\ref{sec:QQ-setup}\@. We will denote this section's expression for this kernel by $\Delta_{ij}^c$, where the ``c'' may stand for ``connected'' or ``Coulomb,'' in reference to the shape of the background worldsheet and to the nature of the interaction potential, respectively. We will denote the solution of the next section by $\Delta_{ij}^d$ with ``d'' standing for ``disconnected''\@. Only after we have both of them at hand will we use Eq.~\eqref{eq:EE-Delta} to relate our answer to the non-Abelian electric field correlator of interest.

To evaluate $\Delta_{ij}^c$, the first step to take is introduce perturbations along all of the AdS${}_5$ coordinates, and consider a string parametrized by
\begin{align}
X^\mu(\tau,\sigma) = (\tau + y_0(\tau,\sigma), \sigma + y_1(\tau,\sigma), y_2(\tau,\sigma) , y_3(\tau,\sigma) , z(\sigma) + y_4(\tau,\sigma), \hat{n}) \, .
\end{align}
We do not consider fluctuations on the $S_5$ coordinates because they are decoupled from the rest at the quadratic level in the Nambu-Goto action, which is all we need to evaluate our correlator. However, this parametrization has redundancies in it, because fluctuations that lie on the tangent space to the worldsheet are not physical deformations, but rather a coordinate reparametrization. This means we can choose our worldsheet coordinates such that we can set $y_0(\tau,\sigma) = 0$, as well as set to zero a certain linear combination of $y_1$ and $y_4$\@. To find it, we need to project $y_1$ and $y_4$ along the directions parallel and perpendicular to the worldsheet. Let $\delta(\tau, \sigma)$ parametrize the fluctuations orthogonal to the worldsheet, and $r(\tau,\sigma)$ describe reparametrizations along the worldsheet. Projecting along the parallel and orthogonal directions to the tangent vector of the worldsheet by means of the metric $g_{\mu \nu}$, one finds that the parallel and perpendicular fluctuations are parametrized by
\begin{align}
    \delta_{\parallel} X^\mu(\tau,\sigma) &= (y_0(\tau,\sigma) ,  r(\tau,\sigma), 0, 0, z'(\sigma) r(\tau,\sigma), 0) \\
    \delta_{\perp} X^\mu(\tau,\sigma) &= (0 ,   \frac{z'(\sigma) \delta(\tau, \sigma) }{\sqrt{z'(\sigma)^2 + f(z(\sigma)) } } , y_2(\tau,\sigma) , y_3(\tau,\sigma) , -  \frac{f(z(\sigma)) \delta(\tau, \sigma) }{\sqrt{z'(\sigma)^2 + f(z(\sigma))} }, 0) \, .
\end{align}
As long as we are in the linear response regime, it is a straightforward exercise to show that the Nambu-Goto action only depends on $\delta_\perp X^\mu$, with no dependence on $\delta_\parallel X^\mu$ after using the background equations of motion.\footnote{Of course, the reparametrization invariance of the worldsheet means that there are degrees of freedom absent in the action beyond the linear response regime, but to determine them one would require more information than just the tangent vector to the surface.}
Therefore, we can describe the perturbed worldsheet by
\begin{align} \label{eq:fluct-parametrization}
X^\mu(\tau,\sigma) = (\tau, \sigma + \frac{z'(\sigma) \delta(\tau, \sigma) }{\sqrt{z'(\sigma)^2 + f(z(\sigma)) } } , y_2(\tau,\sigma) , y_3(\tau,\sigma) , z(\sigma) -  \frac{f(z(\sigma)) \delta(\tau, \sigma) }{\sqrt{z'(\sigma)^2 + f(z(\sigma))} }, \hat{n}) \, .
\end{align}
This will be accurate as long as the fluctuations can be treated perturbatively, which is indeed the case of interest because we only need to evaluate the derivative of the action about the background configuration up to quadratic order in the fluctuations, which means that the corresponding equations of motion we will have to solve are linear.

The next step is to write down the action up to quadratic order and derive the equations of motion for the perturbations. After using the background equations of motion and the boundary conditions, one finds that the linear terms in the fluctuations vanish, and it is then straightforward to show Eq.~\eqref{eq:NG-action} becomes
\begin{align}
    \mathcal{S}_{\rm NG}[\Sigma] = - \frac{\sqrt{\lambda}}{2\pi} \left( S_{\rm NG,c}^{(0)}[z] + S_{\rm NG,c}^{(2),\parallel}[\delta] + S_{\rm NG,c}^{(2),\perp}[y_2] + S_{\rm NG,c}^{(2),\perp}[y_3]  \right) \, ,
\end{align}
where each term is given by
\begin{align}
    S^{(0)}_{\rm NG,c}[z] &= \T \int_{0}^L \!\! d\sigma \frac{\sqrt{z'{}^2 + f} }{z^2} \, , \\
    S^{(2),\parallel}_{\rm NG,c}[\delta] &= \int_{-\T/2}^{\T/2} \!\! d\tau \! \int_{0}^L \!\! d\sigma \left[ \frac{f }{2 z^2 \sqrt{z'{}^2 + f} } \delta'{}^2 - \frac{ \sqrt{z'{}^2 + f}}{2 z^2 f} \dot{\delta}{}^2 \right. \nonumber \\ & \quad \quad \quad \quad \quad \quad \quad \left. + \frac{2 z z' f (\pi T z)^4}{z^4 (z'{}^2 + f )^{3/2}} \delta \delta' -  \frac{ f  ( z'{}^2 f + 1 + 5 (\pi T z)^4  )  }{ z^4 (z'{}^2 + f )^{3/2} } \delta^2 \right] \, , \label{eq:action-delta} \\
    S^{(2),\perp}_{\rm NG,c}[y] &= \int_{-\T/2}^{\T/2} \!\! d\tau \! \int_{0}^L \!\! d\sigma \left[ \frac{f}{2 z^2 \sqrt{z'{}^2 + f}} y'{}^2 - \frac{ \sqrt{z'{}^2 + f}}{2 z^2 f} \dot{y}{}^2 \right] \, . \label{eq:action-y}
\end{align}
In these equations, $\parallel$ and $\perp$ should be distinguished from the meanings of being tangent and perpendicular to the background worldsheet. Rather, they indicate whether the perturbations on the boundary ($z=0$) are in the same plane as the Wilson loop or perpendicular to it.

From the action $S^{(2)}_{{\rm NG},c}$, one can derive the equations of motion for the fluctuations. 
We want to emphasize that if we had kept the redundant fluctuations $y_0(\tau,\sigma), r(\tau,\sigma)$, we would have obtained an action containing them up to quadratic order. However, upon calculating their equations of motion, one finds that they are trivial (they vanish identically), and also do not enter the equations of motion for the rest of the fluctuations up to the linear response level.

Before proceeding to the calculation of the kernel $\Delta_{ij}^c$, we note that from here we can already give formal expressions for the on-shell action at quadratic order in the perturbations. As one can always do for a quadratic action of dynamical variables and their first derivatives, we can integrate the action density by parts to obtain the equation of motion plus a total derivative, which reduces to a boundary term. Using Eq.~\eqref{eq:z-eom} and considering nonzero boundary conditions at the timelike segments of the Wilson loop only, we find (in the limit $\T \to \infty$)
\begin{align}
    S^{(2),\parallel}_{\rm NG,c}[\delta]_{\rm on-shell} &= \frac{\sqrt{f_m}}{2 z_m^2} \int_{-\infty}^\infty \!\! d\tau \left[ \delta'(\tau,\sigma = L) \delta(\tau,\sigma = L) - \delta'(\tau,\sigma = 0) \delta(\tau,\sigma = 0) \right] \, , \\
    S^{(2),\perp}_{\rm NG,c}[y]_{\rm on-shell} &= \frac{\sqrt{f_m}}{2 z_m^2} \int_{-\infty}^\infty \!\! d\tau \left[ y'(\tau,\sigma = L) y(\tau,\sigma = L) - y'(\tau,\sigma = 0) y(\tau,\sigma = 0) \right] \, ,
\end{align}
which conveniently are of the same form. The relative simplification of these expressions is due to the fact that all of the coefficients of $\delta$ and $\delta'$ are evaluated at the boundary $z=0$\@. 

The remaining task is to find the derivatives $\delta'$ and $y'$ that solve the equations of motion derived from Eqs.~\eqref{eq:action-delta} and~\eqref{eq:action-y}, in terms of general boundary conditions on the timelike segments of the Wilson loop, of the form implied by Eq.~\eqref{eq:f-antisymm}\@. Concretely, we seek $\delta'$ and $y'$ whose boundary conditions at the timelike segments of the Wilson loop are given by
\begin{align}
    \delta(\tau,\sigma=L) &= \delta(\tau,\sigma=0) = h^\parallel(\tau) \, , \\
    y(\tau,\sigma=L) &= - y(\tau, \sigma = 0) = h^\perp(\tau) \, .
\end{align}
There is no sign flip in the boundary condition for $\delta$ relative to that of $y$ because the parametrization~\eqref{eq:fluct-parametrization} already takes it into account.

Because the corresponding equations of motion are linear, it is possible to write down the derivative of the solutions at the boundaries in terms of linear response kernels $K^c_\parallel(\tau,\tau')$, $K^c_\perp(\tau,\tau')$\@. Because of how we have parametrized the longitudinal fluctuations, $\delta$ will be an even function of $\sigma$ around $\sigma = L/2$, and $y$ will be odd. Then, we can write
\begin{align}
    \delta'(\tau,\sigma=L) = - \delta'(\tau,\sigma=0) &= \int_{-\infty}^\infty \!\! d\tau' K^c_\parallel(\tau,\tau';L) h^\parallel(\tau') \, , \\
    y'(\tau,\sigma=L) = y'(\tau,\sigma=0) &= \int_{-\infty}^\infty \!\! d\tau' K^c_\perp(\tau,\tau';L) h^\perp(\tau') \, ,
\end{align}
with which the on-shell actions can be written as
\begin{align}
    S^{(2),\parallel}_{\rm NG,c}[h]_{\rm on-shell} &= \frac{\sqrt{f_m}}{z_m^2} \int_{-\infty}^\infty \!\! d\tau \,d\tau' h^\parallel(\tau) K^c_\parallel(\tau,\tau';L) h^\parallel(\tau') \, , \\
    S^{(2),\perp}_{\rm NG,c}[h]_{\rm on-shell} &= \frac{\sqrt{f_m}}{ z_m^2} \int_{-\infty}^\infty \!\! d\tau \,d\tau' h^\perp(\tau) K^c_\perp(\tau,\tau';L) h^\perp(\tau') \, .
\end{align}
Because of the time translational symmetry in the limit $\T \to \infty$, we also have $K^c_\parallel(\tau,\tau';L) = K^c_\parallel(\tau-\tau';L)$, $K^c_\perp(\tau,\tau';L) = K^c_\perp(\tau-\tau';L)$\@. From here, it is clear that by evaluating $K^c_\parallel, K^c_\perp$ we will have all the information we need to evaluate the contribution of each type of fluctuations to $\Delta_{ij}^c$:
\begin{align}
 \Delta^c_{ij}(t_2 - t_1;L) =  \frac{\sqrt{\lambda}}{\pi}  \frac{\sqrt{f_m}}{z_m^2} \left[ \delta_{i1} \delta_{1j} K^c_\parallel(t_2-t_1;L) + (\delta_{i2} \delta_{2j} + \delta_{i3} \delta_{3j}) K^c_\perp(t_2 - t_1;L) \right] \, . \label{eq:Delta-in-terms-of-K}
\end{align}

Finally, to evaluate all the different response kernels $K^c(\tau-\tau')$, because of the time translational invariance, it is most helpful to introduce their Fourier transforms, which for the fluctuations are given by
\begin{align}
     \delta_\omega(\sigma) &= \int_{-\infty}^\infty d\tau \,e^{i\omega \tau} \delta(\tau,\sigma)\, , \\
    y_\omega(\sigma) &= \int_{-\infty}^\infty d\tau \,e^{i\omega \tau} y(\tau,\sigma) \, ,
\end{align}
and for the response kernels by
\begin{align}
     K_{\parallel}^c(\omega, \omega') &= \int_{-\infty}^\infty d\tau \, d\tau' \, e^{i \omega \tau - i \omega' \tau'} K_{\parallel}^c(\tau-\tau') \equiv (2\pi) \delta(\omega - \omega') K_{\parallel}^c(\omega)\, , \\
     K_{\perp}^c(\omega, \omega') &= \int_{-\infty}^\infty d\tau \, d\tau' \, e^{i \omega \tau - i \omega' \tau'} K_{\perp}^c(\tau-\tau') \equiv (2\pi) \delta(\omega - \omega') K_{\perp}^c(\omega)\, ,
\end{align}
where $K_{\perp/\parallel}^c(\omega) = \int_{-\infty}^\infty d\tau e^{i\omega \tau} K_{\perp/\parallel}^c(\tau)$ is the Fourier representation of the response kernel for either type of perturbation. With this, we can simply write down 
\begin{align}
      \delta_\omega'(\sigma = L) &= K_\parallel^c(\omega) \delta_\omega(\sigma=L)\, , \\ 
    y_\omega'(\sigma = L) &= K_\perp^c(\omega) y_\omega(\sigma=L)\, ,
\end{align}
and the problem is reduced to finding the respective response function $K_{\perp/\parallel}^c(\omega)$ at each frequency.

\subsection{Calculation of the time-ordered non-Abelian electric field correlator} \label{sec:EE-calculation-HQ}

After setting up all of the machinery, we now describe the calculation of the response kernels for fluctuations in the configuration where the two timelike segments of the SYM Wilson loop have the same $S_5$ coordinates. We proceed with a greater level of detail for transverse fluctuations, which is arguably the simpler case, in the hope that it will make the longitudinal discussion less cumbersome. We also provide an increased level of detail in the hope that future calculations of fluctuations on top of extremal worldsheets to extract correlation functions from holography may benefit from this discussion.

Furthermore, in anticipation of obtaining results that might require careful regularization, we will also carry out the calculation allowing for more flexibility in the fluctuations than using a single perturbation $h^i$\@. Specifically, we will set boundary conditions on the two timelike segments of the contour $\mathcal{C}$ independently. The purpose of this will be to verify that the contributions to the response kernel proportional to a Dirac delta function in time are not due to (omitted) contact terms in the RHS of Eq.~\eqref{eq:E-insertion-2}\@.

\subsubsection{Transverse fluctuations} \label{sec:transverse-v-fluct}

As we just discussed, our goal now is to solve for $y_\omega(\sigma)$ and extract its derivatives on the boundary. 
Varying $S_{\rm NG,c}^{(2),\perp}$ with respect to $y$ and transforming to the frequency domain, the equation we have to solve is
\begin{equation}
\frac{\partial^2 y_\omega}{\partial \sigma^2}(\sigma) + \frac{z_m^4}{z^4(\sigma)} \frac{\omega^2}{f_m} y_\omega(\sigma) = 0 \, ,
\end{equation}
where $z(\sigma)$ is determined by solving
\begin{equation}
\frac{f}{z^2 \sqrt{f + z'^2}} = \frac{\sqrt{f_{m}}}{z_{m}^2} \iff z' = \pm \sqrt{f} \sqrt{\frac{z_m^4}{z^4} \frac{f}{f_m} -1}
\end{equation}
subject to $z'=0 \iff z = z_m$ and $z(\sigma = L) = z(\sigma = 0)  = 0$\@. In the interval $\sigma \in (0,L/2)$ we take the plus sign (as the worldsheet goes into AdS${}_5$), and the minus sign when $\sigma \in (L/2,L)$\@. The most useful form of the above expression is
\begin{equation} \label{eq:eom-z}
z' = \pm \sqrt{ \frac{f}{f_m} \frac{z_m^4 - z^4}{z^4} } \, .
\end{equation}

To avoid introducing unnecessary numerical uncertainties, the best alternative is to transform the equation for $y_\omega(\sigma)$ into an equation for $y_\omega(z)$, because then we will not need to solve for $z(\sigma)$ explicitly.\footnote{It is actually possible to do this, but because we are able to perform the change of variables the explicit form of the solution becomes unimportant.} Performing the transformation, we have
\begin{align}
&\sqrt{ \frac{f}{f_m} \frac{z_m^4 - z^4}{z^4} } \frac{\partial}{\partial z} \left( \sqrt{ \frac{f}{f_m} \frac{z_m^4 - z^4}{z^4} } \frac{\partial y_\omega}{\partial z} \right) + \frac{z_m^4}{z^4} \frac{\omega^2}{f_m} y_\omega = 0 \, , \nonumber \\
\implies & f\frac{z_m^4 - z^4}{z^4} \frac{\partial^2 y_\omega}{\partial z^2} - \frac{2[ (z_m^4-z^4) + z^4 f ] }{z^5}  \frac{\partial y_\omega}{\partial z} + \frac{z_m^4}{z^4} \omega^2 y_\omega = 0 \, , \nonumber \\
\implies & \frac{\partial^2 y_\omega}{\partial z^2} - \frac{2}{z} \left[ \frac{1}{f} + \frac{z^4}{z_m^4 - z^4} \right]  \frac{\partial y_\omega}{\partial z} + \frac{\omega^2 z_m^4}{(z_m^4 - z^4) f } y_\omega = 0 \, .
\end{align}

At this point, it is useful to introduce a rescaling of the radial AdS coordinate: $\xi = z/z_m \in (0,1)$\@. In terms of this variable, we have
\begin{equation} \label{eq:eom-y}
 \frac{\partial^2 y_\omega}{\partial \xi^2} - \frac{2}{\xi} \left[ \frac{1}{1 - (\pi T z_m)^4 \xi^4 } + \frac{\xi^4}{1 - \xi^4} \right]  \frac{\partial y_\omega}{\partial \xi} + \frac{\omega^2 z_m^2}{(1 - \xi^4) (1 - (\pi T z_m)^4 \xi^4) } y_\omega = 0 \, .
\end{equation}

The same equation applies for both copies of the transformed intervals $\sigma \in (0,L/2)$ and $\sigma \in (L/2,L)$\@. Let us denote the corresponding solutions as a function of $z$ by $y_\omega^L(z)$ and $y_\omega^R(z)$, respectively, where $L,R$ stand for ``left'' and ``right'' sides of the worldsheet. All that we need to specify in order to close the system are the boundary conditions. In terms of the original coordinate $\sigma$, we have
\begin{align}
y_\omega(\sigma = [L/2]^-) = y_\omega(\sigma = [L/2]^+) \, , & & \frac{\partial y_\omega}{\partial \sigma}(\sigma = [L/2]^-) = \frac{\partial y_\omega}{\partial \sigma}(\sigma = [L/2]^+)  \, ,
\end{align}
which, in terms of the $z$ coordinate, transform to
\begin{align}
\lim_{z \to z_m} y_\omega^L(z) = \lim_{z \to z_m} y_\omega^R(z) \, , & & \lim_{z \to z_m}\sqrt{\frac{f}{f_m} \frac{z_m^4 -z^4}{z^4} }  \frac{\partial y_\omega^L}{\partial z} = - \lim_{z \to z_m} \sqrt{\frac{f}{f_m} \frac{z_m^4 -z^4}{z^4} } \frac{\partial y_\omega^R}{\partial z} \, ,
\end{align}
or, equivalently,
\begin{align}
\lim_{\xi \to 1} y_\omega^L(\xi) = \lim_{\xi \to 1} y_\omega^R(\xi) \, , & & \lim_{\xi \to 1} \sqrt{1 - \xi^4} \frac{\partial y_\omega^L}{\partial \xi} = - \lim_{\xi \to 1} \sqrt{1 - \xi^4} \frac{\partial y_\omega^R}{\partial \xi} \, .
\end{align}

To implement these matching conditions, the best way is to do a WKB-type analysis to extract the leading/possibly singular behavior of the solution near the horizon. We can do that analytically by writing
\begin{align}
y_\omega^L(\xi) &= A_L \exp \left( i  \int_0^\xi \frac{\omega z_m \xi'^3 d\xi'}{\sqrt{(1-\xi'^4)(1-(\pi T z_m \xi')^4)}} \right) F^{-}_\omega(\xi) \nonumber \\ & \quad + B_L \exp \left( -i  \int_0^\xi \frac{\omega z_m \xi'^3 d\xi'}{\sqrt{(1-\xi'^4)(1-(\pi T z_m \xi')^4)}} \right) F^{+}_\omega(\xi) \nonumber \\
&\equiv A_L \, y^-_\omega(\xi) + B_L \, y^+_\omega(\xi) \, , \label{eq:yL} \\
y_\omega^R(\xi) &= A_R \exp \left( i  \int_0^\xi \frac{\omega z_m \xi'^3 d\xi'}{\sqrt{(1-\xi'^4)(1-(\pi T z_m \xi')^4)}} \right) F^{-}_\omega(\xi) \nonumber \\ & \quad + B_R \exp \left( -i  \int_0^\xi \frac{\omega z_m \xi'^3 d\xi'}{\sqrt{(1-\xi'^4)(1-(\pi T z_m \xi')^4)}} \right) F^{+}_\omega(\xi) \nonumber \\
&\equiv A_R \, y^-_\omega(\xi) + B_R \, y^+_\omega(\xi) \, , \label{eq:yR}
\end{align}
where $\partial F^{\pm}_\omega/\partial \xi$ is finite at the turning point $\xi = 1$\@. With this decomposition, the matching conditions translate into
\begin{align}
A_L \, y_\omega^-(\xi=1) + B_L \, y_\omega^+(\xi=1) &= A_R \, y_\omega^-(\xi=1) + B_R \, y_\omega^+(\xi=1) \, , \\
i A_L \, y_\omega^-(\xi=1) - i B_L \, y_\omega^+(\xi=1) &= -i A_R \, y_\omega^-(\xi=1) + i B_R \, y_\omega^+(\xi=1) \, .
\end{align}

To solve the system in terms of one of the amplitudes, we need to specify the boundary conditions. To extract the correlation function we are interested in, the natural choice is to prescribe $y^L_\omega(\xi = 0) = - y^R_\omega(\xi = 0)$\@. However, in order to illustrate the nature of contact divergences that will appear in this calculation, we will instead consider the boundary condition $y^R_\omega(\xi = 0) = 0$\@. We can obtain the boundary condition that defines our correlation function (i.e., $y^L_\omega(\xi = 0) = - y^R_\omega(\xi = 0)$) by taking linear superpositions of the boundary condition $y^{L/R}_\omega(\xi = 0) = 0$ and using that the equations we are looking at are symmetric under the exchange of the $L,R$ labels.  With this, it is appropriate to define the response functions $K^{AB}(\omega)$ as the derivative responses $y_\omega'(\sigma)$ on side $A$ due to a unit perturbation on side $B$\@.

Then, the null boundary condition $y^R_\omega(\xi = 0) = 0$ at $\sigma = L$ translates into
\begin{equation}
A_R \, y_\omega^-(\xi=0) + B_R \, y_{\omega}^+(\xi=0) = 0 \, ,
\end{equation}
which, together with the matching conditions at $\xi=1$, fully determine the solution up to an overall constant:
\begin{align}
A_L &= - \frac{y^+_\omega(\xi=1)}{y_\omega^+(\xi=0)} \frac{y^-_\omega(\xi=0)}{y^-_\omega(\xi=1)} A_R \, , \\
B_R &= - \frac{y^-_\omega(\xi=0)}{y^+_\omega(\xi=0)} A_R \, , \\
B_L &= \frac{y^-_\omega(\xi=1)}{y^+_\omega(\xi=1)} A_R \, .
\end{align}

Formally, all that remains is to evaluate the response kernels $K^{RL,c}_{\perp}$ and $K^{LL,c}_{\perp}$\@. We remind the reader that the superscripts are there to make it explicit that they represent partial contributions to the response kernel we want to evaluate, each coming from specific boundary conditions and response locations. These are determined by
\begin{align}
K^{RL,c}_{\perp}(\omega,L) &= -\frac{1}{y_\omega^L(\xi=0)} \lim_{\xi \to 0} \sqrt{\frac{ 1- (\pi T z_m \xi)^4 }{f_m} \frac{1 - \xi^4}{\xi^4} } \frac{1}{z_m} \frac{\partial y_\omega^R}{\partial \xi} \nonumber \\
&= - \frac{1}{z_m \sqrt{f_m}} \frac{1}{y_\omega^L(\xi=0)} \lim_{\xi \to 0} \frac{1}{\xi^2} \frac{\partial y_\omega^R}{\partial \xi} \, , \\
K^{LL,c}_{\perp}(\omega,L) &= \frac{1}{y_\omega^L(\xi=0)} \lim_{\xi \to 0} \sqrt{\frac{ 1- (\pi T z_m \xi)^4 }{f_m} \frac{1 - \xi^4}{\xi^4} } \frac{1}{z_m} \frac{\partial y_\omega^L}{\partial \xi} \nonumber \\
&= \frac{1}{z_m \sqrt{f_m}} \frac{1}{y_\omega^L(\xi=0)} \lim_{\xi \to 0} \frac{1}{\xi^2} \frac{\partial y_\omega^L}{\partial \xi} \, .
\end{align}
Furthermore, choosing the normalization of the mode functions such that $y^{\pm}_\omega(\xi=0) = 1$, and writing the result in terms of the regular functions $F_\omega^{\pm}(\xi)$ whenever possible, we have the response kernels in each case, before subtractions and regularizations, given by
\begin{align}
K^{RL,c}_{\perp}(\omega,L) &=  \frac{1}{z_m \sqrt{f_m}} \frac{\lim_{\xi\to 0} \frac{1}{\xi^2} \left[ \frac{\partial F_\omega^-}{\partial \xi} - \frac{\partial F_\omega^+}{\partial \xi}  \right] }{- \frac{y^{+}_\omega(\xi=1)}{y^-_\omega(\xi=1)} + \frac{y^{-}_\omega(\xi=1)}{y^+_\omega(\xi=1)}  } \, , \\
K^{LL,c}_{\perp}(\omega,L) &=  \frac{1}{z_m \sqrt{f_m}} \frac{\lim_{\xi\to 0} \frac{1}{\xi^2} \left[ \frac{y^{+}_\omega(\xi=1)}{y^-_\omega(\xi=1)} \frac{\partial F_\omega^-}{\partial \xi} - \frac{y^{-}_\omega(\xi=1)}{y^+_\omega(\xi=1)} \frac{\partial F_\omega^+}{\partial \xi}  \right] }{- \frac{y^{+}_\omega(\xi=1)}{y^-_\omega(\xi=1)} + \frac{y^{-}_\omega(\xi=1)}{y^+_\omega(\xi=1)}  } \, .
\end{align}

Note that all of the above expressions are valid for arbitrary $L >0$, and furthermore, all of the discussion in this section holds for arbitrary, complex $\omega$\@.\footnote{This will be useful to enforce the time-ordering prescription.} No approximations have been made. All that remains is to solve the equation for the modes $y_\omega^\pm(\xi)$, or equivalently $F_\omega^\pm(\xi)$, and with that the above expressions can be calculated explicitly.

It is useful to note that the denominators in the expressions for $K_{\perp}^{AB,c}$ can be written in terms of the Wronskian of the differential equation for $y^\pm_\omega$\@. First we observe that
\begin{equation}
    y^+_\omega \frac{\partial y^-_\omega}{\partial \xi} - y^-_\omega \frac{\partial y^+_\omega}{\partial \xi} = W(\xi) \, , 
\end{equation}
where
\begin{align} \label{eq:Wronsk}
    W(\xi) = C \exp \left( 2 \int^\xi \frac{d\xi}{\xi} \left[ \frac{1}{1 - (\pi T z_m \xi)^4} + \frac{\xi^4}{1-\xi^4}  \right] \right) = \frac{C \xi^2}{\sqrt{(1-\xi^4)(1 - (\pi T z_m \xi)^4)}} \, .
\end{align}
The constant $C$ can be fixed by looking at the behavior of $y^{\pm}_\omega$ when $\xi \to 1$\@. The result is
\begin{equation} \label{eq:Wronsk-C}
    C = 2 i \omega z_m y^+_\omega(\xi=1) y^-_\omega(\xi=1) \, ,
\end{equation}
where we have worked under the normalization $y^\pm_\omega(\xi=0) = 1$.

Finally, one can integrate the equation for the Wronskian to derive that
\begin{align}
    \frac{y^-_\omega(\xi=1)}{y^+_\omega(\xi=1)} &= \exp \left( \int_0^1 d\xi \frac{W(\xi)}{y^+_\omega(\xi) y^-_\omega(\xi)} \right) \nonumber \\
    &= \exp \left( 2 i \omega z_m \int_0^1 d\xi \frac{y^+_\omega(\xi=1) y^-_\omega(\xi=1)}{y^+_\omega(\xi) y^-_\omega(\xi)} \frac{ \xi^2}{\sqrt{(1-\xi^4)(1 - (\pi T z_m \xi)^4)}} \right) \, .
\end{align}
Also, note that by construction we have $y^+_\omega(\xi) y^-_\omega(\xi) = F^+_\omega(\xi) F^-_\omega(\xi)$, and all contributions to the response kernels $K_\perp^{RL/LL,c}$ can be written entirely in terms of the regular functions $F^\pm_\omega$\@.
We then find the denominators in the expressions for $K_\perp$ are given by:
\begin{align}
    &- \frac{y^{+}_\omega(\xi=1)}{y^-_\omega(\xi=1)} + \frac{y^{-}_\omega(\xi=1)}{y^+_\omega(\xi=1)} \nonumber \\ &= 2i \sin \left( 2 \omega z_m \int_0^1 d\xi \frac{F^+_\omega(\xi=1) F^-_\omega(\xi=1)}{F^+_\omega(\xi) F^-_\omega(\xi)} \frac{ \xi^2}{\sqrt{(1-\xi^4)(1 - (\pi T z_m \xi)^4)}} \right) \, .
\end{align}
This is consequential because it provides a clean expression to implement the time-ordering prescription to evaluate the correlator.
Concretely, the time-ordering prescription is implemented by taking $\omega \to \omega (1 + i\epsilon)$\@. It is then convenient to define $\phi_\omega(z_m)$ as
\begin{equation}
    \phi_\omega(z_m) \equiv 2 \omega z_m \int_0^1 d\xi \frac{F^+_\omega(\xi=1) F^-_\omega(\xi=1)}{F^+_\omega(\xi) F^-_\omega(\xi)} \frac{ \xi^2}{\sqrt{(1-\xi^4)(1 - (\pi T z_m \xi)^4)}} \, ,
\end{equation}
with which the $i\epsilon$ prescription implies that we can write\footnote{Strictly speaking, one also has to analyze the mode functions and determine explicitly whether $F_\omega^+ F_\omega^-$ gives another $\ml{O}(\epsilon)$ contribution when introducing the prescription. As it turns out, this contribution adds up with the naive one, giving the same overall effect.}
\begin{equation}
    - \frac{y^{+}_\omega(\xi=1)}{y^-_\omega(\xi=1)} + \frac{y^{-}_\omega(\xi=1)}{y^+_\omega(\xi=1)} = 2i \left[ \sin (\phi_\omega(z_m)) + i\epsilon \phi_\omega(z_m) \cos (\phi_\omega(z_m)) \right] \, .
\end{equation}

Using this, and the fact that the mode functions $F_\omega^{\pm}$ satisfy $\frac{\partial^2 F_\omega^{\pm}}{\partial \xi^2}(\xi = 0) = \omega^2 z_m^2 F^{\pm}_\omega(\xi=0)$, an equality that follows from Eq.~\eqref{eq:eom-y}, we get
\begin{align}
K^{RL,c}_{\perp}(\omega,L) &= -i \frac{1}{2z_m \sqrt{f_m}} \frac{\lim_{\xi\to 0} \frac{1}{\xi^2} \left[ \frac{\partial F_\omega^-}{\partial \xi} - \frac{\partial F_\omega^+}{\partial \xi}  \right] }{\sin (\phi_\omega(z_m)) + i\epsilon \phi_\omega(z_m) \cos (\phi_\omega(z_m))  } \nonumber \\
&=  \frac{1}{z_m \sqrt{f_m} } \frac{ \omega z_m F_\omega^+(\xi=1) F_\omega^-(\xi=1) }{\sin(\phi_\omega(z_m)) + i\epsilon \phi_\omega(z_m) \cos(\phi_\omega(z_m))  } \, , \label{eq:Delta-yy-RL-c} \\
K^{LL,c}_{\perp}(\omega,L) &= -i \frac{1}{2z_m \sqrt{f_m}} \frac{\lim_{\xi\to 0} \frac{1}{\xi^2} \left[ e^{-i\phi_\omega(z_m)} \frac{\partial F_\omega^-}{\partial \xi} - e^{i \phi_\omega(z_m) } \frac{\partial F_\omega^+}{\partial \xi}  \right] }{\sin (\phi_\omega(z_m)) + i\epsilon \phi_\omega(z_m) \cos (\phi_\omega(z_m))  } \nonumber \\ 
&= \frac{1}{z_m \sqrt{f_m}} \frac{ \omega z_m F_\omega^+(\xi=1) F_\omega^-(\xi=1) \cos (\phi_\omega(z_m) ) }{\sin(\phi_\omega(z_m)) + i\epsilon \cos(\phi_\omega(z_m))  } \nonumber \\ & \quad -  \frac{1}{4 z_m \sqrt{f_m}} \left[ \frac{\partial^3 F_\omega^-}{\partial \xi^3} +  \frac{\partial^3 F_\omega^+}{\partial \xi^3} \right]_{\xi=0} -  \frac{\omega^2 z_m^2}{\sqrt{f_m}} \lim_{z \to 0} \frac{1}{z} \, , \label{eq:Delta-yy-LL-c}
\end{align}
where we have used our expression for the Wronskian as given by Eqs.~\eqref{eq:Wronsk} and~\eqref{eq:Wronsk-C}\@. The Wronskian is what allowed us to write the difference of the derivatives of $F^+_\omega$ and $F^-_\omega$ purely in terms of $F^{\pm}_\omega$ with no derivatives.

We clearly see that $K_{\perp}^{RL,c}$, $K_{\perp}^{LL,c}$ are different functions. But this is fine, since we only expect them to be equal in the limit $L \to 0$, and up to contact terms. Indeed, the last term in the expression for $K_{\perp}^{LL,c}$ is a divergent term that is exactly of this form. On the other hand, by construction, $K_{\perp}^{RL,c}$ will feature no such contact term contributions, because the variations of the Wilson loop, in the language of Section~\ref{sec:W-loop-AdS}, are always at different values of the parameter $s$\@. While this means that the $RL$ setup to extract the correlator gives a cleaner signal than the $LL$ kernel, where no subtraction for contact terms is required, we shall still calculate both as a consistency check. In what follows, since we have isolated its origin, we will omit the contact term $\frac{\omega^2 z_m^2}{\sqrt{f_m}} \lim_{z \to 0} \frac{1}{z}$ as it does not enter the definition of the correlator from the variations of the Wilson loop~\eqref{eq:EE-corr-from-variations} at $t_1 \neq t_2$\@.

All that remains now is to evaluate the mode functions and substitute the result into the expressions for the response kernels. The equation for the mode functions $F_\omega^{\pm}$ can be found by explicitly substituting $y^\pm_\omega = \exp(\mp i (\cdots) ) F_\omega^\pm$ into the equation of motion for $y$, given by Eq.~\eqref{eq:eom-y}\@. To optimize the notation, we introduce $h \equiv \pi T z_m$ and $\Omega \equiv \omega/(\pi T)$\@.
The equation for $F^\pm_{\omega}$ then reads
\begin{align} 
    \frac{\partial^2 F^\pm_\omega}{\partial \xi^2} &- 2 \left[ \frac{1 - h^4 \xi^8}{\xi(1-\xi^4)(1-h^4\xi^4)} \pm \frac{i \Omega h \xi^3}{\sqrt{(1-\xi^4)(1-h^4 \xi^4)}} \right] \frac{\partial F^\pm_\omega}{\partial \xi} \nonumber \\ & \quad \quad \quad \quad + \left[ \mp \frac{i \Omega h \xi^2}{\sqrt{(1-\xi^4)(1-h^4 \xi^4)}} + \frac{\Omega^2 h^2 (1 - \xi^6) }{(1-\xi^4)(1-h^4 \xi^4)} \right] F^\pm_\omega = 0 \, . \label{eq:F-transverse-eom}
\end{align}
Now we can proceed to study the behavior of the solutions. The defining condition we have to impose is the regularity of $\partial F_\omega^\pm/\partial \xi$ at the turning point, which, in terms of $\partial^2 F^\pm_\omega / \partial \xi^2$ being finite as $\xi \to 1$ requires $\partial F_\omega^\pm(\xi = 1) / \partial \xi  = 0$, which can be seen by analyzing the most divergent pieces of Eq.~\eqref{eq:F-transverse-eom} when $\xi\to1$\@. The near-boundary behavior of the mode functions also requires
\begin{align}
    \frac{\partial F^\pm_\omega}{\partial \xi}(\xi=0) = 0 \, , & & \frac{\partial^2 F^\pm_\omega}{\partial \xi^2}(\xi=0) = \Omega^2 h^2 F^\pm_\omega(\xi=0) \,,
\end{align}
which can be obtained by expanding Eq.~\eqref{eq:F-transverse-eom} in a power series in $\xi$ near $\xi=0$\@.
While the regularity condition at the midpoint is in principle enough to determine the mode function up to an overall normalization, these near-boundary conditions may also be used in a numerical solution of Eq.~\eqref{eq:F-transverse-eom}\@.

Now we want to take the limit $L\to 0$. The process to do so is written out in full detail in Appendix~\ref{sec:App-transverse-connected}\@. We obtain that
\begin{align}
    K^{RL,c}_{\perp}(\omega,L) = \frac{z_m^2}{\sqrt{f_m}} \left( \frac{c_3^{RL}}{L^3} + \frac{c_1^{RL} \omega^2}{L} + \ml{O}(L) \right) \, , \nonumber \\
    K^{LL,c}_{\perp}(\omega,L) = \frac{z_m^2}{\sqrt{f_m}} \left( \frac{c_3^{LL}}{L^3} + \frac{c_1^{LL} \omega^2}{L} + \ml{O}(L) \right) \, ,
\end{align}
where the dominant contribution in the limit $L\to 0$ is determined by
\begin{align}
    c_3^{RL} = c_3^{LL} = \left( \frac{2 \sqrt{\pi} \, \Gamma(7/4)}{3\, \Gamma(5/4)} \right)^2 \approx 1.43554 \, .
\end{align}
We have kept outside the definition of $c_3$ an overall factor of $ \frac{z_m^2}{\sqrt{f_m}}$ (which does depend on $L$) for convenience to translate to the result for $\Delta^c_{ij}$, which has the inverse of this factor in the front (see Eq.~\eqref{eq:Delta-in-terms-of-K} for comparison)\@.
The subleading $1/L$ contribution is different for each case (see Appendix~\ref{sec:App-transverse-connected}), i.e., $c_1^{RL} \neq c_1^{LL}$, but this presents no issue because we anyways do not expect the results to agree beyond the leading term as a function of $L$\@. On the other hand, the leading contribution is the same for both procedures, and it does not receive contributions from contact terms, because the construction of the $RL$ kernel explicitly prevents this.

Having done the above, one finds that when we introduce anti-symmetric perturbations $h_\perp(t)$ as discussed in Section~\ref{sec:W-loop-AdS}, the sum of the response kernels gives a total of
\begin{equation}
    K^{c}_{\perp}(\omega,L) = 2 \frac{z_m^2}{\sqrt{f_m}} \left( \frac{2 \sqrt{\pi} \, \Gamma(7/4)}{3\, \Gamma(5/4)} \right)^2 \left[ \frac{1}{L^3} + \ml{O}(L^{-1}) \right] \, ,
\end{equation}
and therefore the kernel that determines the two-point function for transverse deformations is given by
\begin{equation}
    \Delta^c_{22}(\omega,L) = \Delta^c_{33}(\omega,L) = \frac{2 \sqrt{\lambda}}{\pi} \left( \frac{2 \sqrt{\pi} \, \Gamma(7/4)}{3\, \Gamma(5/4)} \right)^2 \frac{1}{L^3} + \ml{O}(L^{-1}) \, .
\end{equation}
The main feature of this result is that it diverges as $L^{-3}$ when $L \to 0$\@. 

This concludes our calculation of the response functions that determine the linear response of the Nambu-Goto action to transverse deformations on the boundary countour that defines the Wilson loop, for the background configuration that describes a Coulomb-type potential between the heavy quarks. To complete the result, we now move on to calculate the longitudinal one, following the same steps.

\subsubsection{Longitudinal fluctuations}

Having gone through all the machinery in the previous section, we shall give an abbreviated discussion of the calculation for the case of longitudinal fluctuations. First, we note that to obtain the leading behavior that we got in the previous section, it is sufficient to work in the $T=0$ case. This is clear by looking at how $T$ appears in the solution for $z(\sigma)$ and in the action for the fluctuations $\delta(\tau,\sigma)$: After factoring out the overall scale $L$ from $z(\sigma)$, it is manifest that the leading appearance of $T$ is of the order $(\pi T L)^4$\@. It is then clear that, if we find a $1/L^3$ dependence for $\Delta^{c}_{11}(\omega,L)$ in vacuum, this will be the dominant contribution in the limit $L \to 0$ (note that $\Delta$ has mass dimension three)\@.

When $T=0$, the action for the fluctuations reduces to
\begin{equation}
    S^{(2),\parallel}_{\rm NG,c}[\delta] = \int_{-\T}^{\T} \!\! d\tau \! \int_{0}^L \!\! d\sigma \left[ \frac{1 }{2 z^2 \sqrt{z'{}^2 + 1} } \delta'{}^2 - \frac{ \sqrt{z'{}^2 + 1}}{2 z^2} \dot{\delta}{}^2  -  \frac{ 1  }{ z^4 \sqrt{z'{}^2 + f } } \delta^2 \right] \, .
\end{equation}
Furthermore, if we are only after finding the leading behavior of $\Delta^{AB,c}_{11}$, we can even drop the term with time derivatives in this action, because we will be in the regime $\omega^2 L^2 \ll 1 $\@. This will only modify the result by terms that go as $1/L$\@. 

After using the conservation equation for the background worldsheet in vacuum (i.e., the conserved quantity that appears due to there not being any explicit $\sigma$ dependence in the action), which we can write as $z^2 \sqrt{z'{}^2+1} = z_m^2$, one obtains the following equation of motion for the fluctuations:
\begin{equation}
    \frac{\partial^2 \delta}{\partial \sigma^2}(\sigma) + \frac{2}{z(\sigma)^2} \delta(\sigma) = 0 \, .
\end{equation}
As with the transverse fluctuations, we can change variables from $\sigma$ to $\xi = z/z_m$, and rewrite this equation of motion in terms of two domains, one for $ \sigma \in (0,L/2) $, where we will use $\delta^L$, and the other for $ \sigma \in (L/2, L) $, where we will use $\delta^R$\@. The equation of motion for both of them is
\begin{equation} \label{eq:delta-vacuum}
    (1-\xi^4) \frac{\partial^2 \delta}{\partial \xi^2} - \frac{2}{\xi} \frac{\partial \delta}{\partial \xi} + 2 \xi^2 \delta = 0 \, ,
\end{equation}
subject to matching conditions
\begin{align}
\lim_{\xi \to 1} \delta^L(\xi) = \lim_{\xi \to 1} \delta^R(\xi) \, , & & \lim_{\xi \to 1} \sqrt{1 - \xi^4} \frac{\partial \delta^L}{\partial \xi} = - \lim_{\xi \to 1} \sqrt{1 - \xi^4} \frac{\partial \delta^R}{\partial \xi} \, .
\end{align}

Conveniently, the solutions to Eq.~\eqref{eq:delta-vacuum} can be found explicitly:
\begin{equation}
    \delta^{L,R}(\xi) = A_{L,R} \sqrt{1 - \xi^4} + B_{L,R} \frac{\xi^3 \sqrt{1-\xi^4} }{3} {}_2 F_1 \left[ \frac34 , \frac32 , \frac74, \xi^4 \right] \, .
\end{equation}
Then, the matching conditions set
\begin{align}
    B_L = B_R \, , & & 2 ( A_L + A_R ) =  \frac{ \sqrt{\pi} \, \Gamma(7/4) }{3 \, \Gamma(5/4)} ( B_L + B_R ) \, .
\end{align}
We can then define separate response kernels on either side for perturbations on a given side. Following our discussion of transverse fluctuations, we may define, setting $\delta^L(\xi=0) = A_L = 1$ and $\delta^R(\xi=0) = A_R = 0$,
\begin{align}
    K^{RL,c}_\parallel &= \frac{1}{z_m}  \lim_{\xi \to 0} \frac{1}{\xi^2} \frac{\partial \delta^R}{\partial \xi} =  \frac{1}{2z_m}  \frac{\partial^3 \delta^R}{\partial \xi^3} \, , \\
    K^{LL,c}_\parallel &= - \frac{1}{z_m} \lim_{\xi \to 0} \frac{1}{\xi^2} \frac{\partial \delta^L}{\partial \xi} =  \frac{1}{2z_m}  \frac{\partial^3 \delta^L}{\partial \xi^3} \, ,
\end{align}
where in taking the limit we have used that the mode solutions have vanishing first and second derivatives at $\xi=0$ (this can be seen directly from the mode functions, as their dependence on $\xi$ starts at order $\xi^3$)\@. The result is easily found to be
\begin{equation}
    K^{RL,c}_\parallel = - K^{LL,c}_\parallel = \frac{2}{z_m} \frac{3 \, \Gamma(5/4) }{2 \sqrt{\pi} \Gamma(7/4) } + \ml{O}(L) \, .
\end{equation}
As in the case for transverse fluctuations, these two one-sided response kernels have an equal leading order contribution to the $\Delta_{11}^c$ kernel, and no contact term appears at this order.
The symmetrized response kernel $K_\parallel^c$ is then given by
\begin{equation}
    K_\parallel^c(\omega,L) = \frac{4}{z_m} \frac{3 \Gamma(5/4) }{2 \sqrt{\pi} \, \Gamma(7/4) } + \ml{O}(L) \, ,
\end{equation}
which means that the contribution to the two-point function coming from the linearized fluctuations of the Nambu-Goto action is
\begin{equation}
    \Delta_{11}^c(\omega,L) = \frac{4 \sqrt{\lambda}}{\pi} \left( \frac{2 \sqrt{\pi} \, \Gamma(7/4)}{3\, \Gamma(5/4)} \right)^2 \frac{1}{L^3} + \ml{O}(L^{-1}) \, .
\end{equation}
With this, we have calculated the leading contribution as $L \to 0$ of the longitudinal fluctuations.

We can therefore write the complete leading contribution to the two-point function associated to fluctuations on the extremal worldsheet that gives a Coulomb interaction potential between two heavy quarks:
\begin{align} \label{eq:Delta-c-final}
 \Delta^c_{ij}(\omega,L) =  \frac{16 \pi^2 }{\Gamma(1/4)^4}  \frac{\sqrt{\lambda}}{L^3} \left( 2\delta_{i1} \delta_{1j} + \delta_{i2} \delta_{2j} + \delta_{i3} \delta_{3j} \right) + \ml{O}(L^{-1}) \, .
\end{align}

This completes the calculation for the quadratic fluctuations in this background configuration, and it is all we need in order to compare with the result of the next section. However, because this is highly singular as $L \to 0$, we shall perform a numerical check that our result is not exclusive to the large $\T$ limit, and that the same behavior is obtained in the $L\to 0$ limit for a bounded rectangular Wilson loop at fixed $\T$\@.

\subsubsection{Euclidean numerical calculation for variations on a bounded rectangle} \label{sec:EE-calculation-HQ-numerics}

In what follows, we will verify that the above results continue to hold when the temporal extent of the loop is finite. This is not in vain, as when $T>0$ the Euclidean time direction is finite in extent, and therefore it is relevant to study the expectation value of a Wilson loop with finite temporal extent $\hat{T}_E$, to assess definitively whether the temperature can play a role in the expectation value of interest.

To demonstrate this behavior, we calculate the derivative response to transverse perturbations, solving the analogous problem to that in Section~\ref{sec:transverse-v-fluct}, but in Euclidean signature. The background solution on which the perturbations propagate is specified by the action principle
\begin{align}
    S_{{\rm NG}, c, E}^{(0)}[z] &= \int_{0}^{\T_E} \!\! \diff \tau_E \! \int_0^L \!\! \diff \sigma \frac{\sqrt{\left(\frac{\partial z}{\partial \sigma}\right)^2 + f + \frac{1 }{f} \left(\frac{\partial z}{\partial \tau_E}\right)^2 }}{z^2} \nonumber \\
    &= a b  \int_{0}^{1} \diff \bar{\tau}_E \, \int_0^1 \diff \bar{x} \frac{1}{\xi^2} \sqrt{ 1 - \xi^4 + \frac{\xi'{}^2}{b^2} + \frac{\dot{\xi}{}^2}{a^2 (1 - \xi^4)} } \, , \label{eq:NGactionpseudospectral}
\end{align}
where the dot stands for a derivative with respect to the rescaled imaginary time $\bar{\tau}_E$, the prime stands for a derivative with respect to the rescaled spatial coordinate $\bar{x}$, and we have introduced $\bar{\tau}_E = \tau_E/\T_E$, $\bar{x} = \sigma/L$, $\xi = \pi T z$, $a =  \T_E \pi T$, and $b = L \pi T$\@.
We solve for the background worldsheet at four values of $ b \in \{ 1.0 \times 10^{-2}, 5.0 \times 10^{-3}, 2.5 \times 10^{-3}, 1.0 \times 10^{-3} \}$, holding $ a$ fixed at three different values $a \in \{0.1, 0.3, 1.0 \}$\@.

We obtain the numerical solutions to Eq.~\eqref{eq:NGactionpseudospectral} using the pseudospectral method \cite{boyd01}\@.
The pseudospectral method is an elegant way to solve boundary value problems such as the one in Eq.~\eqref{eq:NGactionpseudospectral}, which approximates the continuous differential equation by a finite set of coupled equations.
Specifically, we introduce an ansatz for $\xi$ in terms of the Chebyshev polynomials:
\[
\xi(\bar x, \bar{\tau}_E ) = \sum_{i,j=0}^{N_\text{coll}} c_{ij}T_i\left(2\bar x - 1\right)T_j\left(2\bar{\tau}_E - 1\right),
\]
with $c_{ij}$ unknown coefficients for which we need to solve and $T_i$ denoting the Chebyshev polynomial or order $i$\@. We then plug the ansatz into the equations of motion obtained from the action shown in Eq.~\eqref{eq:NGactionpseudospectral}\@. By evaluating these equations of motion at a finite number of points called \emph{collocation points}, we obtain a set of coupled equations. The number of collocation points is given by $N_{\rm coll}$\@. Any collocation points that lie on the boundary of the problem are constrained using the boundary conditions instead.
This immediately highlights a significant advantage of this pseudospectral method, as in this way boundary conditions are automatically satisfied, which is otherwise nontrivial for other approaches to boundary value problems.

For a general choice of such collocation points, the solution obtained in this way will not converge to the solution of the differential equation as we take the number of collocation points $N_\text{coll}$ to infinity.
If, however, we choose our collocation points to lie on the simultaneous zeroes of $T_{N_\text{coll} + 1}(2\bar x - 1)$ and $T_{N_\text{coll} + 1}(2{\bar \tau}_E - 1)$, the solution obtained is guaranteed to converge to the solution of Eq.~\eqref{eq:NGactionpseudospectral}, where the error goes like $\exp\left(-cN_\text{coll}\right)$, with $c$ some positive constant.
In practice, this means that with this choice of collocation points, the convergence as we take $N_\text{coll} \rightarrow \infty$ is very fast, so that we can achieve impressive precision even with a relatively small number of collocation points.
By varying the number of collocation points, we can also get an estimate of our truncation error.

For linear problems, the procedure described above leads to a set of $(N_\text{coll} + 1)^2$ coupled linear equations, which is exactly the number of unknown coefficients $c_{ij}$ we have, so in this case one can find the solution by matrix inversion.
Here we should note that the matrix that needs to be inverted is often close to singular, requiring us to work with more significant figures than machine precision provides.

A second complication is that the problem defined by Eq.~\eqref{eq:NGactionpseudospectral} is not linear.
Because of this, to find a solution we linearize the equations around a trial solution, and then use the Newton-Raphson method to iteratively update the trial solution until our iteration converges.
This introduces the usual difficulties associated with Newton-Raphson, namely that for certain choices of initial trial solution the iteration might not converge, but if for the first couple of steps in the iteration one uses very small step size in the update, it is generally not hard to reach the correct solution from a reasonably chosen initial trial solution.

\begin{figure}
    \centering
    \includegraphics[width=0.44\textwidth]{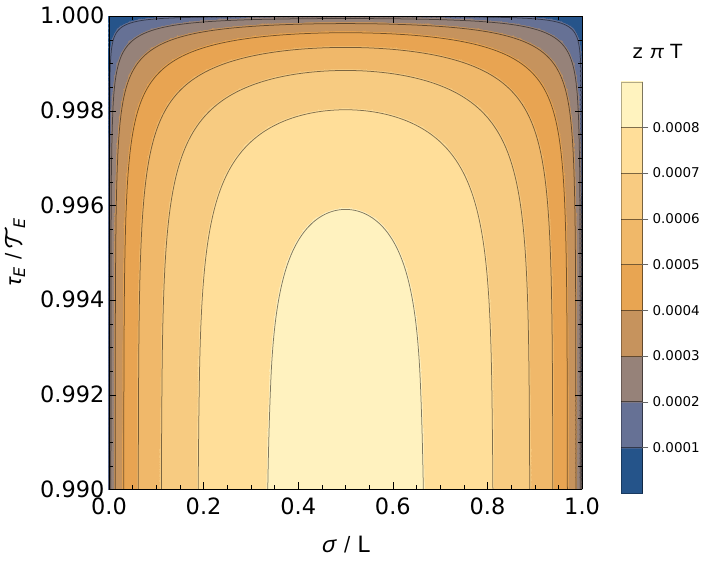}
    \includegraphics[width=0.54\textwidth]{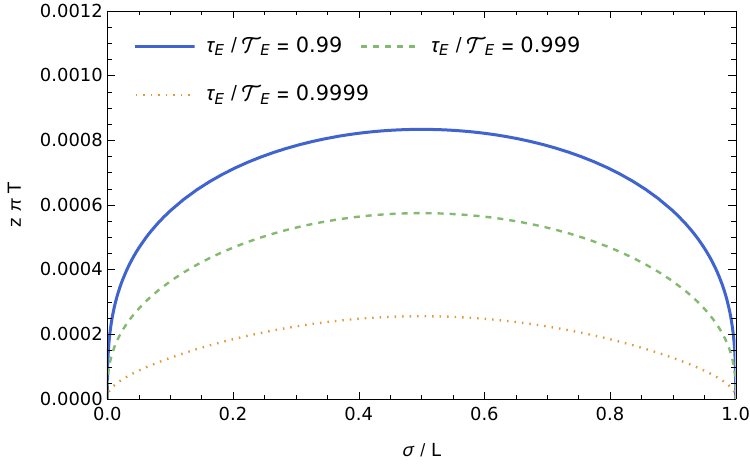}
    \caption{Example of a Euclidean worldsheet configuration hanging from a rectangle of dimensions $\T_E \times L = 0.1 (\pi T)^{-1} \times 10^{-3} (\pi T)^{-1}$ on the boundary of AdS${}_5$ into its radial direction, where a black hole lies at $z = (\pi T)^{-1}$\@. Away from $\tau_E = 0, \T_E$, the solution approaches the stable solution of the 1D problem given by Eq.~\eqref{eq:implicit-wx} in the small $z_m$ branch. Close to the corners it smoothly interpolates between the solution to the effective 1D problem and the boundary conditions specified by the bounded rectangular Wilson loop.}
    \label{fig:background-sample}
\end{figure}

A sample solution for $a = 0.1$ and $b = 10^{-3}$ can be found in Fig.~\ref{fig:background-sample}\@.
We can see that for $\tau_E$ away from the boundary conditions at $\tau_E = 0$ and $\tau_E = \mathcal{T}_E$, the solution agrees with the 1D problem from Eq.~\eqref{eq:implicit-wx}\@.

Once we obtain the background solution, we can introduce perturbations on the boundary and solve for the response functions.
On each of the background surfaces, we will solve for the linear response on one timelike side of the rectangle to perturbations in the boundary conditions on the other side. We use the same decomposition as for the background solution and write the perturbation as 
\begin{equation}
    y(\bar{x},\bar{\tau}_E) = \sum_{i,j=0}^{N_{\rm coll}} d_{ij} T_i(2 \bar{x} - 1) T_j(2 \bar{\tau}_E - 1) \, ,
\end{equation}
where $d_{ij}$ are the unknown coefficients we need to solve for.

Given that the numerical method to solve for the background worldsheet already defines a preferred basis on which to formulate this problem, we will calculate the response functions by mapping a perturbation onto a given basis element $T_n(2\bar{\tau}_E - 1)$, where $T_n$ is a Chebyshev polynomial determining the boundary condition for the transverse fluctuations along the time axis on one side of the rectangular contour, to another basis element $T_m(2\bar{\tau}_E - 1)$ on the other side of the contour. That is to say, given a boundary condition at $\bar{x} = 0$, specified by $y(0,\bar{\tau}_E) = T_n(2\bar{\tau}_E - 1)$, we will want to determine the response at the other side of the contour $y'(\bar{x} = 1, \bar{\tau}_E)$, decomposed in terms of Chebyshev polynomials.

To put this on a concrete mathematical footing, we shall repeat some of the discussions in Section~\ref{sec:HQ-fluct}, keeping in mind that we now work in Euclidean signature with a finite Euclidean time extent. As before, the response kernel of interest is obtained by solving the equations of motion derived from the action for the fluctuations, which can be written as
\begin{align}
    S^{(2),\perp}_{{\rm NG},c,E}[y] &= \int_{0}^{\T_E} \!\! \diff \tau_E \! \int_{0}^L \!\! \diff \sigma  \frac{  \left(\frac{\partial y}{\partial \tau_E}\right)^2  + f \left(\frac{\partial y}{\partial \sigma}\right)^2  + \frac{1}{f} \left(  \frac{\partial z}{\partial \sigma} \frac{\partial y}{\partial \tau_E} - \frac{\partial z}{\partial \tau_E} \frac{\partial y}{\partial \sigma} \right)^2 }{2 z^2 \sqrt{f + \left(\frac{\partial z}{\partial \sigma}\right)^2 + \frac{1 }{f} \left(\frac{\partial z}{\partial \tau_E}\right)^2 }} \,  \nonumber \\
    &= a b \int_{0}^{1} \diff \bar{\tau}_E \int_0^1 \diff \bar{x}  \frac{ \frac{\dot{y}{}^2}{\T_E^2} \left(1 + \frac{\xi'{}^2}{b^2 (1 - \xi^4) } \right) + \frac{y'{}^2}{L^2} \left( 1 - \xi^4 + \frac{\dot{\xi}{}^2}{a^2(1-\xi^4)} \right) - \frac{2 y' \dot{y} \xi' \dot{\xi} }{L \T_E b a (1 - \xi^4) } }{ 2 \xi^2 \sqrt{1 - \xi^4 + \frac{\xi'{}^2}{b^2 } + \frac{\dot{\xi}{}^2}{a^2(1-\xi^4)} } } \, .
\end{align}
Evaluating this action on-shell, with the only non-vanishing boundary conditions being in the temporal segments of the Wilson loop, we get 
\begin{align}
    S^{(2),\perp}_{{\rm NG},c,E}[y]_{\rm on-shell} &= \frac{ab}{2L^2} \int_{0}^{1} \diff \bar{\tau}_E \int_0^1 \diff \bar{x} \frac{\partial}{\partial \bar{x} } \left( \frac{ \left( 1 - \xi^4 + \frac{\dot{\xi}{}^2}{a^2(1-\xi^4)} \right) y y' }{\xi^2 \sqrt{1 - \xi^4 + \frac{\xi'{}^2}{b^2 } + \frac{\dot{\xi}{}^2}{a^2(1-\xi^4)} }} \right) \nonumber \\
    &= \frac{ab}{2L^2} \int_{0}^{1} \diff \bar{\tau}_E \left[ \lim_{\bar{x} \to 1} \frac{y y'}{\xi^2 \sqrt{1 + \frac{\xi'{}^2}{b^2} } } - \lim_{\bar{x} \to 0} \frac{y y'}{\xi^2 \sqrt{1 + \frac{\xi'{}^2}{b^2} } } \right] \, ,
\end{align}
where we have dropped $\xi$ and $\dot{\xi}$ as they vanish at the boundaries defined by the temporal segments of the Wilson loop, where by definition $\xi = 0$ and therefore $\dot{\xi}=0$\@. A posteriori, knowing the solution to the background worldsheet, one can verify that the limits $\ell(\bar{\tau}_E; a,b) \equiv \lim_{{\bar x} \to 0,1} \xi^2 \sqrt{1 + \frac{\xi'{}^2}{b^2} } $ are finite and equal. 
To avoid issues with contact terms, we will extract the quadratic kernel from variations on opposite sides of the contour. That is to say, we will calculate the quadratic kernel $\Delta_{\perp,E}^{c}(\tau_{E1},\tau_{E2};L)$ that appears as a $ \int d\tau_{E1} d\tau_{E2} \, y_L(\tau_{E1}) \Delta(\tau_{E1},\tau_{E2}) y_R(\tau_{E2}) $ contribution to the on-shell action, given by
\begin{equation}
    \Delta_{\perp,E}^{c}(\tau_1,\tau_2;L) = \frac{\sqrt{\lambda}}{\pi} \frac{ab}{ L^2 \T_E^2 } \left( \frac{1}{\ell(\bar{\tau}_{E1})} + \frac{1}{\ell(\bar{\tau}_{E2})} \right) \widetilde{K}^{RL,c}_{\perp, { E}}(\bar{\tau}_{E2},\bar{\tau}_{E1}) \, ,
\end{equation}
where $\widetilde{K}^{RL,c}_{\perp, E}(\tau_{E2},\tau_{E1})$ is defined as the following response kernel:
\begin{equation}
    y'(\bar{x} = 1, \bar{\tau}_E) = - \int_0^1 \diff \bar{\tau}_E' \, \widetilde{K}^{RL,c}_{\perp, { E}}(\bar{\tau}_E,\bar{\tau}_E') y(\bar{x}=0,\bar{\tau}_E') \, .
\end{equation}

\begin{figure}
    \centering
    \includegraphics[width=\textwidth]{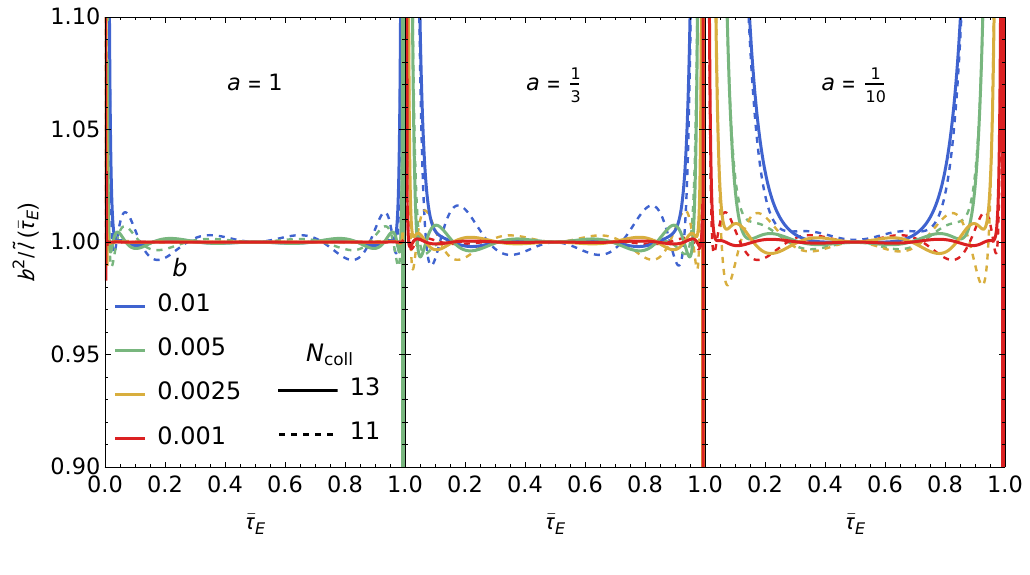}
    \caption{Plots of $b^2/\tilde{\ell}(\bar{\tau}_E)$ for different values of $a, b$ and different number of collocation points $N_{\rm coll}$\@. Here we introduced $\tilde{\ell}(\bar{\tau}_E) = \left( \frac{2 \sqrt{\pi} \, \Gamma(7/4)}{3\, \Gamma(5/4)} \right)^{2} \ell(\bar{\tau}_E)$ so that the limit is rescaled to unity.}
    \label{fig:ell-approach}
\end{figure}

All that remains is to evaluate the response kernel $\widetilde{K}^{RL,c}_{\perp, {E}}$ and the limit $\ell(\bar{\tau}_E)$ in the background solution. Before proceeding, we first discuss what the expected result is. From our analysis of the effective 1D problem (in the limit $\T \to \infty$) in Minkowski signature, we see from Eq.~\eqref{eq:Delta-c-final} that the limit $L \propto b \to 0$ should give us
\begin{equation}
    \Delta_{\perp,E}^{c}(\tau_1,\tau_2;L) = \frac{2\sqrt{\lambda}}{\pi} \left( \frac{2 \sqrt{\pi} \, \Gamma(7/4)}{3\, \Gamma(5/4)} \right)^2 \frac{1}{L^3} \delta(\tau_1 - \tau_2) + \ml{O}(L^{-1}) \, .
\end{equation}
Noting that $\delta(\tau_1 - \tau_2) = \T_E^{-1} \delta(\bar{\tau}_{E1} - \bar{\tau}_{E2})$, and that in the strict limit $\T \to \infty$ we have $\ell = b^2 \left(\frac{2\sqrt{\pi}\, \Gamma(7/4)}{3 \, \Gamma(5/4)} \right)^{-2} $, we expect
\begin{align}
    \lim_{b \to 0} \widetilde{K}^{RL,c}_{\perp, {E}}(\bar{\tau}_E,\bar{\tau}_E') = \delta(\bar{\tau}_{E1} - \bar{\tau}_{E2}) \, \, , & & \lim_{b \to 0} \frac{b^2}{ \ell(\bar{\tau}_E) } = \left( \frac{2 \sqrt{\pi} \, \Gamma(7/4)}{3\, \Gamma(5/4)} \right)^{2} \, \, . \label{eq:numerical-target}
\end{align}
Given these expectations, our numerical evaluation of $\widetilde{K}^{RL,c}_{\perp, { E}}$ and $\ell(\bar{\tau}_E)$ needs only to demonstrate Eq.~\eqref{eq:numerical-target}\@.

\begin{figure}
    \centering
    \includegraphics[width=\textwidth]{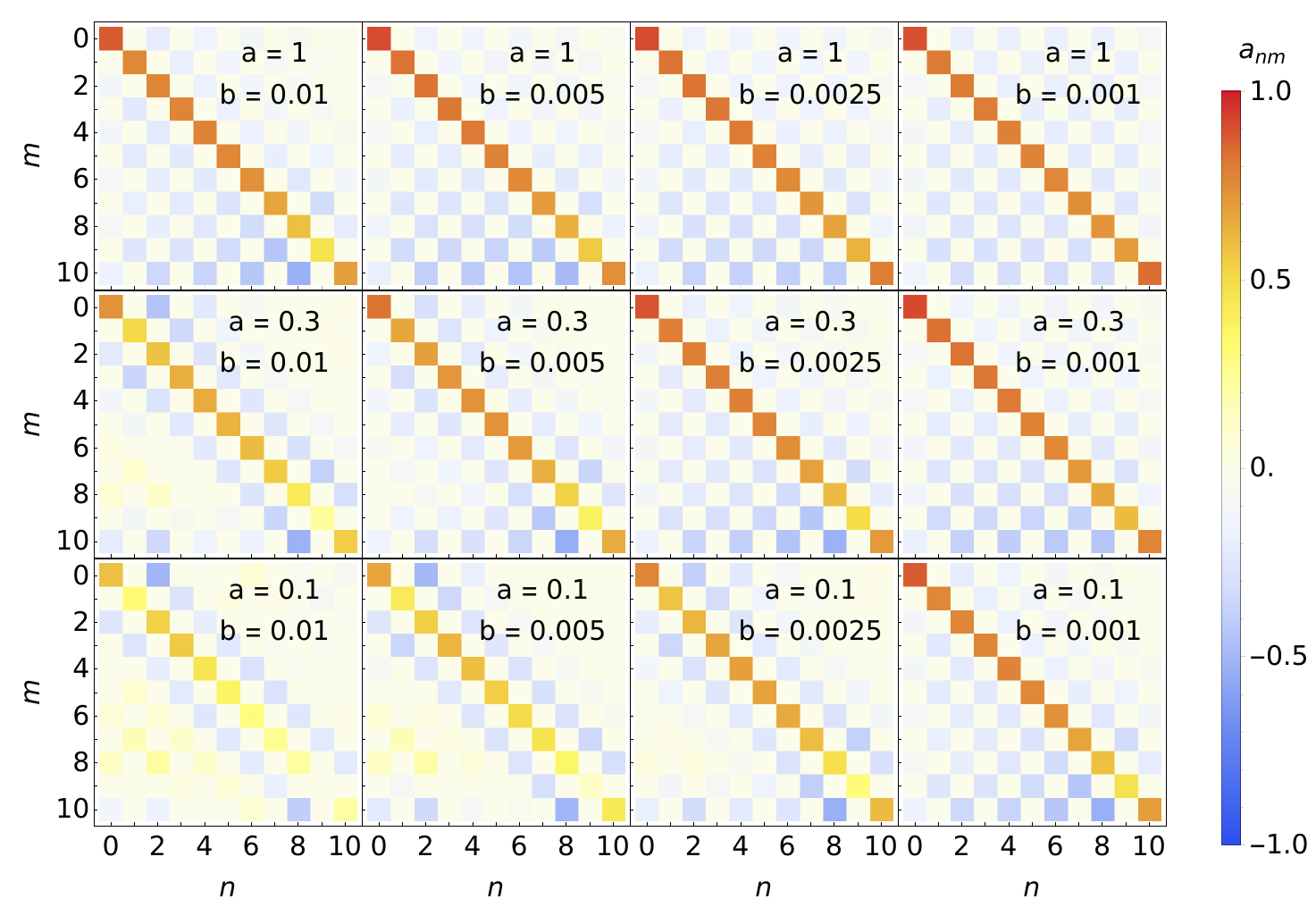}
    \caption{Coefficients $a_m^{(n)}$ as defined in Eq.~\eqref{eq:anm-def}, for $N_{\rm coll} = N_{\rm pols} = 11$\@.}
    \label{fig:amn-approach-10}
\end{figure}

\begin{figure}
    \centering
    \includegraphics[width=\textwidth]{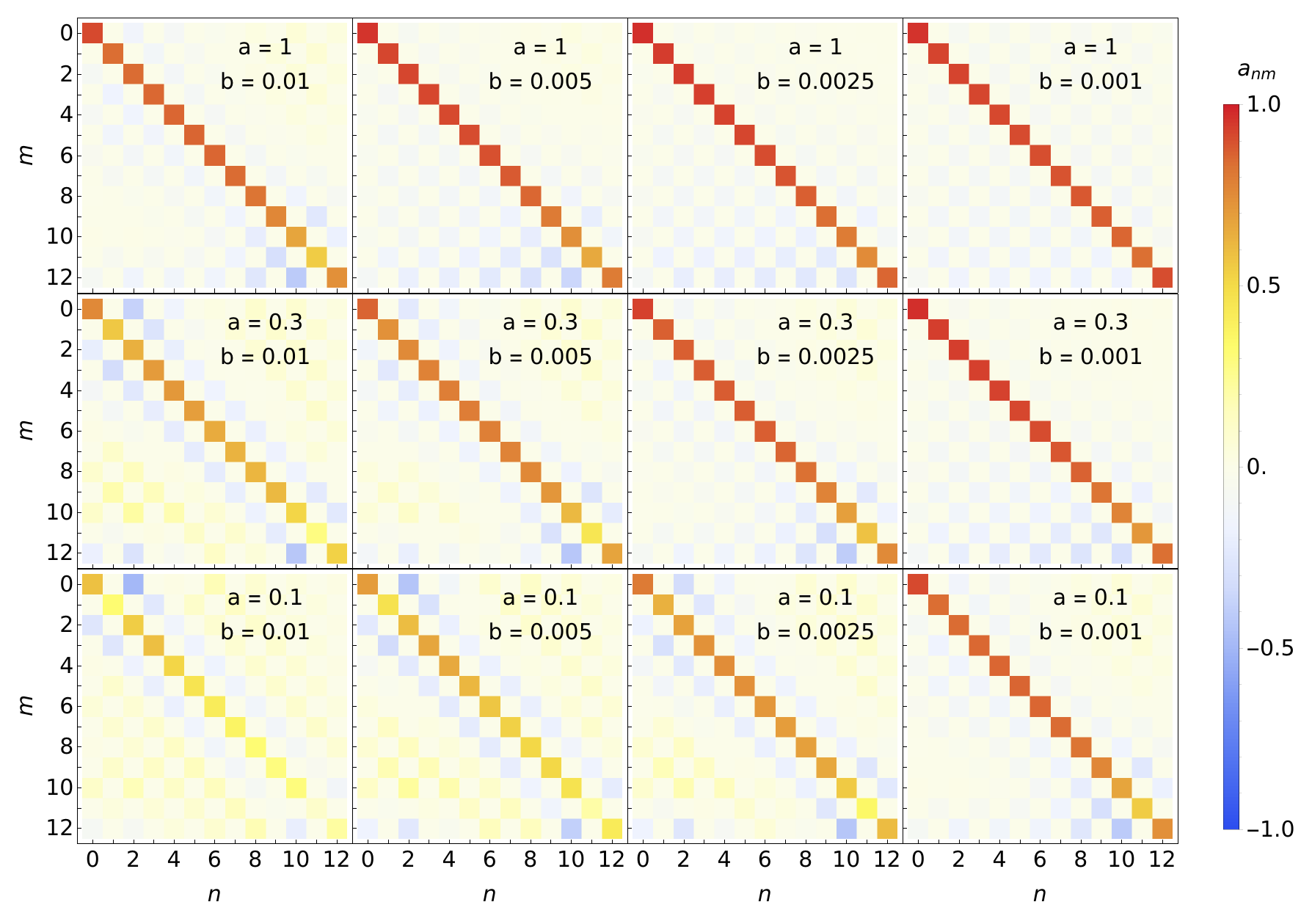}
    \caption{Coefficients $a_m^{(n)}$ as defined in Eq.~\eqref{eq:anm-def}, for $N_{\rm coll} = N_{\rm pols} = 13$\@.}
    \label{fig:amn-approach-12}
\end{figure}

We show results for $b^2/\ell(\bar{\tau}_E)$, normalized by its limiting value in Fig.~\ref{fig:ell-approach}\@. We observe two general trends:
\begin{enumerate}
    \item As $b \to 0$, the solution indeed approaches the limiting value, converging first in the middle region $\bar{\tau}_E \sim 1/2$ and later near the corners. For sufficiently small $b$, the convergence is more strongly dependent on the ratio $b/a$ than on $b$ alone. 
    \item The oscillations in the solution, which are artifacts of a truncated basis, get suppressed as we increase the number of collocation points $N_{\rm coll}$, and the convergence of $b^2/\ell(\bar{\tau}_E)$ as $b \to 0$ is observed even more clearly at large $N_{\rm coll}$\@. By increasing the number of collocation points we would reduce the truncation effects, and the solution would approach the exact profile at each value of $b$, with which the limit $b \to 0$ could be examined even more precisely. However, we will refrain to go further, as we deem the plots in Fig.~\ref{fig:ell-approach} as sufficient evidence for the value of the limit we wished to verify.
\end{enumerate}

Next we test the convergence of $\widetilde{K}^{RL,c}_{\perp, {\rm E}}$ by studying the derivative response $y'$ at $\bar{x}=1$ to a boundary condition specified by a Chebyshev polynomial $T_n(2\bar{\tau}_E - 1)$ at $\bar{x}=0$\@. Concretely, we expand the derivative response in terms of Chebyshev polynomials and find the coefficients $a_m^{(n)}$ of the expansion
\begin{equation} \label{eq:anm-def}
    y_{(n)}'(\bar{x}=1, \bar{\tau}_E) = -\sum_{m=0}^{N_{\rm pols}-1} a_m^{(n)} T_m(2\bar{\tau}_E - 1) \, ,
\end{equation}
and we plot these coefficients for different values of $a$ and $b$\@. $N_{\rm pols}$ is the number of Chebyshev polynomials we use in the spectral approach and it is equal to the number of collocation points $N_{\rm coll}$\@.
The expectation is that $a_n^{(n)} \to 1$ and $a_m^{(n)} \to 0$ if $m \neq n$\@. As we can see from Figs.~\ref{fig:amn-approach-10} and~\ref{fig:amn-approach-12}, as the ratio $b/a$ goes to zero, the convergence $a_m^{(n)} \to \delta_{mn}$ is rather good, and qualitatively improves as we refine the set of basis functions in the pseudospectral method.

In a nutshell, we see that everything in the numerical Euclidean approach is consistent with our previous real time analysis in Section~\ref{sec:transverse-v-fluct}, meaning that the quadratic kernel $\Delta^c$ diverges as $L^{-3}$ when we take $L \to 0$\@. As such, we conclude that the limit $L \to 0$ of this SYM Wilson loop does not describe the physics we wish to capture, because it is dominated by UV contributions that are not in the domain of any low-energy effective description. This is also consistent with our previous discussion that we in fact expect $\langle \hat{\T} W[\mathcal{C}] \rangle_T = 1$ for the SU(3) Wilson line configuration that is relevant to quarkonium. As such, we conclude that we must seek other configurations to describe the Wilson loop that is relevant for quarkonium dynamics in a thermal medium.

\section{The Wilson loop with antipodal \texorpdfstring{$S_5$} {} coordinates} \label{sec:QQ-setup}
Having studied the supersymmetric Wilson loop with constant $\hat{n}$, we now proceed to investigate the other candidate configuration, where we have $\hat{n} = \hat{n}_0$ on one of the timelike segments comprising the loop, and $\hat{n} = -\hat{n}_0$ on the other antiparallel segment of the loop.

This configuration has received less attention in the AdS/CFT studies of heavy quarks, mainly because it does not generate a Coulomb interaction potential between a heavy quark pair. However, while it has been usually less emphasized, it has been discussed in many AdS/CFT studies, starting from the same works that discussed the heavy quark interaction potential~\cite{Maldacena:1998im,Rey:1998ik}\@. As we will also verify momentarily, it has the crucial property that $\langle W_{\rm BPS}[\mathcal{C}] \rangle = 1$ for a contour going from one point to another and coming back to the starting point along the same path. Apart from the fact that one can verify this identity by hand in the CFT, this relation is protected by supersymmetry~\cite{Zarembo:2002an}\@. Moreover, the fact that this configuration might be relevant for the dynamics of a quark-antiquark pair was hinted in Ref.~\cite{Rey:1998ik}, where the first appearance of quark pairs and heavy quark pairs featured antipodal positions on the $S_5$\@.

Given the relative lack of attention that this configuration has received, especially for phenomenological applications, we will try to make our discussions as detailed as possible.

\subsection{Background}

As before, we consider the $\T \to \infty$ limit of the contour $\mathcal{C}$ that defines the Wilson loop from which we can extract the correlator relevant for quarkonium transport. In $\mathcal{N}=4$ SYM, we also have to specify the position on the $S_5$ that the Wilson loop goes over for it to have a dual gravitational description in terms of an extremal surface in AdS${}_5 \times S_5$\@. As discussed in Section~\ref{sec:nhat}, a natural choice is to have two long timelike Wilson lines with antipodal positions on the $S_5$\@. Without loss of generality, we can describe the distance between the $S_5$ coordinates on the two boundary segments by a large circle angular coordinate $\phi \in [0,2\pi)$, and then we may substitute $d\Omega_5^2$ into the metric shown in Eq.~\eqref{eq:Schwarzschild-AdS} by $d\phi^2$\@.

First we study whether there is an extremal surface connecting the two boundary segments that can be described by
\begin{align}
X^\mu(\tau,\phi) = (\tau, 0, 0 , 0, z(\phi), \hat{n}(\phi)) \, ,
\end{align}
with which the Nambu-Goto action reads:
\begin{equation} \label{eq:NG-z-of-phi}
    \mathcal{S}_{\rm NG}[\Sigma] = - \frac{\sqrt{\lambda}}{2\pi} \int d\tau \, d\phi \frac{\sqrt{z'{}^2+ f z^2}}{z^2} \, ,
\end{equation}
where now  we define $z' = \partial z/\partial \phi$\@. The action of the resulting extremal surface should be compared with the action of two disconnected worldsheets falling into the black hole. If we find a positive\footnote{This is due to the overall minus sign in the definition of $\mathcal{S}_{\rm NG}$\@.} regularized action by extremizing the action~\eqref{eq:NG-z-of-phi}, then it will be the preferred configuration, as it will be the one of the lowest energy. On the other hand, if the energy of the configuration we find by extremizing the action~\eqref{eq:NG-z-of-phi} is higher than that of two disconnected worldsheets (i.e., if their regularized action is negative), then the dynamically favored configuration will be the ``trivial'' one, given by the two disconnected worldsheets. We note that no spatial separation between the two timelike Wilson lines is necessary in this configuration, as the angular separation on the $S_5$ already provides the surface $\Sigma$ a non-vanishing coordinate region for it to extend itself over.

Therefore, let us calculate the extremal surfaces that can be derived from Eq.~\eqref{eq:NG-z-of-phi}\@. Like its counterpart discussed in Section~\ref{sec:HQ-setup}, this action has a conserved quantity due to the explicit independence of the action on $\phi$,
\begin{equation}
    \frac{f(z)}{\sqrt{z'{}^2+ f(z) z^2}} = \sqrt{\frac{f(z_m)}{z_m^2}} \, ,
\end{equation}
which allows us to find an implicit solution for $z(\phi)$ by direct integration
\begin{equation}
\label{eqn:arc_length}
    \int_0^{z(\phi)}  \frac{dz}{\sqrt{z_m^2 f(z) - z^2 f(z_m)}} \sqrt{\frac{f(z_m)}{f(z)}} = \phi \, ,
\end{equation}
where we have chosen one of the timelike Wilson lines to lie at $\phi = 0$\@. The above expression determines the worldsheet configuration up to its maximal radial value $z_m$\@. The value of $z_m$ may then be related to (half of) the total angular distance spanned by the worldsheet $\Delta \phi$ by replacing the upper limit $z(\phi)$ with $z_m$ in Eq.~\eqref{eqn:arc_length}:
\begin{equation} \label{eq:delta-phi-zm}
    \Delta \phi = \int_0^{z_m}  \frac{dz}{\sqrt{z_m^2 f(z) - z^2 f(z_m)}} \sqrt{\frac{f(z_m)}{f(z)}}  \, .
\end{equation}

\begin{figure}
    \centering
    \includegraphics[width=0.4\textwidth]{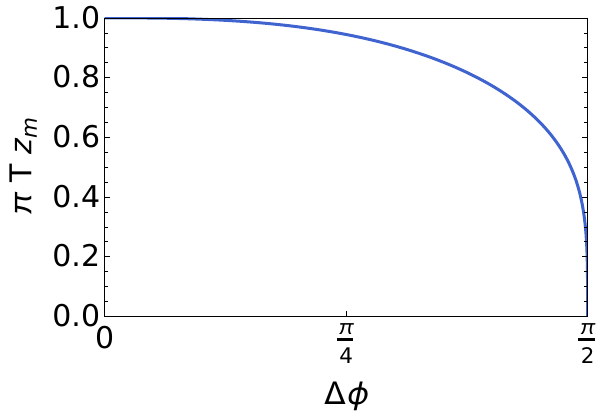}
    \includegraphics[width=0.4\textwidth]{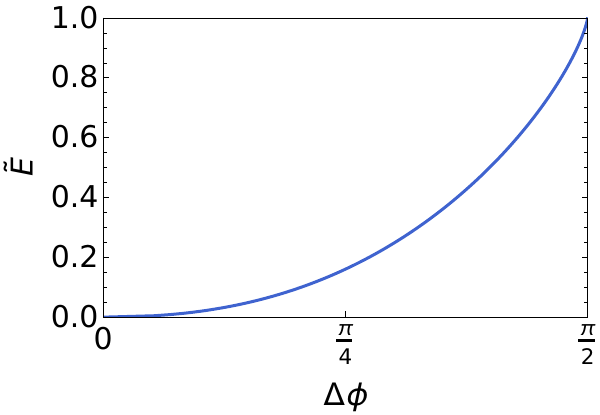}
    \caption{Left: the relation $\Delta \phi ( \pi T z_m )$ that determines the angular distance spanned on the $S_5$ by the connected configuration that reaches a maximal AdS radial coordinate $z = z_m$\@. Right: dimensionless configuration energy $\widetilde{E}$ for the extremal worldsheet that is described by a connected configuration $z = z(\phi)$\@.}
    \label{fig:S5-background}
\end{figure}

One can then evaluate the regularized action $\mathcal{S}_{\rm NG}[\Sigma] - \mathcal{S}_0$ with this solution. A straightforward calculation, which we write out in detail in the next equation, gives the energy of the configuration as
\begin{align}
    \mathcal{S}_{\rm NG}[\Sigma] - \mathcal{S}_0 &= - \frac{\sqrt{\lambda} \T }{2\pi} \int_0^\pi d\phi \frac{\sqrt{z'{}^2+ f z^2}}{z^2} + \frac{\sqrt{\lambda} \T}{\pi} \int_0^{(\pi T)^{-1}} \frac{dz}{z^2} \nonumber \\ 
    &= - \frac{\sqrt{\lambda} \T }{\pi} \int_0^{z_m} \frac{dz}{z^2} \left( \frac{1}{ \sqrt{1 - \frac{z^2 f(z_m)}{z_m^2 f(z)} }} - 1 \right) + \frac{\sqrt{\lambda} \T}{\pi} \int_{z_m}^{(\pi T)^{-1}} \frac{dz}{z^2} \nonumber \\
    &= - \frac{\sqrt{\lambda} \T}{\pi} (\pi T) \frac{1}{\pi T z_m} \left[ \int_0^1 \frac{du}{u^2} \left( \frac{1}{\sqrt{1 - u^2 \frac{1- (\pi T z_m)^4 }{1 - (\pi T z_m)^4 u^4} }} - 1 \right) - (1 - (\pi T z_m) ) \right] \nonumber \\
    &\equiv - \sqrt{\lambda} \T T \widetilde{E}(\pi T z_m) \, .
\end{align}
where $\widetilde{E}(\pi T z_m)$ is a function of a single variable that characterizes the configuration energy as a function of the $S_5$ angular separation $2\Delta \phi$ of the two boundary timelike Wilson lines, determined by Eq.~\eqref{eq:delta-phi-zm}\@. We plot this quantity in Fig.~\ref{fig:S5-background}, i.e., we plot the energy of the configuration in units of the temperature times $\sqrt{\lambda}$, together with the relation that determines $\Delta\phi = \Delta \phi(\pi T z_m)$ as prescribed by Eq.~\eqref{eq:delta-phi-zm}\@. We see that for worldsheets that can be parametrized by the functions $z(\phi)$, and therefore are connected, the map $\Delta \phi = \Delta \phi (\pi T z_m)$ is one to one, and that at any $\Delta \phi > 0$, their energy is strictly greater than that of two disconnected, radially infalling worldsheets each hanging from their respective boundary toward the bulk of AdS${}_5$\@. 

Crucially, this means that for our present purpose, which is to find the minimal energy configuration for $\Delta \phi = \pi/2$, the relevant extremal surface is that of two disconnected (at least at times $|t| \ll \T/2$) worldsheets hanging radially into the bulk of AdS${}_5$\@. The alternative solution, which is energetically disfavored, is a surface that lies at $z=0$ and can be parametrized by time $t \in (-\T/2,\T/2)$ and the angle $\phi \in (0 , \pi)$\@. It is interesting to note that in the strict limit $T = 0$, all connected configurations have the same energy as the radially infalling solution (i.e., zero), for any value of $z_m$\@. However, since we are interested in the physics in the presence of a thermal plasma, there is no ambiguity in terms of which solution to choose.

Therefore, the solution presently relevant to our purpose is parametrized by two disjoint surfaces
\begin{align}
X_L^\mu(\tau,z) = (\tau, 0, 0 , 0, z, -\hat{n}_0) \, ,
\end{align}
and
\begin{align}
X_R^\mu(\tau,z) = (\tau, 0, 0 , 0, z, \hat{n}_0) \, ,
\end{align}
where $z \in (0, (\pi T)^{-1} )$ is an independent coordinate in this description, and plays the role of one of the worldsheet coordinates.
We note that the solution we found above also applies if the two timelike sides of the contour $\mathcal{C}$ are at nonzero spatial separation $L>0$ (provided they remain at antipodal positions $\Delta \phi = \pi/2$ on the $S_5$), because allowing for a $\phi$-dependent $y_1$ coordinate in the connected background solution can only increase the configuration energy. 

In summary, this configuration features two worldsheets that fall into the black hole, which intuitively represents the propagation of two unbound heavy quarks, with their interactions being screened by the thermal medium~\cite{Friess:2006rk,Liu:2006ug,casalderrey2014gauge}\@. By construction, this configuration has $E^d = 0$ (after subtracting the energy associated to the heavy quark masses), as required to satisfy $\langle W_{\rm BPS}[\mathcal{C}] \rangle = 1$\@.

\subsection{Fluctuations}

To calculate the response kernel of interest for this configuration $\Delta_{ij}^d$, we study the dynamics of fluctuations on top of the background worldsheet we just found. In consistency with the preceding discussion, we work in the limit $\T \to \infty$ (concretely, $|t_1|,|t_2|\ll \T$), which also simplifies the calculations because of the time translational invariance. In this setup, whether $\T$ is finite or infinite
does not affect the final result, as long as the time domain of interest is covered by the timelike Wilson lines, since the parts of the timelike Wilson lines that are out of the time domain of interest cancel due to $U U^{-1} = \mathbbm{1}$\@.\footnote{Up to operator ordering subtleties that do not affect this conclusion. We discuss this in Appendix~\ref{sec:App-W-ordering}\@.} As such, taking $\T \to \infty$ is not an approximation, but rather, it is manifestly equal to the starting point.

As discussed in Section~\ref{sec:setup}, the appropriate boundary deformations of the contour $\mathcal{C}$ to evaluate the chromoelectric field correlator in the pure SU($N_c$) gauge theory are the antisymmetric ones, as shown in Eq.~\eqref{eq:f-antisymm}\@. By the same argument as in that Section, these are also the only nontrivial deformations from which we can extract the desired correlation function in our present setup. The reason is that symmetric boundary deformations of the contour $\mathcal{C}$ preserve the value of the (supersymmetric) Wilson loop $W_{\rm BPS}[\mathcal{C}] = 1$, which is also a consequence of having each antiparallel timelike Wilson line with antipodal positions on the $S_5$\@.

Having made these remarks, we now proceed to introduce perturbations on top of the background worldsheet to enable us to evaluate the path functional derivatives on the boundary. Compared to the setup in the previous section, the parametrization of the perturbed worldsheet here is remarkably simpler:
\begin{align}
    X^\mu_L(t,z) &= (t, - y_1(t,z), - y_2(t,z), - y_3(t,z), z, -\hat{n}_0) \, , \\
    X^\mu_R(t,z) &= (t, + y_1(t,z), + y_2(t,z), + y_3(t,z), z, + \hat{n}_0) \, ,
\end{align}
where we have already set to zero the fluctuations corresponding to reparametrization invariance, $y_0$ and $y_4$\@. We have also already incorporated the fact that, because of the antisymmetry of the boundary conditions we will use, the solutions for the fluctuations on either surface will be equal but with opposite signs. 

The action, up to quadratic order for the fluctuations, takes the form
\begin{align}
    \mathcal{S}_{\rm NG}[\Sigma] - \mathcal{S}_0[\mathcal{C};\hat{n}] = - 2 \times \frac{\sqrt{\lambda}}{2\pi} \left[ S_{{\rm NG},d}^{(2),\perp}[y_1] + S_{{\rm NG},d}^{(2),\perp}[y_2] + S_{{\rm NG},d}^{(2),\perp}[y_3] \right] \, ,
\end{align}
where we have subtracted the action corresponding to the energy of two heavy quarks at rest, which is incidentally equal to the background action in this case. The form of the action for the fluctuations is the same for all components and is given by
\begin{align}
    S_{{\rm NG},d}^{(2),\perp}[y] = \int_{-\infty}^\infty \!\! dt \int_0^{(\pi T)^{-1}} \!\!\!\!\!\! dz \left[ \frac{f}{2z^2} y'{}^2 - \frac{1}{2z^2 f} \dot{y}{}^2 \right] \, ,
\end{align}
where a prime denotes the derivative with respect to $z$ and a dot represents the derivative with respect to $t$\@.

Integrating by parts, and using the equations of motion, we can evaluate the on-shell action as
\begin{align} \label{eq:S-fluct-thermal-on-shell}
    S_{{\rm NG},d}^{(2),\perp}[y]_{\rm on-shell} = { \frac12} \int_{-\infty}^\infty dt \left[ \lim_{z \to (\pi T)^{-1} } \frac{f y'(t, z ) y(t, z )}{z^2} - \lim_{z \to 0} \frac{f y'(t, z ) y(t, z )}{z^2} \right] \, ,
\end{align}
which involves two explicitly nontrivial limits. Let us first focus on the first limit, where $z \to (\pi T)^{-1}$\@. While it is tempting to conclude that it is zero because $f(z=(\pi T)^{-1}) = 0$, we actually have to solve for the mode functions first and verify that its product with $y' y$ indeed goes to zero. In the following Section~\ref{sec:EE-calculation-QQ}, after selecting appropriate boundary conditions at the black hole horizon, we will confirm that this is the case. Therefore, we shall drop this term in the remainder of this section. 

The second limit, where $z \to 0$, is even more subtle, because it can be manifestly divergent if $y'$ does not go to zero fast enough. Nonetheless, after solving for the mode functions and investigating them at small $z$, we shall see that it contains a $1/z$ divergent term of the form discussed around Eq.~\eqref{eq:F-insertion-2} and in the footnote.\footref{fn:delta-disc} That is to say, it is generated by the contact term that comes from evaluating the variational derivatives with respect to the path deformations $f^\mu$ at the same point. As we will see later, this contribution has exactly the same form and value as that calculated in the previous Section~\ref{sec:EE-calculation-HQ}\@. As such, we justify it to simply subtract the second limiting term from our final result. Furthermore, using that we will find the relevant mode functions satisfy $y'(t,z=0) = 0$, we can conclude by repeated use of the L'Hopital's rule that
\begin{align}
    S_{{\rm NG},d}^{(2),\perp}[y]_{\rm on-shell} = - \frac{1}{ 4} \int_{-\infty}^\infty dt \frac{\partial^3 y(t, z = 0)}{\partial z^3}  y(t, z=0) \, .
\end{align}
As such, the response function we will need to calculate has a third derivative. The other way of distributing the three derivatives gives terms of the form $ \frac{\partial^2 y}{\partial z^2} \frac{\partial y}{\partial z}$, which vanish because the mode functions satisfy $y'(t,z=0) = 0$\@.

Let us now define the response function we will calculate. As in the case of the calculation of Section~\ref{sec:EE-calculation-HQ}, we identify the value of the fluctuation on the boundary with that of the contour deformation $y(t,z=0) = h^\perp(t)$\@. As such, we introduce the response kernel
\begin{align} \label{eq:K-perp-d}
    \frac{\partial^3 y}{\partial z^3} (t, z = 0) = -\int_{-\infty}^\infty dt' K_\perp^d(t-t') h^\perp(t') \, ,
\end{align}
with which we find
\begin{align}
    S_{{\rm NG},d}^{(2),\perp}[y]_{\rm on-shell} = { \frac14} \int_{-\infty}^\infty dt \, dt' \, h^\perp(t) K_\perp^d(t-t') h^\perp(t') \, .
\end{align}
With this, the correlation function we seek is determined by
\begin{equation} \label{eq:Delta-disconnected}
    \Delta^d_{ij}(t_2 - t_1) =  \frac{\sqrt{\lambda}}{{ 2}\pi} \delta_{ij} K^d_\perp(t_2-t_1) \, .
\end{equation}
As such, all that remains to be done is to evaluate the response function $K^{d}_{\perp}$\@.

\subsection{Calculation of the time-ordered non-Abelian electric field correlator} \label{sec:EE-calculation-QQ}

To calculate the response function $K^{d}_{\perp}$, we proceed by varying the action $S_{{\rm NG},d}^{(2),\perp}$ with respect to $y$ to obtain its equations of motion, and then transform to the frequency domain. Then, introducing $\xi = \pi T z$, the equation we want to solve is 
\begin{equation} \label{eq:eom-y-th}
 \frac{\partial^2 y_\omega}{\partial \xi^2} - \frac{2}{\xi}  \frac{1+\xi^4}{1 - \xi^4}   \frac{\partial y_\omega}{\partial \xi} + \frac{\omega^2}{(\pi T)^2} \frac{1}{(1 - \xi^4)^2} y_\omega = 0 \, ,
\end{equation}
which is actually equivalent to the one found by Ref.~\cite{Casalderrey-Solana:2006fio} to calculate the heavy quark diffusion coefficient in strongly coupled $\mathcal{N}=4$ SYM theory. For the benefit of the reader, we note that in their notation, the independent variable that parametrizes the worldsheet is $u = \xi^2$\@. 

To find the solutions to Eq.~\eqref{eq:eom-y-th}, we proceed as in Ref.~\cite{Casalderrey-Solana:2006fio} to factor out the highly oscillatory piece that is generated close to the black hole event horizon. This can be done by the same WKB-type analysis we carried out in the calculation of the response functions in Section~\ref{sec:EE-calculation-HQ}\@. Using the same notation, we introduce
\begin{equation} \label{eq:mode-functions-yF}
    y^{\pm}_\omega(\xi) = (1-\xi^4)^{\pm \frac{i \Omega}{4} } F^{\pm}_{\omega}(\xi) \, ,
\end{equation}
where the prefactor $(1-\xi^4)^{\pm {i \Omega}/{4} }$ is obtained by direct integration of the WKB phase in Eqs.~\eqref{eq:yL} and~\eqref{eq:yR}\@. To facilitate comparison, we have written $y_\omega^\pm$ in the way of Eq.~\eqref{eq:mode-functions-yF} such that the resulting mode functions are the same as in Ref.~\cite{Casalderrey-Solana:2006fio}\@. With this definition, $F^\pm_\omega$ satisfies
\begin{align} \label{eq:F-thermal}
    \frac{\partial^2 F^\pm_\omega}{\partial \xi^2} - 2 \left[ \frac{1 + \xi^4}{\xi(1-\xi^4)} \pm \frac{i \Omega \xi^3}{1-\xi^4} \right] \frac{\partial F^\pm_\omega}{\partial \xi} + \left[ \mp \frac{i \Omega \xi^2}{1-\xi^4} + \frac{\Omega^2 (1 - \xi^6) }{(1-\xi^4)^2} \right] F^\pm_\omega = 0 \, .
\end{align}

The purpose of extracting the highly oscillatory phase from $y_\omega^\pm(\xi)$ is to get an equation for $F_\omega^\pm(\xi)$ such that a regular solution can be found at $\xi = 1$\@. This condition must be imposed by hand, because Eq.~\eqref{eq:F-thermal} has two independent solutions: one regular at the horizon, and the other oscillating twice as fast as the solutions for $y_\omega^\pm(\xi)$\@.
Examining the differential equation for $F_\omega^\pm$ and demanding regularity at the horizon, we find this implies
\begin{align}
    & \lim_{\xi \to 1} (1-\xi^4) \left[ - 2 \left[ \frac{1 + \xi^4}{\xi(1-\xi^4)} \pm \frac{i \Omega \xi^3}{1-\xi^4} \right] \frac{\partial F^\pm_\omega}{\partial \xi} + \left[ \mp \frac{i \Omega \xi^2}{1-\xi^4} + \frac{\Omega^2 (1 - \xi^6) }{(1-\xi^4)^2} \right]  F^\pm_\omega \right] = 0 \nonumber \\
    & \implies \frac{1}{F^\pm_\omega(\xi=1)} \frac{\partial F^\pm_\omega(\xi=1)}{\partial \xi} = \mp \frac{i \Omega}{4} \frac{1 \pm \frac{3 i \Omega}{2} }{1 \pm \frac{i \Omega}{2}} \label{eq:F-horizonder} \, ,
\end{align}
where the last condition fully determines the mode solution, up to an overall normalization. This condition allows one to find numerical solutions to Eq.~\eqref{eq:F-thermal} ensuring regularity at the horizon.

The other input required to determine the correlation function is the boundary conditions, i.e., the prescription to select the appropriate linear combination of the mode functions that determines the response kernel. Because we have extended our contour to infinity by taking the limit $\T \to \infty$, the boundary conditions are determined by the time-ordering prescription $\omega \to \omega (1 + i\epsilon)$, which is a consequence of the aspects discussed in Section~\ref{sec:ends-matching}\@. For concreteness, let us focus on the case $\omega > 0$\@. This is without loss of generality, because we are calculating a time-ordered correlation, and so, the full result will be immediately obtained by taking $\omega \to |\omega|$\@.

With these preliminaries, we can now analyze the mode functions~\eqref{eq:mode-functions-yF} under an infinitesimal complex rotation $\omega \to \omega (1 + i\epsilon)$\@. To select or discard a solution, we need to know whether one of the modes generates a divergent limit in the action~\eqref{eq:S-fluct-thermal-on-shell}\@. In particular, whether the limit
\begin{align} \label{eq:limit-onshell-thermal}
     \lim_{z \to (\pi T)^{-1} } \frac{f y'(t, z ) y(t, z )}{z^2}
\end{align}
exists for the mode solutions $y^{\pm}_\omega$ when the $i\epsilon$ prescription is taken into account. The reason why this particular limit is relevant is because of its explicit appearance in the expression for the on-shell action~\eqref{eq:S-fluct-thermal-on-shell}, which must be finite for the action to be at a well-defined extremum.

As discussed before, $F_\omega^\pm$ is regular and finite at the horizon $\xi=1$, and by inspecting the differential equation~\eqref{eq:F-thermal} that defines it, it is also analytic in $\omega$\@. As such, no singularity will appear in $F_\omega^\pm(\xi=1)$ by rotating the frequency $\omega$ by a small amount from the real axis into the complex plane. Therefore, the deformation by $i\epsilon$ affects the result predominantly through the WKB factor $\exp( \pm \frac{i\Omega - \Omega \epsilon}{4} \ln (1 - \xi^4) ) $\@. However, this means that $y^+_\omega$ will grow as $e^{\epsilon \Omega |\ln (1-\xi^4) |/4 }$ close to the horizon for $\omega > 0$, and therefore, substituting the mode function $y_\omega^+$ into Eq.~\eqref{eq:limit-onshell-thermal} leads to a divergent limit. We then conclude that we must keep only $y^-_\omega$ as the allowed mode solution for $\omega > 0$\@. By extension, the mode solution to be kept at arbitrary $\omega$ is $y^-_{|\omega|}$\@.

Now we can evaluate the response function $K_\perp^d(\omega)$ by means of substituting $y_\omega^-$ into Eq.~\eqref{eq:K-perp-d}\@. Given what we just showed, the WKB factor of $y_\omega^-$ goes like $\exp( - \frac{i\Omega - \Omega \epsilon}{4} \ln (1 - \xi^4) ) $, which goes to zero for $\omega >0$ as $\xi \to 1$\@. Consequently, the first term of the on-shell action~\eqref{eq:S-fluct-thermal-on-shell} vanishes, and we are left with the second term only. By direct inspection of the mode equation~\eqref{eq:F-thermal}, we see that regularity of $F^\pm_\omega$ requires
\begin{align}
    \frac{\partial F^\pm_\omega}{\partial \xi}(\xi=0) = 0 \, , & & \frac{\partial^2 F^\pm_\omega}{\partial \xi^2}(\xi=0) = \Omega^2 F^\pm_\omega(\xi=0) \, , \label{eq:boundary-behavior-F}
\end{align}
which verifies our earlier claim that $y'(t,z=0) = 0$\@.

The other claim of the previous section that we have yet to verify is that the (divergent) contact terms are the same as in the heavy quark interaction potential case. To show this, we may write the unregularized response kernel ${K}_{0\perp}^d$ from the second term in Eq.~\eqref{eq:S-fluct-thermal-on-shell}, and find
\begin{align}
    {K}_{0\perp}^d(\omega) &= - \frac{1}{y^-_{|\omega|}(z=0)} \lim_{z\to 0} \frac{2}{z^2} \frac{\partial y^-_{|\omega|}}{\partial z} = - \frac{1}{y^-_{|\omega|}(z=0)} \lim_{z\to 0} \frac{1}{z} \frac{\partial^2 y^-_{|\omega|}}{\partial z^2}\,,
\end{align}
where we have used L'Hopital's rule to obtain the second equality. By virtue of the second condition in Eq.~\eqref{eq:boundary-behavior-F}, $\frac{\partial^2 y^-_\omega}{\partial z^2}(z=0) = \omega^2 y^-_\omega(z=0)$, we can add and subtract the divergent part to get the unregularized response kernel as
\begin{align}
    {K}_{0\perp}^d(\omega) &=  - \frac{1}{y^-_{|\omega|}(z=0)} \lim_{z\to 0} \frac{\partial^3 y^-_{|\omega|}}{\partial z^3} - \omega^2 \lim_{z\to 0} \frac{1}{z} \, . \label{eq:K-thermal-unreg}
\end{align}
By comparison with Eq.~\eqref{eq:Delta-yy-LL-c}, and keeping in mind that the $z_m^2/\sqrt{f_m}$ in that expression is cancelled by the relative prefactor in the definition of $\Delta_{ij}^c$, it is clear that the nature of the last term in Eq.~\eqref{eq:K-thermal-unreg} is that of a contact term. 

Therefore, switching back to $\xi = \pi T z$, we may write the regularized response kernel (which is the one that enters the chromoelectric field correlator) as
\begin{equation} \label{eq:K-perp-d-sol}
    K_\perp^d(\omega) = -  \frac{(\pi T)^3}{y^-_{|\omega|}(\xi=0)} \frac{\partial^3 y^-_{|\omega|}}{\partial \xi^3}(\xi=0) = -  \frac{(\pi T)^3}{F^-_{|\omega|}(0)} \frac{\partial^3 F^-_{|\omega|}}{\partial \xi^3}(0) \, .
\end{equation}
The last equality follows from direct inspection of the mode functions shown in Eq.~\eqref{eq:mode-functions-yF}, and the fact that the prefactor $(1-\xi^4)^{- i\Omega/4}$ has no effect in the result because all of its first three derivatives with respect to $\xi$ vanish.

In terms of $\Delta_{ij}^d$, the result is
\begin{align}
    \Delta_{ij}^d(\omega) = - \delta_{ij} \frac{\sqrt{\lambda} \pi^2 T^3}{{ 2} F^-_{|\omega|}(0)} \frac{\partial^3 F^-_{|\omega|}}{\partial \xi^3}(0) \, .
\end{align}
This result may be evaluated numerically after plugging the boundary condition shown in Eq.~\eqref{eq:F-horizonder} to the differential equation~\eqref{eq:F-thermal}, which defines $F_\omega^-$\@.\footnote{Contrary to what one might hope, the third derivative with respect to $\xi$ may not be evaluated directly from the differential equation~\eqref{eq:F-thermal}\@. If one takes another derivative and $\xi \to 0$, one ends up with an identity.} We plot the result for a general temperature in the next section, where we give the final result of our calculation.

\begin{figure}
    \centering
    \includegraphics[width=\textwidth]{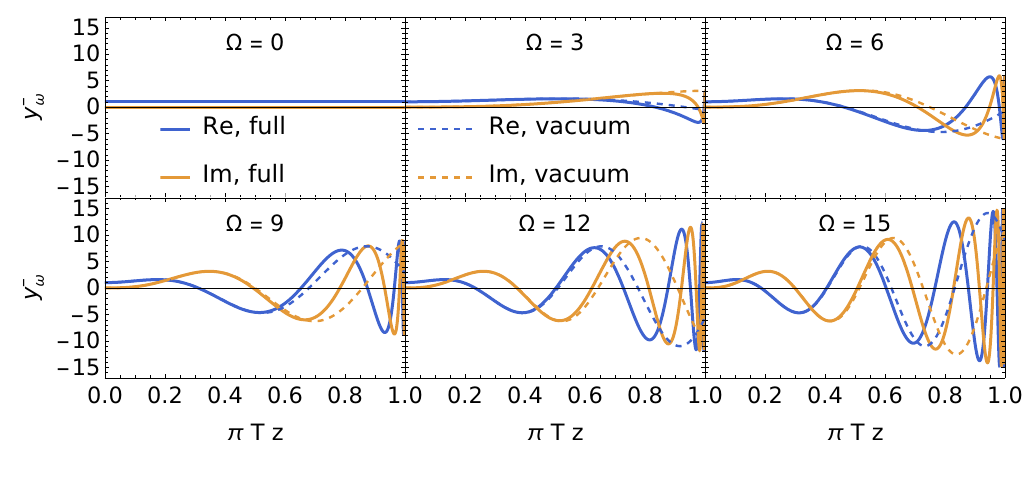}
    \caption{Solid lines: real (blue) and imaginary (orange) parts of the mode solution $y_\omega^{-}$ at selected values of $\Omega = \omega / (\pi T)$\@. Dashed lines: real (blue) and imaginary (orange) parts of the mode solution $y_\omega^{-}$ for $T = 0$. The arguments of the vacuum solutions are rescaled by $(\pi T)^{-1}$, which is the position of the event horizon in black hole AdS, to allow for a clean visual comparison at the same physical value of the AdS${}_5$ radial coordinate $z$\@.}
    \label{fig:mode-functions}
\end{figure}

Before proceeding further, it is also instructive to evaluate the zero temperature limit of our expression. To do so, it is most convenient to go back to the original AdS coordinate $z$ in the mode equation for $y_\omega$\@. Then, setting $T=0$, Eq.~\eqref{eq:eom-y-th} becomes
\begin{align}
    \frac{\partial^2 y_\omega}{\partial z^2} - \frac{2}{z} \frac{\partial y_\omega}{\partial z} + \omega^2 y_\omega = 0
\end{align}
and the solutions are relatively simpler:
\begin{align}
    y_\omega^\pm(z) = \left( 1 \pm i \omega z \right) e^{ \mp i \omega z} \, .
\end{align}
We choose the sign labelling so that the solutions match their finite-temperature counterparts in the limit $T\to 0$\@. For visual clarity, we show the mode solutions as a function of $z$ for a range of frequencies rescaled by the temperature in Fig.~\ref{fig:mode-functions}\@. The fact that we have an explicit expression then allows us to evaluate Eq.~\eqref{eq:K-perp-d-sol} explicitly, obtaining
\begin{equation}
    K_\perp^d(\omega)_{T=0} = - 2 i |\omega|^3 \, .
\end{equation}
We note that the cubic power in the frequency is exactly what one expects from dimensional analysis of the correlation function we are interested in from the field theory perspective. In terms of $\Delta_{ij}^d$, we have
\begin{equation}
    \Delta_{ij}^d(\omega)_{T = 0} = - \delta_{ij} \frac{i \sqrt{\lambda}} {\pi} |\omega|^3 \, .
\end{equation}

In summary, we have obtained the response kernel $K_\perp^d$ that determines the on-shell Nambu-Goto action up to second order in the contour deformations of the Wilson loop expectation value that is dual to it. Contrary to the results we found in Section~\ref{sec:HQ-setup}, these are well-behaved and so provide a quantitative description of the dynamics of in-medium quarkonium. From our discussion in Section~\ref{sec:nhat}, the background configuration is also well-founded. Therefore, we conclude that this is the $\mathcal{N}=4$ observable that most closely resembles the analogous QCD correlation function, and will use it as the $\mathcal{N}=4$ result for the quarkonium transport coefficients.\footnote{Insofar as $\mathcal{N}=4$ SYM is a different theory than QCD, analogous results are, in the end, all we can get.} We give the expressions and plots for the chromoelectric field correlator that we extract from this holographic calculation in the next section, where we also discuss its implications as a baseline for phenomenological applications.

\section{Results and applications} \label{sec:results-applications}

\subsection{Main result} \label{sec:result}

Since we have calculated $\Delta_{ij}^{c,d}(\omega,L)$ for each configuration, we are ready to conclude and give the expression for the analog object in $\mathcal{N}=4$ SYM to the time-ordered correlator of chromoelectric fields dressed by Wilson lines in QCD\@.

Inspecting the behavior of $\Delta_{ij}^{c}(\omega,L)$ as $L \to 0$, we conclude we must discard the configuration with constant $\hat{n}$ on the two timelike segments of the Wilson line on the AdS boundary, as it does not have a sensible limit and does not satisfy $\langle W_{\rm BPS}[\ml{C}]\rangle = 1$, which is required to give the Wilson line configuration between the two non-Abelian electric fields the interpretation of an adjoint Wilson line. On the other hand, the configuration where the two timelike Wilson lines have antipodal positions on the $S_5$ does provide a sensible answer with perturbations that modify the energy of the worldsheet by a finite amount, and moreover, it fulfils all of the expectations that we require for the standard QCD Wilson loop. This is further substantiated by our discussion in Subsection~\ref{sec:nhat}, where we advanced that the appropriate description for quarkonium transport in medium is given by the Wilson loop where $\hat{n}$ takes antipodal positions on the $S_5$ for the two timelike segments. As such, we use the latter one and obtain:
\begin{align}
     \frac{g^2}{N_c} [g_E^{\T}]^{\mathcal{N}=4}_{ij}(\omega) =  \delta_{ij}  \frac{ (\pi T)^3 \sqrt{\lambda}}{{ 4}\pi} \left( \frac{-i}{F^-_{|\omega|}(0)} \frac{\partial^3 F^-_{|\omega|}}{\partial \xi^3}(0) \right) \, .
\end{align}

\begin{figure}
    \centering
    \includegraphics[width=0.49\textwidth]{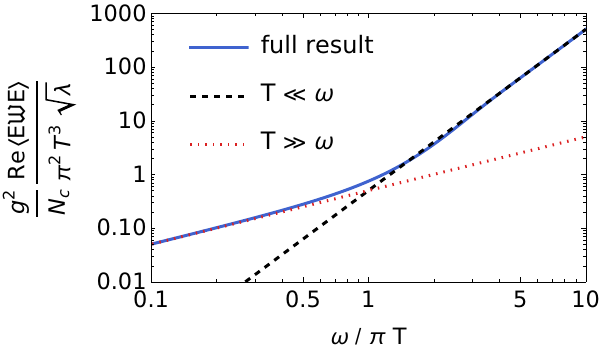}
    \includegraphics[width=0.49\textwidth]{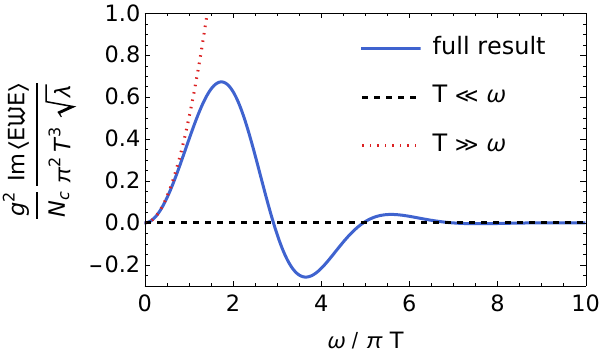}
    \caption{Real part (left) and imaginary part (right) of the non-Abelian electric field correlator of interest. The finite temperature result is shown in solid lines, and the zero temperature limit is shown in black dashed lines. The leading low-frequency limit is shown in red dotted lines. As before, the arguments of the functions at zero temperature have been rescaled by $\pi T$ to have a clean visual comparison.}
    \label{fig:correlator}
\end{figure}

This is the main result of this paper. We plot this in Fig.~\ref{fig:correlator}\@. Furthermore, its zero-temperature limit is given by
\begin{align}
     \frac{g^2}{N_c} [g_E^{\T}]^{\mathcal{N}=4}_{ij}(\omega)_{T = 0} = \delta_{ij}\frac{ \sqrt{\lambda}}{{ 2}\pi} |\omega|^3  \, .
\end{align}

The other limit of interest is the low-frequency limit, which can be extracted analytically by solving the mode equation~\eqref{eq:F-thermal} up to linear order in $\Omega$. The algebraic steps necessary to do this are the same as those in the heavy quark diffusion coefficient calculation~\cite{Casalderrey-Solana:2006fio}\@. The result is
\begin{align}
     \frac{g^2}{N_c} [g_E^{\T}]^{\mathcal{N}=4}_{ij}(\omega) = \delta_{ij} \frac{ \sqrt{\lambda} (\pi T)^3 }{{ 2}\pi} \left[ \frac{|\omega|}{\pi T}  + i \frac{\omega^2}{(\pi T)^2} + O \! \left( \left(\frac{\omega}{\pi T}\right)^3 \right) \right] \, ,
\end{align}
where we have kept one higher order in $\omega/T$ than that explicitly shown in Ref.~\cite{Casalderrey-Solana:2006fio}\@. The details of how this expansion was carried out can be found in Appendix~\ref{sec:App-omega-expansion}\@.

With the expression for the correlation function in hand, we can now use it to describe how a heavy quark-antiquark pair will propagate through the thermal $\mathcal{N}=4$ SYM plasma. Specifically, we can calculate the spectral function that determines the transition rates of in-medium quarkonium within a potential non-relativistic EFT description as shown in Section~\ref{sec:Quarkonia}, and draw the phenomenological implications thereof for strongly coupled plasmas. This is what we will first show in the next section. As an additional application of our result, we will compare our result to its weak-coupling limit in $\mathcal{N}=4$ SYM, and lay out the case for computing subleading corrections, both at weak and strong coupling.

\subsection{Applications}
\label{sec:applications}

Having calculated the non-Abelian electric correlator of interest, we can now discuss the physical consequences and applications of our result. We shall focus on two pieces of theoretical understanding we can gain from here. First, we will discuss the immediate consequences of our result for in-medium quarkonium dynamics in a strongly coupled $\mathcal{N}=4$ supersymmetric Yang-Mills plasma. After that, we will discuss how a comparison with the weak coupling expansion of the same correlator may be used together with our strong coupling result to advance our understanding of the always elusive non-perturbative dynamics of gauge theories at intermediate coupling.

\subsubsection{Evaluation of chromoelectric field spectral function}

\begin{figure}
    \centering
    \includegraphics[width=0.8\textwidth]{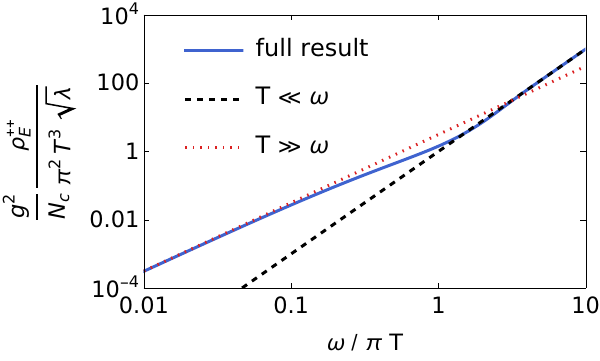}
    \caption{Spectral function for quarkonium transport. Only the positive frequency domain is shown, as $\rho_E^{++}$ vanishes for $\omega < 0$\@.}
    \label{fig:Spectral}
\end{figure}

We are now ready to evaluate the spectral function that encodes important information of the plasma relevant for quarkonium transport, by using the relations introduced in Section~\ref{sec:Quarkonia}, as dictated by our $\mathcal{N}=4$ SYM result. Specifically, we use
\begin{equation}
    [g_{E}^{++}]^>(\omega) = {\rm Re} \left\{ [g_E^{{\mathcal{T}}}]^{\mathcal{N}=4}(\omega) \right\} + \frac{1}{\pi} \int_{-\infty}^\infty d p_0 \, \mathcal{P} \left( \frac{1}{p_0} \right) {\rm Im} \left\{ [g_E^{{\mathcal{T}}}]^{\mathcal{N}=4}(\omega + p_0) \right\} \, .
\end{equation}
In general, the above equation would be an integral expression that can only be evaluated numerically. However, in $\mathcal{N}=4$ Yang-Mills theory, the fact that one obtains $[g_E^{{\mathcal{T}}}]^{\mathcal{N}=4}(\omega)$ by selecting modes using the time-ordering prescription $\omega \to \omega(1 + i\epsilon)$ gives us more analytic control. As we show in Appendix~\ref{sec:App-analytic-rot}, this property allows us to prove that
\begin{equation} \label{eq:analytic-prop-gE}
    \frac{1}{\pi} \int_{-\infty}^\infty d p_0 \, \mathcal{P} \left( \frac{1}{p_0} \right) {\rm Im} \left\{ [g_E^{{\mathcal{T}}}]^{\mathcal{N}=4}(\omega + p_0) \right\} = {\rm sgn}(\omega) {\rm Re} \left\{ [g_E^{{\mathcal{T}}}]^{\mathcal{N}=4}(\omega) \right\} \, ,
\end{equation}
and consequently, the correlation function that enters the quantum and classical quarkonium time evolution equations is given by
\begin{equation}
\label{eqn:gE++}
    [g_{E}^{++}]^>(\omega) = 2 \theta(\omega) {\rm Re} \left\{ [g_E^{{\mathcal{T}}}]^{\mathcal{N}=4}(\omega) \right\} \, .
\end{equation}
One more step gives us an explicit expression for the spectral function
\begin{equation}
    \rho_E^{++}(\omega) = 2 \theta(\omega) \big(1 - e^{- \omega/T} \big) {\rm Re} \left\{ [g_E^{{\mathcal{T}}}]^{\mathcal{N}=4}(\omega) \right\} \, ,
\end{equation}
which we plot in Fig.~\ref{fig:Spectral}\@. This function is manifestly neither even nor odd, as expected from the evidence coming from the perturbative calculations~\cite{Binder:2021otw,Scheihing-Hitschfeld:2022xqx}\@. 

One immediate implication of our results, which may already be seen from Fig.~\ref{fig:correlator} is that the transport coefficients introduced in the Quantum Brownian motion limit~\cite{Brambilla:2016wgg,Brambilla:2017zei,Eller:2019spw,Brambilla:2020qwo,Brambilla:2021wkt} of the open quantum system approach to in-medium quarkonium, namely, the analogs to
\begin{align}
\kappa_{\rm adj} &= \frac{T_F g^2}{3 N_c} {\rm Re} \int dt\, \big\langle \hat{\ml{T} }E^a_i(t) \mathcal{W}^{ab}(t,0) E^b_i(0) \big\rangle_T \\
\gamma_{\rm adj} &= \frac{T_F g^2}{3N_c} {\rm Im} \int dt\, \big\langle \hat{\ml{T}} E^a_i(t) \mathcal{W}^{ab}(t,0) E^b_i(0) \big\rangle_T \,,
\end{align}
vanish in strongly coupled $\mathcal{N}=4$ supersymmetric Yang-Mills theory. Explicitly,
\begin{equation}
    \kappa_{\rm adj}^{\mathcal{N}=4 } = \gamma_{\rm adj}^{\mathcal{N}=4 } = 0 \, .
\end{equation}
The quantity $\gamma_{\rm adj}$ represents the mass shift of quarkonium states inside a plasma. The result $\gamma_{\rm adj}^{\mathcal{N}=4 } = 0$ is consistent with a recent lattice QCD study~\cite{Bala:2021fkm}\@.

In the quantum optical limit where the quarkonium time evolution can be effectively described by a Boltzmann equation as shown in Eq.~\eqref{eqn:rate}, it is the finite frequency part of the chromoelectric field correlator that enters the quarkonium dissociation and recombination rates. Because the argument of $[g_E^{++}]^>$ is negative in Eq.~\eqref{eqn:disso}, our result in Eq.~\eqref{eqn:gE++} indicates that the dissociation rate of a small-size quarkonium state in a strongly coupled QGP vanishes. Using Eq.~\eqref{eqn:kms}, we also see that the recombination rate in Eq.~\eqref{eqn:reco} also vanishes in the same limit.

\subsubsection{Towards intermediate couplings in \texorpdfstring{$\mathcal{N}=4$} {} SYM}

A natural question we can ask is how the calculation result in the strong coupling limit compares with that in the weak coupling limit. We also want to understand if they allow for an interpolation at intermediate couplings.

At weak coupling, the non-Abelian electric field correlation function we have set out to calculate can be evaluated directly using the standard real-time perturbation theory in the Schwinger-Keldysh formalism. In the large $N_c$ limit of $\mathcal{N}=4$ SYM, it reads
\begin{equation}
    \frac{g^2}{N_c} [g_E^{\T}]_{ij}(\omega) = \delta_{ij} \frac{\lambda}{6\pi} |\omega|^3 \coth \! \left( \frac{|\omega|}{2T} \right) \, .
\end{equation}
It is apparent that the small $\omega/T$ limit of the strong coupling and weak coupling results is different: in the weakly coupled case it goes as $|\omega|^2$, while for the strongly coupled limit it is linear in $|\omega|$\@. That is to say, for the range of frequencies that is sensitive to thermal effects, the physics at weak and strong coupling is different. Note that the $|\omega|^2$ behavior displayed above implies that the transport coefficients $\kappa_{\rm adj}^{\ml{N}=4}$ and $\gamma_{\rm adj}^{\ml{N}=4}$ vanish at leading order in perturbation theory. At NLO, however, $\kappa_{\rm adj}^{\mathcal{N}=4}$ is known to be nonzero~\cite{Caron-Huot:2008dyw}, and is actually equal to $\kappa_{\rm fund}^{\ml{N}=4}$ up to this order in perturbation theory. Therefore, it seems that the most interesting thermal physics lies in the intermediate coupling regime. However, a first approximation to this regime via interpolation between weakly and strongly coupled results would require to calculate the first nonvanishing contributions from both sides, which is a challenging computation we do not undertake in this work.

\begin{figure}
    \centering
    \includegraphics[width=0.8\textwidth]{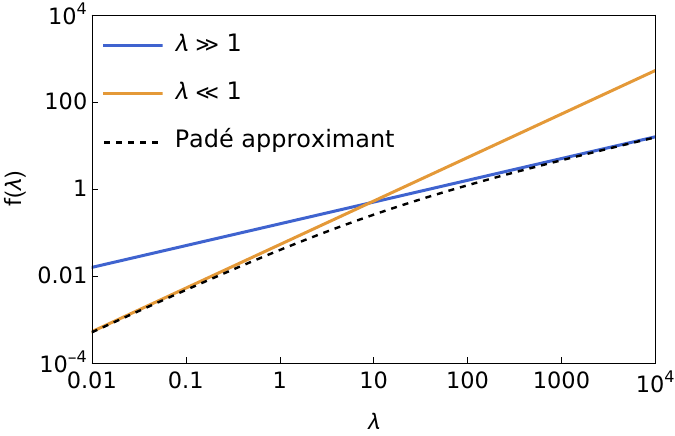}
    \caption{Coupling dependence of the time-ordered chromoelectric correlator in vacuum $T=0$. The solid lines depict the information currently available at weak and strong coupling, and the dashed line is the lowest order Pad\'e approximant consistent with both asymptotic behaviors.}
    \label{fig:interpolation-SYM}
\end{figure}

On the other hand, in the $T=0$ limit, the frequency dependence of both results agrees: both are proportional to $|\omega|^3$\@. This is unsurprising given that $\mathcal{N}=4$ supersymmetric Yang-Mills theory is a conformal field theory, and there is thus no other scale available to give rise to a different behavior. As such, in vacuum we have 
\begin{equation}
    \frac{g^2}{N_c} [g_E^{\T}]_{ij}(\omega) = \delta_{ij} f(\lambda) |\omega|^3 \, ,
\end{equation}
where
\begin{equation}
    f(\lambda) \approx \begin{cases} \dfrac{\lambda}{6\pi} & \lambda \ll 1 \\ & \\  \dfrac{\sqrt{\lambda}}{{ 2}\pi} & \lambda \gg 1 \end{cases} \,\,\, .
\end{equation}
We plot both limits in Fig.~\ref{fig:interpolation-SYM}, together with the Pad\'e approximant of order $[2/1]$ in $\sqrt{\lambda}$ that interpolates between the two limits. Such an interpolation constitutes, at most, an educated guess of the result for the chromoelectric correlation function in the intermediate coupling regime around $\lambda \sim \ml{O}(10)$\@. As it should be clear from the comparison, we only expect the result to be valid asymptotically. 

Nonetheless, such a comparison may provide valuable insight into what the behavior of the correlator is at intermediate couplings. In fact, Fig.~\ref{fig:interpolation-SYM} is just the first step towards a more complete understanding of the intermediate coupling regime, as the tools to make progress on either limiting case are already available. At weak coupling, what is required is a next-leading order calculation analogous to what has already been done for QCD in Ref.~\cite{Binder:2021otw}, but this time for $\mathcal{N}=4$ SYM\@. At strong coupling, one would have to evaluate the quantum corrections to the string worldsheet action in the so-called semiclassical expansion, which is tantamount to a 1-loop calculation of the fluctuation fields on the worldsheet~\cite{Drukker:2000ep}\@. Both are necessary steps towards a more complete understanding of the correlator, which are along the path that we want to follow in the future (in the hope that the convergence of the series is comparable to that of other observables in $\mathcal{N}=4$ SYM, e.g., the thermodynamic pressure~\cite{Du:2021jai})\@.

\section{Conclusions}
\label{sec:conclusions}

In this paper we studied the gauge-invariant correlation function of two chromoelectric fields connected via an adjoint Wilson line at finite temperature. In QCD, this correlation function determines the in-medium dynamics of small-size quarkonium states. However, in light of the tools provided by the gauge/gravity duality at strong coupling, and of the impressive success that holography has had in describing the properties of the quark-gluon plasma, we studied this correlation function in $\mathcal{N}=4$ supersymmetric Yang-Mills theory. 
By expressing the correlation of two non-Abelian electric fields at different times connected via Wilson lines as a variation of a fundamental Wilson loop, we were able to calculate it by evaluating the variations of the holographic dual description of the Wilson loop, which is given by an extremal surface in ${\rm AdS}_5 \times S_5$\@.

We examined two candidates for the $\mathcal{N}=4$ Yang-Mills description of the relevant Wilson loop using the AdS/CFT correspondence. We considered two worldsheet configurations: one with a constant $S_5$ coordinate on the two timelike segments of the Wilson loop on the AdS boundary and the other with antipodal $S_5$ coordinates. In our calculations, we first obtained the background worldsheet for each configuration, which is an extremal surface that hangs from the Wilson loop contour on the AdS boundary. Then we introduced deformations of the Wilson loop contour and studied how the perturbations propagate on the worldsheet. Finally we obtained the correlator of our interest by calculating the derivative response function on one side of the timelike segments of the Wilson loop caused by the deformation on the other.

We showed that the non-Abelian electric field correlator obtained from the configuration with a constant $S_5$ coordinate diverges and does not give a sensible result for the correlator of our interest. This is related to the fact that this configuration does not satisfy our expectation about the pure gauge Wilson loop $\langle W[\ml{C}] \rangle = 1$, where $\ml{C}$ denotes a Wilson loop consisting of two antiparallel timelike Wilson lines on top of each other.
On the other hand, the configuration with antipodal $S_5$ coordinates is well-motivated from a dynamical perspective, as it describes the propagation of two heavy probes with conjugate (opposite) charges. Furthermore, it is consistent with the properties required for a dual description of the pure gauge Wilson loop according to the prescriptions that have been proposed for this type of loop, and leads to a finite and sensible result for the correlator. Following these observations, we conjectured that the pure gauge adjoint Wilson line in the strong coupling limit of $\mathcal{N}=4$ SYM may be described by two antiparallel 1/2 BPS Wilson lines with antipodal values of the $S_5$ coordinate. Further verification of this proposal, such as evaluating this configuration for a finite temporal extent (as opposed to the limit we considered in this paper, where we let $\T \to \infty$), is an interesting endeavor that we hope to undertake in the near future.

Our results indicate that the two transport coefficients in the Lindblad equation for quarkonium in-medium dynamics in the quantum Brownian motion limit, as defined through the chromoelectric field correlator we studied, vanish in a strongly coupled $\ml{N}=4$ SYM plasma. Furthermore, the correlation functions $[g_E^{\pm\pm}]^>$ that determine quarkonium dissociation and recombination in the Boltzmann equation that is valid in the quantum optical limit also vanishes in a strongly coupled $\ml{N}=4$ SYM plasma. This means the in-medium dynamics of small-size quarkonium states is trivial at leading order in the multipole expansion, which is used to simplify the evolution equations in both the quantum Brownian motion and quantum optical limits. The reason behind this triviality can be multi-fold. Firstly, in the strongly coupled, $N_c \to \infty$ limit of $\ml{N}=4$ SYM theory, the energy gap between a heavy quark-antiquark pair in the color singlet and an octet pair, as extracted from the expectation value of a rectangular Wilson loop, is infinitely large. This means that it would take an infinite amount of energy to break the bound color singlet pair, which a QGP at finite temperature does not have. The inverse process, i.e., recombination, cannot happen either, because the process would release an infinite amount of energy, which cannot be absorbed by the local degrees of freedom of a QGP at finite constant temperature. Secondly, the validity of both the quantum Brownian motion and quantum optical limits may break down when the plasma is strongly coupled, which should be further investigated in the future. Given the fact that some properties of the QGP at low temperatures can be well described by a strongly coupled $\ml{N} = 4$ SYM plasma, going beyond these limits might prove of paramount relevance for phenomenology.

Furthermore, it is unclear to what extent the vanishing result may be a consequence of the leading description in $\sqrt{\lambda}$\@. This motivates going beyond the leading strong coupling limit in $\ml{N}=4$ SYM by accounting for effects of quantum fluctuations on the worldsheet on the gravitational side of the duality. 
It also motivates calculating the subleading terms in a weak-coupling perturbative expansion in order to have an improved Pad\'e approximation of the complete correlation function, both for the large frequency regime as well as for the low frequency regime. Once calculated, these results would provide insight into the intermediate coupling regime, and therefore improve our description of quarkonium dynamics by setting the coupling at phenomenologically relevant values.

On the QCD side, our results further advance the need for a non-perturbative determination of the chromoelectric correlator that determines in-medium quarkonium dynamics. Current and past estimates of the transport coefficients rely on spectral function reconstructions determined either from the correlator that describes heavy quark diffusion or from the heavy quark interaction potential dynamics. Neither provides direct access to the correlation function we have discussed in this paper. Given that the results for the heavy quark diffusion coefficient $\kappa$ and the quarkonium transport coefficient $\kappa_{\rm adj}$ differ in $\ml{N}=4$ SYM, the expectation is that this will also be the case in QCD\@. Therefore, a non-perturbative calculation of the quarkonium transport coefficients has the potential to make a significant contribution to phenomenological studies of in-medium quarkonium.

Finally, from a data-driven perspective, one should try to extract this correlation from phenomenological studies and experimental measurements by applying Bayesian analysis techniques, where one uses some ansatz for the correlation that is well-motivated from both weak-coupling and strong-coupling studies, varies the parameters in the ansatz and compares the calculation results of quarkonium nuclear modification factors and elliptic flow coefficients with experimental data. The Bayesian analysis is then applied to systematically find the best set of parameters in describing the data and estimate the parameters' uncertainties. 
For example, it will be particularly instructive to see how data from 200 GeV Au-Au collisions, particularly from the sPHENIX program, can be used to better constrain the finite frequency dependence of the correlator via a Bayesian analysis.
We expect data coming from these collision energies will be sensitive to its finite frequency dependence, because the prevailing temperature regime in this experiment is of low temperature comparable to
the energy level splittings of quarkonia states.
All these studies will deepen our understanding of quarkonium in-medium dynamics and the relevant transport properties of the QGP\@. 

\acknowledgments

We are grateful for useful comments from Krishna Rajagopal. We are particularly grateful for invaluable discussions with Hong Liu. This work is supported by the U.S.~Department of Energy, Office of Science, Office of Nuclear Physics under grant Contract Number DE-SC0011090\@. XY is also supported in part by the U.S.~Department of Energy, Office of Science, Office of Nuclear Physics, Inqubator for Quantum Simulation (IQuS) under Award Number DOE (NP) Award DE-SC0020970\@.

\newpage

\appendix

\section{KMS relations of chromoelectric correlators} \label{sec:App-KMS}

In this Appendix we will verify the KMS relation between $[g_{E}^{++}]^>(t)$ and $[g_{E}^{++}]^<(t)$, as introduced in the main text in Eqs.~\eqref{eq:gE++>} and~\eqref{eq:gE++<}\@. For the proof of how the time-reversal symmetry relates $[g_E^{++}]$ and $[g_E^{--}]$, we refer the reader to Appendix A of~\cite{Binder:2021otw}\@. 

The following proof adds an omission that two of us made in  Appendix A of~\cite{Binder:2021otw}, where we did not pay enough attention to the effects of the adjoint color charge on the thermal average of the medium. We hope that the subsequent discussion makes these aspects more transparent.

We begin by noting that an adjoint Wilson line in the interaction picture can be written in terms of time-evolution operators as
\begin{equation}
    \mathcal{W}^{ab}_{[t_f,t_i]} = e^{i H t_f} \left[e^{- i (H - g A_0^c(0) [T_{\rm Adj}^c] ) (t_f - t_i) }\right]^{ab} e^{-i H t_i} \, , \label{eq:W-as-T-evol}
\end{equation}
where $H$ is the QGP Hamiltonian, and one may interpret $ H \delta^{ab} - g A_0^c(0) [T_{\rm Adj}^c]^{ab} $ as the total Hamiltonian when there is a point color charge in the adjoint representation at the position ${\bs x} = 0$\@.

With this, the Wightman correlator $[g_{E}^{++}]^>(t)$ can be written as
\begin{align}
    [g_{E}^{++}]^>(t) &= \frac{1}{Z} {\rm Tr}_{\mathcal{H}} \left[ E_i^a(t) \W^{ac}(t,+\infty) \W^{cb}(+\infty,0) E_i^b(0) e^{-\beta H} \right] \nonumber \\
    &= \frac{1}{Z} {\rm Tr}_{\mathcal{H}} \left[ e^{i H t} E_i^a(0) \left[e^{- i (H - g A_0^c(0) [T_{\rm Adj}^c] ) t}\right]^{ab} E_i^b(0) e^{-\beta H} \right] \, .
\end{align}
The KMS conjugate of this correlator is obtained by shifting $t \to t -i\beta$\@. We therefore obtain
\begin{align}
    [g_{E}^{++}]^<(t) &= [g_{E}^{++}]^>(t - i\beta) \nonumber \\
    &= \frac{1}{Z} {\rm Tr}_{\mathcal{H}} \left[ e^{i H t}  E_i^a(0) \left[e^{- i (H - g A_0^c(0) [T_{\rm Adj}^c] ) (t - i\beta) }\right]^{ab} E_i^b(0) \right] \, . \label{eq:KMS-proof-step}
\end{align}
We can explicitly see in this expression that the thermal ensemble is now determined by the total Hamiltonian $ H \delta^{ab} - g A_0^c(0) [T_{\rm Adj}^c]^{ab} $ instead of just $H$\@.

The equivalence with Eq.~\eqref{eq:gE++<} is then verified by using that
\begin{equation}
    \left[e^{- i (H - g A_0^c(0) [T_{\rm Adj}^c] ) (t - i\beta) }\right]^{ab} = e^{- i H t} \mathcal{W}^{ac}_{[t, t_f]} e^{-\beta H} \mathcal{W}^{cd}_{[t_f - i\beta, t_f]} \mathcal{W}^{db}_{[t_f, 0]} \, ,
\end{equation}
which follows from using Eq.~\eqref{eq:W-as-T-evol} repeatedly. It is interesting to note that this equation holds for any value of $t_f$\@. Plugging this expression into Eq.~\eqref{eq:KMS-proof-step} one obtains
\begin{equation}
    [g_{E}^{++}]^<(t) = \frac{1}{Z} {\rm Tr}_{\mathcal{H}} \left[   E_i^a(t) \mathcal{W}^{ac}_{[t, t_f]} e^{-\beta H} \mathcal{W}^{cd}_{[t_f - i\beta, t_f]} \mathcal{W}^{db}_{[t_f, 0]} E_i^b(0) \right] \, ,
\end{equation}
which is equivalent to Eq.~\eqref{eq:gE++<} as displayed in the main text when $t_f \to +\infty$\@.

The KMS relation between $[g_{E}^{--}]^>(t)$ and $[g_{E}^{--}]^<(t)$ then follows from using the one we just verified above and their relation to $[g_{E}^{++}]^>(t)$ and $[g_{E}^{++}]^<(t)$ through time-reversal respectively.

\section{Operator ordering aspects of Wilson loops} \label{sec:App-W-ordering}

\subsection{Time-ordered products of Wilson lines}

In this work, we deal with the calculation of the time-ordered correlator
\begin{align}
    [g_E^{{\mathcal{T}}}](t) = \langle \hat{\mathcal{T}} E_i^a(t) \mathcal{W}^{ab}_{[t,0]} E_j^b(0) \rangle_T \, ,
\end{align}
where $\mathcal{W}^{ab}_{[t,0]}$ is an adjoint Wilson line, that is written in terms of fundamental Wilson lines as
\begin{equation} \label{eq:app-adj-as-fund}
    \W^{ab}_{[t_2,t_1]} = \frac{1}{T_F} {\rm Tr}_{\rm color}  \left[ \hat{\T} \, T^a_F U_{[t_2,t_1]} T^b_F U_{[t_2,t_1]}^\dagger \right] \, ,
\end{equation}
where the time-ordering symbol is necessary to preserve the explicit ordering of operators in an adjoint Wilson line.

For concreteness, we write both adjoint and fundamental lines below:
\begin{align}
    \W^{ab}_{[t_2,t_1]} &= \left[ {P} \exp \left( i g \int_{t_1}^{t_2} dt\, A_0^c(t) T_{\rm Adj}^c \right) \right]^{ab} \, , \\
    U_{[t_2,t_1],ij} &= \left[ {P} \exp \left( i g \int_{t_1}^{t_2} dt\, A_0^c(t) T_{\rm Fund}^c \right) \right]_{ij} \, ,
\end{align}
where the difference is the representation of the SU($N_c$) generator matrices. $[T_{\rm Fund}^a]_{ij} \equiv [T_{ F}^a]_{ij}$ are the generators of the fundamental representation, normalized in the conventional way ${\rm Tr}[T_{F}^a T_{F}^b] = T_F \delta^{ab} $ with $T_F = 1/2$, and $[T_{\rm Adj}^a]^{bc} = -i f^{abc}$, where $f^{abc}$ are the structure constants of the group $[T^a , T^b] = i f^{abc} T^c$\@.

Note that the operator ordering in the adjoint Wilson line is not the ``natural'' one in terms of fundamental Wilson lines, because the operator products are not ordered in the same way as the matrix products that contract color indices. In this sense, Eq.~\eqref{eq:app-adj-as-fund} without the time-ordering symbol is only indicative of the color product structure, but is not an explicit expression in terms of how the gauge field operators $A_0^a(t)$ are ordered. This is specified by the symbol $\hat{\T}$, but by itself does not provide an explicit method to calculate it. One explicit way to evaluate it is given by its path integral representation, to which we will explain in a moment. Before doing that, however, it is useful to discuss how this time ordering of fundamental Wilson lines appears from the dynamics of two (coincident) point color charges, which we take to be in the fundamental and anti-fundamental representations of SU($N_c$)\@.

To accomplish this, let us collectively denote the ``colors'' of the $Q \bar{Q}$ pair by $\big( Q\bar{Q} \big)_{ij}$, with $i$ being the index of the quark in the fundamental representation, and $j$ the index of the anti-quark in the anti-fundamental representation. The dynamics are given by
\begin{align}
    \frac{d}{dt} \big( Q\bar{Q} \big)_{ij} &= \left[ i g A_0^a(t) \big[ T^a_{\rm Fund} \big]_{ii'} \delta_{jj'} + i g A_0^a(t) \big[ T^a_{\rm Anti-Fund} \big]_{jj'} \delta_{ii'} \right] \big( Q\bar{Q} \big)_{i'j'} \nonumber \\
    &= \left[ i g A_0^a(t) \big[ T^a_{F} \big]_{ii'} \delta_{j'j} - i g A_0^a(t) \big[ T^a_{F} \big]_{j'j} \delta_{ii'} \right] \big( Q\bar{Q} \big)_{i'j'} \, ,
\end{align}
where we have used that $ \big[T^a_{\rm Anti-Fund}\big]_{ij} = - [T^a_{\rm Fund}]_{ji}$\@.

Formally, we can write the solution to this equation as
\begin{equation}
    \big( Q\bar{Q} \big)_{ij} (t) = \mathcal{W}_{i i_0,j_0 j}(t) \big( Q\bar{Q} \big)_{i_0j_0}(t=0) \, ,
\end{equation}
where $\mathcal{W}_{i i_0,j_0 j}(t)$ obeys the same equation as $\big( Q\bar{Q} \big)_{ij}$
\begin{equation} \label{eq:App-def-W-eqn}
    \frac{d}{dt} \mathcal{W}_{i i_0,j_0 j} = \left[ i g A_0^a(t) \big[ T^a_{F} \big]_{ii'} \delta_{j'j} - i g A_0^a(t) \big[ T^a_{F} \big]_{j'j} \delta_{ii'} \right] \mathcal{W}_{i' i_0,j_0 j'} \, ,
\end{equation}
with $\mathcal{W}_{i i_0,j_0 j}(t=0) = \delta_{i i_0} \delta_{j j_0}$ as the initial condition. Note that, by construction, we have
\begin{equation}
    \mathcal{W}_{i i_0,j_0 j} = \hat{\T} \big( \big[U_{[t,0]}\big]_{i i_0} \big[ U_{[t,0]}^\dagger \big]_{j_0 j} \big) \, .
\end{equation}
A quick way to see this is to note that if $A_0^a$ were ordinary numbers, then the time ordering would be irrelevant and the Wilson lines in the last expression would be completely decoupled from each other. However, because $A_0^a(t)$ are in principle non-commuting operators, we have to keep track of the fact that $A_0^a(t)$ is always inserted to the left of operators $A_0^a(t')$ when $t > t'$\@. This is due to $A_0^a(t)$ being to the left of $\W$ in its defining differential equation~\eqref{eq:App-def-W-eqn}\@. Therefore, what we get out of $\mathcal{W}_{i i_0,j_0 j}$ is a fundamental and an anti-fundamental Wilson line put together, with their operators time-ordered.

A consistency check is to verify that we can get the adjoint Wilson line from $\mathcal{W}_{i i_0,j_0 j}$\@. Indeed, we can consider
\begin{align}
    \frac{1}{T_F} [T^a_F]_{j i} \mathcal{W}_{i i_0,j_0 j} [T^b_F]_{i_0 j_0} = \frac{1}{T_F} {\rm Tr}_{\rm color}  \left[ \hat{\T} \, T^a_F U_{[t_2,t_1]} T^b_F U_{[t_2,t_1]}^\dagger \right] \, ,
\end{align}
and get, after contracting the color indices and using the Lie Algebra of the group,
\begin{align}
    \frac{d}{dt} \big( [T^a_F]_{j i} \mathcal{W}_{i i_0,j_0 j} [T^b_F]_{i_0 j_0} \big) = i g A_0^c(t) \big( - i f^{cad} \big) \big( [T^d_F]_{j' i'} \mathcal{W}_{i' i_0,j_0 j'} [T^b_F]_{i_0 j_0} \big) \, ,
\end{align}
which is exactly the defining equation for an adjoint Wilson line:
\begin{equation}
    \frac{d}{dt} \W^{ab} = i g A_0^c(t) \big( - i f^{cad} \big) \W^{db} \, .
\end{equation}

It is also direct to see from here that the Wilson loop we consider in the main text satisfies
\begin{align}
    \langle \hat{\T} W[\mathcal{C}] \rangle &= \frac{1}{N_c} \langle {\rm Tr}_{\rm color} \left[ \hat{\T} U_{[t,0]} U_{[t,0]}^\dagger \right] \rangle \nonumber \\
    &= \frac{1}{N_c} \langle \hat{\T} \big( \big[U_{[t,0]} \big]_{i i_0}  \big[ U_{[t,0]}^\dagger \big]_{i_0 i} \big) \rangle \nonumber \\
    &= \frac{1}{N_c} \langle \W_{i i_0 , i_0 i} \rangle \, ,
\end{align}
and contracting the indices in the previous equations, we see that
\begin{equation}
    \frac{d}{dt} \W_{i i_0 , i_0 i} = 0 \, ,
\end{equation}
and therefore, given the initial condition $\mathcal{W}_{i i_0,j_0 j}(t=0) = \delta_{i i_0} \delta_{j j_0}$, we conclude that
\begin{align}
    \langle \hat{\T} W[\mathcal{C}] \rangle &= \frac{1}{N_c} \langle \W_{i i_0 , i_0 i}(t=0) \rangle \nonumber \\
    &= \frac{1}{N_c} \delta_{i i_0} \delta_{i_0 i} = 1 \, ,
\end{align}
as claimed in the main text.

It is worth noting that this is self-evident from the path integral formulation. Indeed, collectively denoting the field content of the theory by $\varphi$, one can calculate time-ordered (vacuum) correlation functions as
\begin{equation}
    \langle \hat{\T} {O}_1 \ldots {O}_n \rangle = \frac{1}{{Z}} \int D\varphi\, e^{i S[\varphi]} {O}_1[\varphi] \ldots {O}_n[\varphi] \, ,
\end{equation}
which means that for our pair of fundamental Wilson lines we have
\begin{align}
    \langle \hat{\T} \big( \big[U_{[t,0]} \big]_{i i_0}  \big[ U_{[t,0]}^\dagger \big]_{i_0 i} \big) \rangle &= \frac{1}{{Z}} \int D\varphi\, e^{i S[\varphi]} \big[U_{[t,0]} \big]_{i i_0}  \big[ U_{[t,0]}^\dagger \big]_{i_0 i} \nonumber \\
    &= \frac{1}{{Z}} \int D\varphi\, e^{i S[\varphi]} = 1 \, .
\end{align}
The step to the last line is achieved because, inside the path integral, the Wilson lines are just SU$(N_c)$ unitary matrices that are inverses of each other. The fact that we get one out of this is evidently consistent with our previous discussion.

\subsection{Standard products of Wilson lines}

In other contexts, it is also possible for the Wilson lines to have different operator orderings. For instance, we could consider a different kind of Wilson loop, without a time-ordering symbol:
\begin{equation}
    \langle W[\mathcal{C}] \rangle = \frac{1}{N_c} \langle U_{[t,0]} U^\dagger_{[t,0]} \rangle = 1\,,
\end{equation}
which also equals unity. The reason for that here, however, is that the operators $U_{[t,0]}$ and $U_{[t,0]}^\dagger$ are inverses of each other, and should be interpreted as written, with the operator products appearing in the same way as the color products.

Interestingly, the path integral description of this object is less simple than for the time-ordered loop. The reason for this is that inserting complete bases of states along the operator products to convert the expectation value into a path integral requires following the time contour defined by the explicit operator ordering in the correlation function. In the time-ordered case, operators are, by definition, arranged further to the left at later times, and hence it is sufficient to insert complete bases of states that span the $[0,t]$ time interval once. However, for the loop considered in this section, one has to insert complete bases of states along the interval $[0,t]$ once in the forward direction (for $U$), and once in the backward direction (for $U^\dagger$)\@.

In the context of the AdS/CFT correspondence, this type of operator ordering is realized by using the gauge/gravity duality for each segment of the path integral contour and imposing appropriate matching conditions~\cite{Skenderis:2008dg,Skenderis:2008dh,vanRees:2009rw}\@. Both the heavy quark diffusion coefficient calculation~\cite{Casalderrey-Solana:2006fio} and the jet quenching parameter claculation~\cite{DEramo:2010wup} implicitly have this feature. This is in contrast to the calculation presented herein, which does not require to match the background solution across manifolds that have different segments of the Schwinger-Keldysh contour as their boundaries.

\section{Calculation of the transverse kernel with constant $S_5$ coordinate} \label{sec:App-transverse-connected}

In this section we evaluate the correlation kernels $K_{\perp}^{AB,c}$\@. The first step is to take the limit $\xi \to 0$, as this is part of the definition of the correlator at any $L$\@. We use the notation $F_\omega^\pm$ with the understanding that its dependence on $\omega$ will often be through $\Omega = \omega/ (\pi T)$\@. We start from the expressions~\eqref{eq:Delta-yy-RL-c} and~\eqref{eq:Delta-yy-LL-c}\@.

The relevant limit corresponds to taking $z_m \to 0$, which is equivalent to taking $L\to 0$ at fixed $T$\@. Furthermore, because we have an explicit factor of $z_m^{-3}$, we have to calculate the series expansion of the rest of the expression up to order $z_m^3$, so that in the end we get a result of the form
\begin{equation}
    \frac{c_3}{L^3} + \frac{c_2}{L^2} + \frac{c_1}{L} + c_0 \, .
\end{equation}

Let us examine this term by term. We first note that we need to evaluate all terms in this expression up to cubic power in $z_m$\@. In particular, we have to evaluate
\begin{equation}
    \phi_\omega(z_m) \equiv 2 \omega z_m \int_0^1 d\xi \frac{F^+_\omega(\xi=1) F^-_\omega(\xi=1)}{F^+_\omega(\xi) F^-_\omega(\xi)} \frac{ \xi^2}{\sqrt{(1-\xi^4)(1 - (\pi T z_m \xi)^4)}}
\end{equation}
at least up to cubic order in $z_m$\@. As with the rest of this expression, this requires to solve for $F^{\pm}_\omega$ up to $\ml{O}(h^3)$, where $h = \pi T z_m$\@. However, by simple inspection, one quickly realizes that the structure of the solution, up to $\mathcal{O}(h^3)$, is of the form
\begin{equation}
    F^-_\omega(\xi) = F^{(0)}_\omega(\xi) + i \Omega h F^{(1)}_\omega(\xi) + (i \Omega h)^2 F^{(2)}_\omega(\xi) + (i \Omega h )^3 F^{(3)}_\omega(\xi) + \mathcal{O}(h^4) \, ,
\end{equation}
where $F^{(i)}_\omega(\xi)$ are real functions of $\xi$\@. This means that
\begin{equation}
    F^-_\omega(\xi) F^+_\omega(\xi) = |F^-_\omega(\xi)|^2 = \left(F_\omega^{(0)}(\xi) \right)^2 + \Omega^2 h^2 \left( \left(F_\omega^{(1)}(\xi) \right)^2 - 2 F_\omega^{(0)}(\xi) F_\omega^{(2)}(\xi) \right) + \mathcal{O}(h^4) \, ,
\end{equation}
which implies that the product $F^-_\omega(\xi) F^+_\omega(\xi)$ is an even function of $h$ (similarly for $F^+_\omega + F^-_\omega$)\@. This in turn implies that the whole object is an odd function of $h$\@. Moreover, since no terms in the power series up to $\mathcal{O}(h^3)$ involve the temperature explicitly, we have that both correlators are of the form
\begin{equation}
    \frac{c_3}{z_m^3} + \frac{c_1 \omega^2}{z_m} \, 
\end{equation}
and because of Eq.~\eqref{eq:zm-solutions}, $z_m = \frac{3 \Gamma(5/4)}{2\sqrt{\pi} \Gamma(7/4) } L + \mathcal{O}(L^5)$, with which
\begin{equation}
    D_{AB}^{yy}(\omega,L\to 0) \approx \frac{\tilde{c}_3}{L^3} + \frac{\tilde{c}_1 \omega^2}{L} \, .
\end{equation}
In terms of the time coordinate, both terms only contribute to the infinitesimal neighborhood of $t-t' = 0$, and therefore we anyways expect a divergence. The leading nontrivial dependence on $\omega/T$ appears at linear order in $L$, which is to say, from the $\mathcal{O}(h^4)$ corrections from each term. 

For future reference, note that $F_{\omega}^{(0)}$, $F_{\omega}^{(1)}$, and $F_{\omega}^{(2)}$ can be determined explicitly:
\begin{align}
    F_{\omega}^{(0)} &= 1 \, ,\\
    F_{\omega}^{(1)} &= \frac{\xi^3}{3} {}_2 F_1 \! \left( \frac12 , \frac34 , \frac74, \xi^4  \right) + \frac{\sqrt{1-\xi^4}-1}2 \, ,\\
    F_{\omega}^{(2)} &= \frac{\xi^7}{21} {}_p F_q \! \left( \left\{ 1, \frac54 \right\}, \left\{ \frac{11}4 \right\} , \xi^4 \right) + \frac{ \Gamma(5/4) \left( -3\xi^4 + 4\xi^3 -12 \xi^2 + 6 - 6\sqrt{1-\xi^4} \right)}{6 \Gamma(1/4)} \nonumber \\ & \quad + \frac{  E({\rm asin}(\xi), -1) - F({\rm asin}(\xi), -1 )  }{2 \Gamma(1/4) \Gamma(5/4) }   \Big[ \sqrt{2 \pi^3} - 4  \Gamma(5/4)^2 \left( 1 +  E({\rm asin}(\xi), -1) \right) \nonumber \\ & \quad \quad\quad \quad\quad \quad\quad \quad\quad \quad\quad \quad \quad \quad\quad \quad\quad \quad\quad \quad\quad  + 4 \Gamma(5/4)^2  F({\rm asin}(\xi), -1 )    \Big] \, ,
\end{align}
where ${\rm asin}(x) = \arcsin(x)$, $E(x,y) = {\rm EllipticE}(x,y)$, and $F(x,y) = {\rm EllipticF}(x,y)$\@.

Let us evaluate these numbers, and the first nontrivial correction from $T$-dependent effects. Let us then organize the calculation in powers of $h$, up to $\mathcal{O}(h^4)$\@. We define
\begin{equation}
    G_\omega(\xi) = |F^-_\omega(\xi)|^2 = G_\omega^{(0)}(\xi) + \Omega^2 h^2 G_\omega^{(2)}(\xi) + \Omega^4 h^4 G_\omega^{(4)}(\xi) + \mathcal{O}(h^6) \, ,
\end{equation}
where
\begin{align}
    G_\omega^{(0)}(\xi) &= \left(F^{(0)}(\xi) \right)^2 = 1 \, , \\
    G_\omega^{(2)}(\xi) &= \left(F_\omega^{(1)}(\xi) \right)^2 - 2 F_\omega^{(0)}(\xi) F_\omega^{(2)}(\xi) \, , \\
    G_\omega^{(4)}(\xi) &= \left(F_\omega^{(2)}(\xi) \right)^2 - 2 F_\omega^{(1)}(\xi) F_\omega^{(3)}(\xi) + 2 F_\omega^{(0)}(\xi) F_\omega^{(4)}(\xi) \, ,
\end{align}
For notational brevity, it is also useful to define
\begin{equation}
    H_\omega(\xi) = \frac{F^+_\omega(\xi=1) F^-_\omega(\xi=1)}{F^+_\omega(\xi) F^-_\omega(\xi)} = H_\omega^{(0)}(\xi) + \Omega^2 h^2 H_\omega^{(2)}(\xi) + \Omega^4 h^4 H_\omega^{(4)}(\xi) + \mathcal{O}(h^6) \, ,
\end{equation}
where, in terms of $G^{(n)}$, we have
\begin{align}
    H_\omega^{(0)}(\xi) &= \frac{G_\omega^{(0)}(\xi=1)}{G_\omega^{(0)}(\xi)} = 1 \, , \\
    H_\omega^{(2)}(\xi) &= G^{(2)}_\omega(\xi=1) - G^{(2)}_\omega(\xi) \, , \\
    H_\omega^{(4)}(\xi) &= \left( G^{(2)}_\omega(\xi) \right)^2 - G^{(2)}_\omega(\xi) G^{(2)}_\omega(\xi=1) - G^{(4)}_\omega(\xi) + G^{(4)}_\omega(\xi=1) \, .
\end{align}
This means that we can expand $\phi_\omega(z_m)$ as
\begin{equation}
    \phi_\omega(z_m) = h \Omega \phi_\omega^{(1)} + h^3 \Omega^3 \phi_\omega^{(3)} + h^5 \Omega^5 \phi_\omega^{(5)} \,,
\end{equation}
where
\begin{align}
    \phi_\omega^{(1)} &= 2 \int_0^1 \frac{\xi^2 d\xi}{\sqrt{1-\xi^4}} =  \frac{2 \sqrt{\pi} \, \Gamma(7/4)}{3\, \Gamma(5/4)} \, , \\
    \phi_\omega^{(3)} &= 2 \int_0^1 \frac{\xi^2 d\xi}{\sqrt{1-\xi^4}} H^{(2)}_\omega(\xi) \, , \\
    \phi_\omega^{(5)} &= 2 \int_0^1 \frac{\xi^2 d\xi}{\sqrt{1-\xi^4}} \left( \frac{\xi^4}{2 \Omega^4} + H_\omega^{(4)}(\xi) \right) \, .
\end{align}
Here $G_\omega^{(4)}(\xi)$, $H_\omega^{(4)}(\xi)$ and $\phi_\omega^{(5)}$ have a nontrivial dependence on $\omega$\@.

We can absorb the $i\epsilon$ into the definition of $\omega$ and rotate $\omega \to \omega(1+i\epsilon)$ at the end. It means we can expand the trigonometric functions in the definitions of the correlators and proceed without obstacle. However, because the structures are slightly different, we proceed separately for each correlator.

\subsection{Calculating \texorpdfstring{$K^{RL,c}_{\perp}(\omega,L)$} {} }

Recall that
\begin{align}
    K^{RL,c}_{\perp}(\omega,L) = \frac{\omega}{\sqrt{f_m}} \frac{ F_\omega^+(\xi=1) F_\omega^-(\xi=1) }{\sin(\phi_\omega(z_m)) } \, .
\end{align}
This means that, expanding up to $\mathcal{O}(h)$, we have
\begin{align}
    & K^{RL,c}_{\perp}(\omega,L) \nonumber \\ & = \frac{\omega}{\sqrt{f_m} }  \left[ \frac{1}{h \Omega \phi^{(1)}_\omega} + h \Omega \left( \frac{G^{(2)}_\omega(\xi=1)}{\phi^{(1)}_\omega} + \frac{\left(\phi^{(1)}_\omega \right)^3 - 6 \phi^{(3)}_\omega}{6 \left(\phi^{(1)}_\omega \right)^2} \right) \right. \nonumber \\
    & \quad + \left. h^3 \Omega^3 \left( \frac{ 360 G^{(4)}_\omega(\xi=1) \left(\phi^{(1)}_\omega \right)^2 + 60 G^{(2)}_\omega(\xi=1) \left(\phi^{(1)}_\omega \right)^4 + 7 \left(\phi^{(1)}_\omega \right)^6}{360 \left(\phi^{(1)}_\omega \right)^3} \right. \right. \nonumber \\
    &\quad + \left. \left. \frac{- 360 G^{(2)}_{\omega}(\xi=1) \phi^{(1)}_\omega \phi^{(3)}_\omega  + 60 \left(\phi^{(1)}_\omega \right)^3 \phi^{(3)}_\omega + 360 \left(\phi^{(3)}_\omega \right)^2 - 360 \phi^{(1)}_\omega \phi^{(5)}_\omega }{360 \left(\phi^{(1)}_\omega \right)^3 } \right) + \mathcal{O}(h^5) \right] 
\end{align}

As a function of $L$, there will be one further contribution coming from the mapping $z_m(L)$, which receives corrections of $\mathcal{O}(h^4)$ at small $h$. These will only contribute at $\mathcal{O}(L)$ from the $1/h^3$ term.

We can get the terms proportional to $1/h^3$ and $1/h$ explicitly, because we can solve for $F_\omega^-$ up to $\mathcal{O}(h^2)$ explicitly. Then, writing
\begin{equation}
    K^{RL,c}_{\perp}(\omega,L) = \frac{z_m^2}{\sqrt{f_m}} \left( \frac{c_3^{RL}}{L^3} + \frac{c_1^{RL} \omega^2}{L} + L \, \omega^4 f_{RL}(\omega/T) + \mathcal{O}(L^3) \right) \, ,
\end{equation}
we have
\begin{equation}
    c_3^{RL} = \frac{1}{\left(\frac{3\Gamma(5/4)}{2\sqrt{\pi}\Gamma(7/4)} \right)^3 \times 2 \int_0^1 \frac{\xi^2 d\xi}{\sqrt{1-\xi^4}}} = \left( \frac{2 \sqrt{\pi} \, \Gamma(7/4)}{3\, \Gamma(5/4)} \right)^2 \approx 1.43554002209 \, ,
\end{equation}
and
\begin{align}
    c_1^{RL} &= \frac{1}{\frac{3\Gamma(5/4)}{2\sqrt{\pi}\Gamma(7/4)}} \left[ \frac{1 - \frac{4\pi^3}{\left(3 \Gamma(-3/4) \Gamma(5/4)\right)^2 } }{\frac{2 \sqrt{\pi} \Gamma(7/4)}{3 \Gamma(5/4)}} + \frac16 \frac{2 \sqrt{\pi} \, \Gamma(7/4)}{3\, \Gamma(5/4)} - \frac{2 \int_0^1 \frac{\xi^2 d\xi}{\sqrt{1-\xi^4}} H^{(2)}_\omega(\xi)}{\left(  \frac{2 \sqrt{\pi} \Gamma(7/4)}{3 \Gamma(5/4)} \right)^2}  \right] \nonumber \\
    &\approx 0.49022320139 \, .
\end{align}

The last term has two new contributions that would have to be determined numerically: $G_\omega^{(4)}(\xi=1)$ and $\phi_\omega^{(5)}$\@. We could continue this process forever as well.

Now, because all of the $1/L$ divergences come from the vacuum part, we can calculate the vacuum-subtracted contribution to the correlator without expanding in powers of $z_m$ to get the $T$-dependent part. We achieve this by numerically solving for $F_\Omega$ using the methods discussed in Section~\ref{sec:EE-calculation-HQ-numerics}\@. This extraction gives
\begin{equation}
    K^{RL,c}_{\perp}(\omega,L) - K^{RL,c}_{\perp}(\omega,L)_{T=0} = (-0.20898059) \times \frac{z_m^2}{\sqrt{f_m}} \left[ (\pi T)^4 L + \mathcal{O}((\pi T L)^3) \right] \, ,
\end{equation}
with no frequency dependence at this order in $\pi T L$\@.
    
\subsection{Calculating \texorpdfstring{$K^{LL,c}_{\perp}(\omega,L)$} {} }

Let us introduce the notation $\bar{K}^{LL,c}_{\perp}= K^{LL,c}_{\perp} + \frac{\omega^2 z_m^2}{\sqrt{f_m}} \lim_{z\to 0} \frac{1}{z^2}$, so that the divergent piece from the contact term is automatically subtracted. Then
\begin{align}
    \bar{K}^{LL,c}_{\perp}(\omega,L) = \frac{\omega}{\sqrt{f_m}} \frac{ F_\Omega^+(\xi=1) F_\Omega^-(\xi=1) }{\tan(\phi_\omega(z_m)) } -  \frac{1}{4 z_m \sqrt{f_m}} \left[ \frac{\partial^3 F^-_\omega}{\partial \xi^3} + \frac{\partial^3 F^+_\omega}{\partial \xi^3} \right]_{\xi=0} \, .
\end{align}

We again proceed to expand up to $\mathcal{O}(h)$\@. The result is
\begin{align}
    & \bar{K}^{LL,c}_{\perp}(\omega,L) \nonumber \\ & = \frac{\omega}{\sqrt{f_m}}  \left[ \frac{1}{h \Omega \phi^{(1)}_\omega} + h \Omega \left( \frac{G^{(2)}_\omega(\xi=1)}{\phi^{(1)}_\omega} - \frac{\left(\phi^{(1)}_\omega \right)^3 + 3 \phi^{(3)}_\omega}{3 \left(\phi^{(1)}_\omega \right)^2} \right) \right. \nonumber \\
    & \quad + \left. h^3 \Omega^3 \left( \frac{ 45 G^{(4)}_\omega(\xi=1) \left(\phi^{(1)}_\omega \right)^2 - 15 G^{(2)}_\omega(\xi=1) \left(\phi^{(1)}_\omega \right)^4 - \left(\phi^{(1)}_\omega \right)^6}{45 \left(\phi^{(1)}_\omega \right)^3} \right. \right. \nonumber \\
    &\quad + \left. \left. \frac{- 45 G^{(2)}_{\omega}(\xi=1) \phi^{(1)}_\omega \phi^{(3)}_\omega  - 15 \left(\phi^{(1)}_\omega \right)^3 \phi^{(3)}_\omega + 45 \left(\phi^{(3)}_\omega \right)^2 - 45 \phi^{(1)}_\omega \phi^{(5)}_\omega }{45 \left(\phi^{(1)}_\omega \right)^3 } \right) + \mathcal{O}(h^5) \right] \nonumber \\
    & \quad - \frac{1}{2 z_m \sqrt{f_m}} \left[ - h^2 \Omega^2 \frac{\partial^3 F^{(2)}_\omega}{\partial \xi^3} + h^4 \Omega^4 \frac{\partial^3 F^{(4)}_\omega}{\partial \xi^3} + \mathcal{O}(h^6) \right]_{\xi=0} \, .
\end{align}
We need the same numbers and functions as before to evaluate this, and we can similarly write
\begin{equation}
    \bar{K}^{LL,c}_{\perp}(\omega,L) = \frac{z_m^2}{\sqrt{f_m}} \left( \frac{c_3^{LL}}{L^3} + \frac{c_1^{LL} \omega^2}{L} + L \, \omega^4 f_{LL}(\omega/T) + \mathcal{O}(L^3) \right) \, .
\end{equation}
The resulting numbers are again calculable. We have
\begin{equation}
    c_3^{LL} = \frac{1}{\left(\frac{3\Gamma(5/4)}{2\sqrt{\pi}\Gamma(7/4)} \right)^3 \times 2 \int_0^1 \frac{\xi^2 d\xi}{\sqrt{1-\xi^4}}} = \left( \frac{2 \sqrt{\pi} \, \Gamma(7/4)}{3\, \Gamma(5/4)} \right)^2 \approx 1.43554002209 \, ,
\end{equation}
and
\begin{align}
    c_1^{LL} &= \frac{1}{\frac{3\Gamma(5/4)}{2\sqrt{\pi}\Gamma(7/4)}} \left[ \frac{1 - \frac{4\pi^3}{\left(3 \Gamma(-3/4) \Gamma(5/4)\right)^2 } }{\frac{2 \sqrt{\pi} \Gamma(7/4)}{3 \Gamma(5/4)}} - \frac13 \frac{2 \sqrt{\pi} \, \Gamma(7/4)}{3\, \Gamma(5/4)} - \frac{2 \int_0^1 \frac{\xi^2 d\xi}{\sqrt{1-\xi^4}} H^{(2)}_\omega(\xi)}{\left(  \frac{2 \sqrt{\pi} \Gamma(7/4)}{3 \Gamma(5/4)} \right)^2}  \right] \nonumber \\ & \quad + \frac12 \frac{1}{\frac{3\Gamma(5/4)}{2\sqrt{\pi}\Gamma(7/4)}} \left[ 1 + \frac{ 2\sqrt{\pi} \Gamma(7/4) + 3(-1 + 2 E(-1) - 2K(-1) ) \Gamma(5/4)}{3 \Gamma(5/4) } \right] \nonumber \\
    &\approx 1.20799321244 \, ,
\end{align}
where $E(-1) = {\rm EllipticE}(-1)$ and $K(-1) = {\rm EllipticK}(-1)$ are Elliptic integrals.

We can calculate the vacuum-subtracted contribution to the correlator as before. The result is explicitly the same as for the $RL$ case up to leading order in $\pi T L$:
\begin{equation}
    \bar{K}^{LL,c}_{\perp}(\omega,L) - \bar{K}^{LL,c}_{\perp}(\omega,L)_{T=0} = (-0.20898059) \frac{z_m^2}{\sqrt{f_m}} \left[ (\pi T)^4 L + \mathcal{O}((\pi T L)^3) \right] \, .
\end{equation}
Higher order terms in $L$ may differ because the $RL$ and $LL$ response functions are only guaranteed to agree up to contact terms in the limit $L \to 0$\@.
    
\section{Analytic aspects of the correlation functions}

\subsection{Expansion of \texorpdfstring{$F^{-}_\Omega$} {} in powers of \texorpdfstring{$\Omega$} {}} \label{sec:App-omega-expansion}

Consider the defining equation for $F^-_\omega(\xi)$, given by Eq.~\eqref{eq:F-thermal}:
\begin{align} \label{eq:F-thermal-app}
    \frac{\partial^2 F^-_\omega}{\partial \xi^2} - 2 \left[ \frac{1 + \xi^4}{\xi(1-\xi^4)} - \frac{i \Omega \xi^3}{1-\xi^4} \right] \frac{\partial F^-_\omega}{\partial \xi} + \left[  \frac{i \Omega \xi^2}{1-\xi^4} + \frac{\Omega^2 (1 - \xi^6) }{(1-\xi^4)^2} \right] F^-_\omega = 0 \, ,
\end{align}
and instead of attempting to find a solution for arbitrary $\Omega$, let us expand the solution in powers of $\Omega = \frac{\omega}{\pi T}$\@. To that end, we write
\begin{equation}
    F_\omega^-(\xi) = F^{(0)}(\xi) + i\Omega F^{(1)}(\xi) + (i\Omega)^2 F^{(2)}(\xi) + \mathcal{O}(\Omega^3) \, ,
\end{equation}
and solve Eq.~\eqref{eq:F-thermal-app} order by order in $\Omega$ with boundary conditions determined by
\begin{align}
    F^-_\omega(0) = 1 \, \, , & & \frac{\partial_\xi F_\omega^-(1)}{F_\omega^-(1)} = \frac{i \Omega}{4} \frac{1 - \frac{3i\Omega}{2}}{1 - \frac{i\Omega}{2}} \, .
\end{align}
The solutions can then be found order by order:
\begin{align}
    F^{(0)}(\xi) &= 1 \, , \\
    F^{(1)}(\xi) &= \frac{1}{4} \left[ \ln \left( (1+\xi)^2 (1+\xi^2) \right) - 2 \arctan(\xi) \right] \, , \\
    F^{(2)}(\xi) &= \frac{1}{8} \left[ \left( \arctan(\xi) - \ln(1+\xi) - 4 \right) \left( \arctan(\xi) - \ln(1+\xi) \right) \right. \nonumber \\ & \quad \quad \left. - \left( \arctan(\xi) - \ln(1+\xi) + 2 \right) \ln(1+\xi^2) + \frac{1}{4} \left(\ln(1+\xi^2) \right)^2 \right] \, ,
\end{align}
and with them, one can easily evaluate the input needed for the correlation function:
\begin{equation}
    \frac{-i}{F^-_{|\omega|}(0)} \frac{\partial^3 F^-_{|\omega|}}{\partial \xi^3}(0) = 2 |\Omega| + 2 i \Omega^2 + \mathcal{O}(\Omega^3) \, .
\end{equation}
Higher order terms may be obtained by solving the differential equation~\eqref{eq:F-thermal-app} up to higher powers of $\Omega$\@.

\subsection{Consequences of the pole positions of the time-ordered correlator} \label{sec:App-analytic-rot}

In this section we shall prove that $[g^\T_E]$ satisfies Eq.~\eqref{eq:analytic-prop-gE}\@.

Up to overall factors, and setting the normalization $F^\pm_\omega(\xi=0) = 1$, the time-ordered correlator we obtained is given by
\begin{equation}
    G(\omega) = -i \frac{\partial^3 F^-_{|\omega|} }{\partial \xi^3}\bigg|_{\xi=0} \, ,
\end{equation}
which we obtained by shifting $\omega \to \omega(1 + i\epsilon)$, which is essentially a prescription to avoid potential poles along the real $\omega$ axis.

In order to prove that $[g^\T_E]$ satisfies Eq.~\eqref{eq:analytic-prop-gE}, we will take a seemingly disconnected starting point, which nonetheless will allow us to prove our claim. To begin, consider the integral
\begin{equation}
    I(\omega) = \int_0^\infty d p_0 \frac{2 \omega G(p_0) }{p_0^2 - \omega^2 + i\epsilon} \, .
\end{equation}
Specifically, we will prove that ${\rm Im} \left\{ I(\omega) \right\} = 0 $\@.

Note that this integral only involves $\omega > 0$\@. Then, because we constructed $G(\omega)$ by shifting potential poles on the positive real axis towards the lower half of the complex plane, there is no obstruction to Wick-rotate the integral onto the positive imaginary axis. This is possible because $F_\omega^-$ itself is an analytic function, provided we handle the potential UV divergences properly. We will deal with the potential large $\omega$ divergences in the next subsection.

After doing the Wick rotation, we get
\begin{equation}
    I(\omega) = - i \int_0^\infty dp_E \frac{2 \omega G(i p_E) }{p_E^2 + \omega^2 } \, .
\end{equation}
Then, by observing that $G(i p_E) = - i \frac{\partial^3 F^-_{i p_E} }{\partial \xi^3}(0) $, and inspecting Eq.~\eqref{eq:F-thermal}, we see that $F^-_{i p_E}$ is a real function (it solves a differential equation with real coefficients and real boundary conditions)\@. Therefore, $I(\omega)$ is a real function, and hence
\begin{equation}
    {\rm Im} \left\{ I(\omega) \right\} = 0 \, .
\end{equation}
It is now straightforward to manipulate this expression into what we want to prove:
\begin{align}
    {\rm Im} \left\{ I(\omega) \right\} &= {\rm Im} \left\{ \int_{-\infty}^\infty \frac{G(p_0)  \, \diff p_0}{p_0 - \omega (1 - i\epsilon) } \right\} \nonumber \\
    &=   \int_{-\infty}^\infty \diff p_0 \left[ \frac{{\rm Im} \left\{ G(\omega) \right\}}{p_0 - \omega} - \pi {\rm sgn}(\omega) \delta(\omega - p_0)  {\rm Re} \left\{ G(p_0) \right\}  \right] \, ,
\end{align}
with which
\begin{align}
    {\rm Im} \left\{ I(\omega) \right\} = 0 \quad \implies \quad {\rm sgn}(\omega) {\rm Re} \left\{ G(\omega) \right\} = \int_{-\infty}^\infty \frac{\diff p_0}{\pi p_0}  {\rm Im} \left\{ G(\omega + p_0) \right\} \, . \label{eq:condition-C6}
\end{align}
    
We have also verified this relation numerically for the vacuum-subtracted correlation functions (i.e., for $\Delta G(\omega) = G(\omega) - G(\omega)_{T = 0}$), where all integrals are convergent. For the vacuum part, where the integrals are UV-divergent because of the $\omega^3$ power-law behavior, we present a proof in the next subsection.

\subsection{UV divergent pieces in the Wick rotation}

To be sure that we have the correct expression for all contributions in the above, we may work out the contributions proportional to $\omega^3$ in the correlator independently. Because the integrals are divergent, and we do not have a natural regulator that respects all of the AdS symmetries, we will use Lorentz covariance of the boundary theory and the fact that $\mathcal{N}=4$ SYM is a conformal field theory (CFT)\@. Once firmly on the side of the boundary theory, we may use all of the standard dimensional regularization machinery to calculate the integrals.

At $T=0$ we have restored Lorentz covariance of the boundary theory, and therefore we can obtain the same correlation function but with the Wilson lines at an angle with the time axis by applying boosts. This, plus the fact that the theory is a CFT means that $G(\omega) \propto |\omega|^3 $ may be derived by integrating a momentum-space two-point function of a massless particle:
\begin{equation}
    G(\omega) = \# \int \frac{d^3 {k}}{(2\pi)^3} \frac{i \omega^2 }{\omega^2 - {\bf k}^2 + i\epsilon} \, .
\end{equation}
Now, we simply verify the right hand side of Eq.~\eqref{eq:condition-C6} mode by mode, i.e.,
\begin{equation}
    {\rm sgn}(\omega) {\rm Re} \left\{ \frac{i \omega^2 }{\omega^2 - {\bf k}^2 + i\epsilon} \right\} = \int_{-\infty}^\infty \frac{\diff p_0}{\pi p_0} {\rm Im} \left\{ \frac{i (\omega + p_0)^2 }{(\omega + p_0)^2 - {\bf k}^2 + i\epsilon} \right\} \, .
\end{equation}
This identity is indeed satisfied, because we may write the numerator of the integrand on the right hand side as $(\omega + p_0)^2 = (\omega + p_0)^2 - {\bf k}^2 + {\bf k}^2$: the first two terms then cancel the denominator, leaving their contribution as the Cauchy principal value integral of $1/p_0$, which vanishes. The last term gives a contribution that can be cast in the form of a Dirac delta function by means of
\begin{equation}
    \int_{-\infty}^\infty dx\,\mathcal{P} \left( \frac{1}{(x-1) (x^2 - a^2)} \right) = \frac{\pi^2}{2} \delta(|a| - 1) \, ,
\end{equation}
and the left hand side may be immediately seen to be proportional to a Dirac delta function $\delta(\omega^2 - {\bf k}^2)$\@. Verifying that the coefficients match is straightforward.

Therefore, the zero-temperature piece of $[g^\T_E]$ satisfies Eq.~\eqref{eq:analytic-prop-gE}, as does the thermal contribution. Thus, we have verified the claim presented in the main text.

\bibliographystyle{jhep}
\bibliography{main.bib}

\end{document}